\documentclass[journal]{IEEEtran}
\usepackage{booktabs} % For formal tables
\usepackage{tabularx}
\usepackage{color}
\usepackage[strict]{changepage}
\usepackage{calc}
\usepackage{mathtools,lipsum}
\usepackage{amsmath}
\usepackage{amssymb}
\usepackage{mathrsfs}
\usepackage{upgreek}
\usepackage{graphicx}
\usepackage{algorithmicx}
\usepackage{setspace}
\usepackage{algorithm}
\usepackage{enumitem}
\usepackage[noend]{algpseudocode}
\usepackage{caption}
\usepackage{subcaption}
\usepackage{amsthm, times, amsfonts, cite}

\newtheorem{theorem}{Theorem}
\newtheorem{lemma}{Lemma}
\newtheorem{corollary}{Corollary}

\usepackage[T1]{fontenc}
\usepackage[latin9]{inputenc}
\usepackage{array}
\usepackage{float}
\usepackage{amsmath}
\usepackage{amssymb}
\usepackage{graphicx}

\usepackage{amsmath}

\begin{document}

\title{
Multi-tier Caching Analysis in CDN-based Over-the-top Video Streaming Systems}

\author{Abubakr O. Al-Abbasi, Vaneet Aggarwal and  Moo-Ryong Ra \thanks{A. O. Al-Abbasi and V. Aggarwal are with Purdue University, West Lafayette, IN 47907, USA, email: \{aalabbas,vaneet\}@purdue.edu. M. Ra is with AT\&T Research Labs, Bedminster, NJ 07921, USA, email: mra@research.att.com. 
This work has been presented in part at the IEEE Infocom Workshop, Apr 2018 \cite{edgeCacheINFWKSHP017}.

}}

	\maketitle

\begin{abstract}
Internet video traffic has been been rapidly increasing and  is further expected to increase with the emerging 5G applications such as higher definition videos, IoT and augmented/virtual reality applications. As end-users consume video in massive amounts and in an increasing number of ways, the content distribution network (CDN) should be efficiently managed to improve the system efficiency. The streaming service can include multiple caching tiers, at the distributed servers and the edge routers, and efficient content management at these locations affect the quality of experience (QoE) of the end users. 

In this paper, we propose a model for video streaming systems, typically composed of a centralized origin server, several CDN sites, and edge-caches located closer to the end user. We comprehensively consider different systems design factors including the limited caching space at the CDN sites, allocation of CDN for a video request, choice of different ports (or paths) from the CDN and the central storage, bandwidth allocation, the edge-cache capacity, and the caching policy. We focus on minimizing a performance metric, stall duration tail probability (SDTP), and present a novel and efficient algorithm accounting for the multiple design flexibilities. The theoretical bounds with respect to the SDTP metric are also analyzed and presented. The implementation on a virtualized cloud system managed by Openstack demonstrate that the proposed algorithms can significantly improve the SDTP metric, compared to the baseline strategies. %We further implement an online version of our proposed algorithm on small-scale video streaming system in a real cloud environment to validate our results.

%However, current CDN infrastructures are designed such that origin storage servers are not well protected from bursty traffic, due to unicast configuration. To avoid the so-called thundering herd problem, where many multiple clients request very hot content at the same time from various locations, multi-cast can be enabled to reduce the traffic at the backbone stoage nodes and improve the total system efficiency.
%We take one step further and implement a small-scale video streaming systems 
%in a real cloud environment and validate our results.
\end{abstract}

%\begin{IEEEkeywords}
%	Video Streaming,  Distributed Storage Systems, Content Distribution Network, Caching, Two-stage Probabilistic Scheduling, Bandwidth Allocation. \end{IEEEkeywords}

\section{Introduction}
\label{sec:Intro}

Over-the-top video streaming, e.g., Netflix and YouTube, has been dominating the global IP traffic in recent years. The traffic will continue to grow due to the introduction of even higher resolution video formats such as 4K on the horizon. As end-users consume video in massive amounts and in an increasing number of ways, service providers need flexible solutions in place to ensure that they can deliver content quickly and easily regardless of their customer's device or location. More than $50\%$ of over-the-top video traffic are now delivered through content distribution networks (CDNs) \cite{rep62}. Even though multiple solutions have been proposed for improving congestion in the CDN system, managing the ever-increasing traffic requires a fundamental understanding of the system and the different design flexibilities (control knobs) to make the best use of the limited hardware resources. This is the focus of this paper.

%One of the approach that is used is caching of the video content. This can be enabled using content distribution networks (CDNs). 

%CDNs enable caching of the video content at distributed servers. 

%This paper proposes a CDN-based video streaming system with multiple caching tiers, that optimizes user QoE metrics. 

% It is shown in \cite{rep62} that video streaming applications in North America now represents $62\%$ of the Internet traffic, and this figure will continue to grow due to the introduction of even higher resolution video formats such as 4K on the horizon. With the growing popularity of video services, increased congestion and latency related to retrieving content from remote datacenters can lead to degraded end customer experience. Service and content providers often seek to mitigate such performance issues by employing caching at the network edge and by pushing content closer to their customers using content distribution networks (CDNs). More than $50\%$ of over-the-top video traffic are now delivered through CDNs \cite{rep62}.

%[T1] Netflix. Netflix OpenConnect Appliance Deployment Guide, April 2015. Version 3.7. Version 3.7.
%[T2] Cache Content-Selection Policies for Streaming Video Services, INFOCOM 2016

The service providers typically use two-tier caching approach to improve the quality of streaming services \cite{li2017collaborative33,rosario2018service,perez2011elastic}. In addition to the  distributed cache servers provided by the CDN,  the edge router can also have a cache so that some videos could be stored in this cache and gets the advantage of the proximity to end-users.
%
% The key idea of this cache is that portion of the video content could be stored in caches closer to clients, since the edge routers are closer to the clients and, hence, they give better advantage.
However, there are many edge routers which imply that the hot content could be stored at multiple edge routers. There is an additional cache at the distributed cache servers (in CDN) from which data can be obtained if not already at the edge router.
%
%\textcolor{red}{ The number of distributed cache servers are fewer than the number of edge routers, thus aiding the redundancy of content and efficient utilization of storage. \textbf{?????}}
Such multi-tier caching is related to fog computing where the caching could be distributed at multiple locations in the network \cite{rosario2018service}. We also assume that the edge cache can help provide advantages similar to multicasting. If another user on the edge router is already consuming the file, the part of the video already downloaded is sent directly to the new user and the later part is sent to multiple users who requested the content on the same edge router. This paper aims to analyze two-tiered caching in video streaming systems.

%Caching of video content has to address a number of crucial challenges that differ from caching of web objects, see for instance \cite{7524619} and the references therein. First, video streaming services such as Netflix \cite{T1} often adopt a proactive caching strategy, which consciously pushes video files into local caches during off-peak hours, while cache content is updated according to changes in predicted demand. Due to the correlation in user preferences within different regions \cite{7524619}, it calls for new solutions that take into account both regional and global popularity of video files, for jointly optimizing cache content and performance. Second, video files are significantly larger in size than web objects. In order to minimize congestion and latency, caching of video files must be optimized together with network resource allocation and request scheduling, which however, is currently under-explored. Finally, while recent work have considered video-streaming over distributed storage systems \cite{7524619}, they normally focus on network performance metrics similar to those considered by web object caching (e.g., packet delay and cache hit rate), rather than Quality-of-Experience (QoE) metrics that are more relevant to end user experience in over-the-top video streaming. 

In this paper, we consider an architecture of streaming system with a Virtualized Content Distribution Network (vCDN) \cite{8109330,7936595}.  The main role of this CDN infrastructure is not only to provide users with lower response  time and higher bandwidth,  but also to distribute the load (especially during peak time) 
across many edge locations.  The infrastructure consists of a remote datacenter that stores complete original video data and multiple CDN sites (i.e., distributed cache servers) that only have part of those data and are equipped with solid state drives (SSDs) for high throughput. In addition, we assume that a second caching tier is located at the edge routers. A user request for video content not served from the edge cache is directed to a distributed cache. If it is still not completely served, the remaining part of the request is directed to the remote datacenter (as shown in  Fig. \ref{fig:sysModel}). Multiple parallel connections are established between the distributed cache server and the edge router, as well as between the distributed cache servers and the origin server, to support multiple video streams simultaneously. Our goal is to develop an optimization framework and QoE metrics that service providers (or infrastructure) could use to answer the following questions: How to quantify the impact of multi-tier video caching on end user experience? What is the best video multi-tier caching strategy for CDN? How to optimize QoE metrics over various ``control knobs"? Are there enough benefits to justify the adoption of proposed solutions in practice?

It has been shown that, in modern cloud applications such as Facebook, Bing, and Amazon's retail platform, the long tail of latency is of a major concern, with $99.9$th percentile response times that are, orders of magnitude worse than the mean \cite{T1,T2}. Thus, this paper considers a QoE metric, called the stall duration tail probability (SDTP), which measures the likelihood of end users suffering a worse-than-expected stall duration, and develop a holistic optimization framework for minimizing the overall SDTP over joint caching content placement, network resource optimization and user request scheduling.  SDTP, denoted by Pr$(\Gamma^{(i)}>\sigma)$,  measures the probability that the stall duration $\Gamma^{(i)}$ of video $i$ is greater than a pre-defined threshold $\sigma$. Despite resource and load-balancing mechanisms, large scale storage systems evaluations show that there is a high degree of randomness in delay performance \cite{JK13}. In contrast to web object caching and delivery, the video chunks in the latter part of a video do not have to be downloaded much earlier than their actual play time to maintain the desired QoE, making SDTP highly dependent on the joint optimization with resource management and request scheduling in CDN-based video streaming. %hence multiple parallel streams help achieve better QoE. 
%Hence, the QoE metric of SDTP becomes crucial. 

Quantifying SDTP with multi-tier cache/storage is an open problem. Even for single-chunk video files, the problem is equivalent to minimizing the download tail latency, which is still an open problem \cite{MDS-Queue}. The key challenge arises from the difficulty of constructing and analyzing a scheduling policy that (optimally) redirects each request based on {\em dependent system and queueing dynamics} (including cache content, network conditions, request queue status) on the fly. To overcome these challenges, we propose a novel two-stage, probabilistic scheduling approach, where each request of video $i$ is (i) processed by cache server $j$ with probability $\pi_{i,j}$ and (ii) assigned to video stream $v$ with probability $p_{i,j,v}$. The two-stage, probability scheduling allows us to model each cache server and video stream as separate queues, and thus, to characterize the distributions of different video chunks' download time and playback time. Further, the edge caching policy plays a key role in the system design. This paper proposes an adaption of least-recently-used (LRU) caching mechanism \cite{cacheChes02}, where each file is removed from the edge cache if it has not been requested again for a time that depends on the edge router and the file index.  By optimizing these probabilities and the edge-cache parameters, we quantify SDTP through a closed-form, tight upper bound for CDN-based video streaming with arbitrary cache content placement and network resource allocation. We note that the analysis in this paper is fundamentally different from those for distributed file storage, e.g., \cite{Xiang:2014:Sigmetrics:2014,Yu_TON}, because the stall duration of a video relies on the download times of all its chunks, rather than simply the time to download the last chunk of a file. Further, since video chunks are downloaded and played sequentially, the download times and playback times of different video chunks are highly correlated and thus jointly determine the SDTP metric.

This paper proposes a holistic optimization framework for minimizing overall SDTP in CDN-based video streaming. To the best of our knowledge, this is the first framework for multi-tier caching to jointly consider all key design degrees of freedom, including bandwidth allocation among different parallel streams, multi-tier cache content placement and update, request scheduling, and the modeling variables associated with the SDTP bound. An efficient algorithm is then proposed to solve this non-convex optimization problem. In particular, the proposed algorithm performs an alternating optimization over the different dimensions, such that each sub-problem is shown to have convex constraints and thus can be efficiently solved using the iNner cOnVex
Approximation (NOVA)  algorithm proposed in \cite{scutNOVA}. 
%The proposed algorithm is shown to converge to a local optimal. 
The proposed algorithm is implemented in a virtualized cloud system managed by Openstack~\cite{openstack}. The experimental results demonstrate significant improvement of QoE metrics as compared to the considered baselines.

%{\color{blue} Talk about our holistic optimization framework, different decision variables involved, our optimization algorithm. Highlight the novelty of this optimization by comparing it with prior work.} To the best of our knowledge, this is the first paper to jointly consider all these degrees of freedom for QoE optimization in video caching and SDN.

%{\color{blue} Discuss implementation details, simulation results, and key insights learned.} 

The main contributions of this paper can be summarized as follows:
%\begin{itemize}
%\item We propose SDTP and quantify it for CDN ...
%\item A holistic optimization framework is developed to ...
%\item We implement the proposed solutions ...
%\end{itemize}

\begin{itemize}[leftmargin=0cm,itemindent=.3cm,labelwidth=\itemindent,labelsep=0cm,align=left]
\item We propose a novel framework for analyzing CDN-based over-the top video streaming systems with the use of multiple caching tiers and multiple parallel streams between nodes. A novel two-stage  probabilistic scheduling policy is proposed  to assign each user request to different cache servers and parallel video streams. Further, the edge router uses an adaptation of LRU, and the distributed cache servers cache partial files. 

%analyze the bounds on SDTP for the CDN-based over-the top video streaming systems through a new two-stage probabilistic scheduling policy for assigning each user request to different cache servers and parallel video streams. 
%the choice of server and the parallel streams. Further, the cache placement is optimized. 

\item The distribution of (random) download time of different video chunks are analyzed. Then, using ordered statistics, we quantify the playback time of each video segment. %and derive an analytical upper bound on SDTP for arbitrary cache content placement and the parameters of the two-stage probabilistic scheduling.

\item Multiple recursive relations are set up to compute the stall duration tail probability. We first relate the compute of download time of each chunk to the play time of each chunk, Since the play time depends not only on download of the current chunk, but also on the previous chunks. Second, stall duration must account for whether the file has been requested by anyone within a window time of a certain size to get advantage of edge cache. If it had been requested, the stall duration is a function of the time of the last request and the stall duration at that time. In the steady state analysis, this will lead to a recursion. This analysis has been used to derive an analytical upper bound on SDTP for arbitrary distributed cache content placement, parameters of edge cache, and the parameters of the two-stage probabilistic scheduling (Appendix \ref{sec:DownLoadPlay}).

%the random variable corresponding to the playback time of each video segment is characterized. This is further used to give a bound on the SDTP. 

\item  A holistic optimization framework is developed to optimize a weighted sum of SDTP of all video files over the request scheduling probabilities, distributed cache content placement, the bandwidth allocation among different streams, edge cache parameters, and the modeling parameters in SDTP bound. An efficient algorithm is provided to decouple and solve this non-convex optimization (Section \ref{sec:SDTP}).
\item To better understand the SDTP and how it relates to the QoE of users, we correlate this metric to a well-known QoE metric (called mean stall duration). Since the optimal point for the mean stall duration is not the same as that of the SDTP, we optimize a convex combination of the two metrics and show how the two QoE metrics can be compromised based on the point on the curve that is appropriate for the clients  (Appendix \ref{sec:mean}).

\item 
The algorithm is implemented on a virtualized cloud system managed by Openstack. The simulation and trace-based results validate our theoretical analysis with the implementation and analytical results being close, thus demonstrating the efficacy of our proposed
algorithm. The QoE metric is shown to have significant improvement as compared to competitive  strategies (Appendix \ref{sec:num}).
\end{itemize}

The rest of this paper is organized as follows. Section \ref{sec:RelWork}
provides related work for this paper. In Section \ref{sec:SysMod}, we describe
the system model used in the paper with a description of
CDN-based Over-the-top video streaming systems. Section  \ref{sec:StallTailProb}  provides an upper bound on the mean stall duration. Section \ref{sec:SDTP} formulates the  QoE optimization problem as a weighted sum of all SDTP of all files and proposes the iterative algorithmic solution of this problem.   Experimental results are provided in Section \ref{sec:num}. Section \ref{sec:conc} concludes the paper.

%Section \ref{sec:DownLoadPlay} derives expressions for the download and play times of the chunks from storage cache servers as well as origin datacenter, which are used in  Section \ref{sec:StallTailProb}  to find an upper bound on the mean stall duration

%
% {\bf Abubakr, 1. Revise the organization para, 2. Since the Intro is very big, figure out what of the old stuff could be removed?}
\section{Related Work}\label{sec:RelWork}

Video on Demand services and Live TV Content from cloud servers have been studied widely \cite{lee2013vod,Aggarwal:2011:Infocom,Aggarwal:multimedia:2013,jointContent,Hu2016}. 
The placement of content and resource optimization over the cloud servers have been considered. In \cite{jointContent}, authors utilize the social
information propagation pattern to improve the efficiency of
social video distribution. Further, they used replication and user request dispatching mechanism in the cloud content
delivery network architecture to reduce the system operational
cost, while maintaining the averaged service latency. However, \cite{jointContent}  only considers video download. The benefits of delivering videos at the edge network is shown in \cite{Hu2016}. Authors show that bringing videos at the edge network can significantly improve the content item delivery performance, in terms of improving quality experienced by
users as well as reducing content item delivery costs.
To the best of our knowledge, reliability of content over the cloud servers have not been considered for video streaming applications. There are novel challenges to characterize and optimize the QoE metrics at the end user. Adaptive streaming algorithms have also been considered for video streaming \cite{chen2012amvsc,wang2013ames,Garcia016} \cite{7939148} which are beyond the scope of this paper and are left for future work. 

Mean latency and tail latency have been characterized in \cite{Xiang:2014:Sigmetrics:2014,Yu_TON,al2018mean} and \cite{Jingxian,Tail_TON}, respectively, for a system with multiple files using probabilistic scheduling. However, these papers consider only file downloading rather than video streaming.  This paper considers CDN-based video streaming. We note that file downloading can follow as a special case of streaming, which makes our model more general. Additionally,  the metrics for video streaming do not only account for the end of the download of the video but also for the download of each segment. Hence, the analysis for the content download cannot be extended to the video streaming directly and the analysis approach in this paper is very different from the prior work in the area of file downloading.

More recently, the authors of \cite{Abubakr_TON} considered video-streaming over distributed storage systems. However, caching placement optimization and is not considered. Also, caching at the edge level is not considered. Moreover, only a single stream between each storage server and edge node is assumed and hence neither the two-stage probabilistic scheduling nor bandwidth allocation were considered. Similarly, there is no edge cache in \cite{alabbasi2018fasttrack}. Thus, the analysis and the problem formulation here is an extension of that in \cite{Abubakr_TON,alabbasi2018fasttrack}. 

% In \cite{edgeCacheINFWKSHP017}, cloud-based cache-aided video streaming was studied. However, neither online algorithm nor edge-cache parameter optimization were considered. Further, implementation on real cloud environment was not conducted.  

\section{System Model}
\label{sec:SysMod}
\subsection{Target System}

Our work is motivated by the architecture of a production system with a Virtualized Content Distribution Network (vCDN).
Such services, for instance, include video-on-demand (VoD), live linear streaming services 
(also referred to as over-the-top video streaming services), 
firmware over the air (FOTA) Android updates to mobile devices, etc.  
The main role of this CDN infrastructure is not only to provide users with lower response 
time and higher bandwidth,  but also to distribute the load (especially during peak time) 
across many edge locations. Consequently, the core backbone network will have 
reduced network load and better response time. 
The origin server has original data and CDN sites have only part of those data. Each CDN site is composed of multiple cache servers each of which is typically implemented as a VM backed by 
multiple directly attached solid state drives (SSDs) for higher throughput. The cache servers store video segments and a typical duration of each segment 
covers $5-11$ seconds of playback time.

Further, the typical vCDN architecture includes an additional cache at the edge, called edge cache. This edge cache allows for saving some recently accessed videos. This cache can also help multicasting content to another user connected to the same edge router. One of the typical policy that is used in edge cache is based on least-recently-used (LRU) caching policy \cite{cacheChes02}. In this paper, we will consider a modification of this strategy to weigh the eviction policy of contents dependent on their weight, placement, and access rates and thus can be optimized.

When a client such as VoD/LiveTV app requests a certain content, 
it goes through multiple steps. First, it sees whether the content is in edge cache. If so, the content is directly accessed from the edge cache. Then, it sees whether the content has been requested by someone connected to the same edge router and is being sent to them. In this case, the content already received at the edge router is sent to the user and the remaining content is passed as received (equivalent to a multicast stream setup). If the content cannot be obtained in the two steps, the client then contacts \emph{CDN manager}, 
choose the best CDN service
to use and retrieve a fully qualified domain name (FQDN). 
Fourth, with the acquired FQDN, it gets a cache server's IP address from 
a content routing service (called iDNS). Then we use the IP address to connect 
to one of the cache servers. %\footnote{While CDN system architecture can have two layers (or more) of cache servers, i.e., smaller number of parent cache servers and more number of child cache servers, in this paper, we abstract our infrastructure in a  per-site basis without loss of generality.}
The cache server will directly serve the incoming request if it has data in its local 
storage (cache-hit). If the requested content is not on the cache server (i.e., cache-miss), 
the cache server will fetch the content from the origin server and then serve 
the client. %\footnote{Some popular contents can be prefetched from the origin server a priori but we did not consider this aspect in this paper.}. 
%Further, we assume that  popular contents are stored in the edge-cache for some time, that can be optimized, and hence future requests will enjoy better QoE metrics. Thus, if these popular files are requested again with a window size of a certain length, the requests for these files are directly served from the edge-cache.

In the rest of this section, we will present a generic mathematical model  applicable to not only our vCDN system but also other video streaming systems 
that implement CDN-like two-tier caching structure. %-tier closer to the users.
%~\cite{
%youtube - https://www.youtube.com/watch?v=w5WVu624fY8, fb-haystack-osdi10}. 

\subsection{System Description}

\begin{figure}[t]
\centering\includegraphics[scale=0.25]{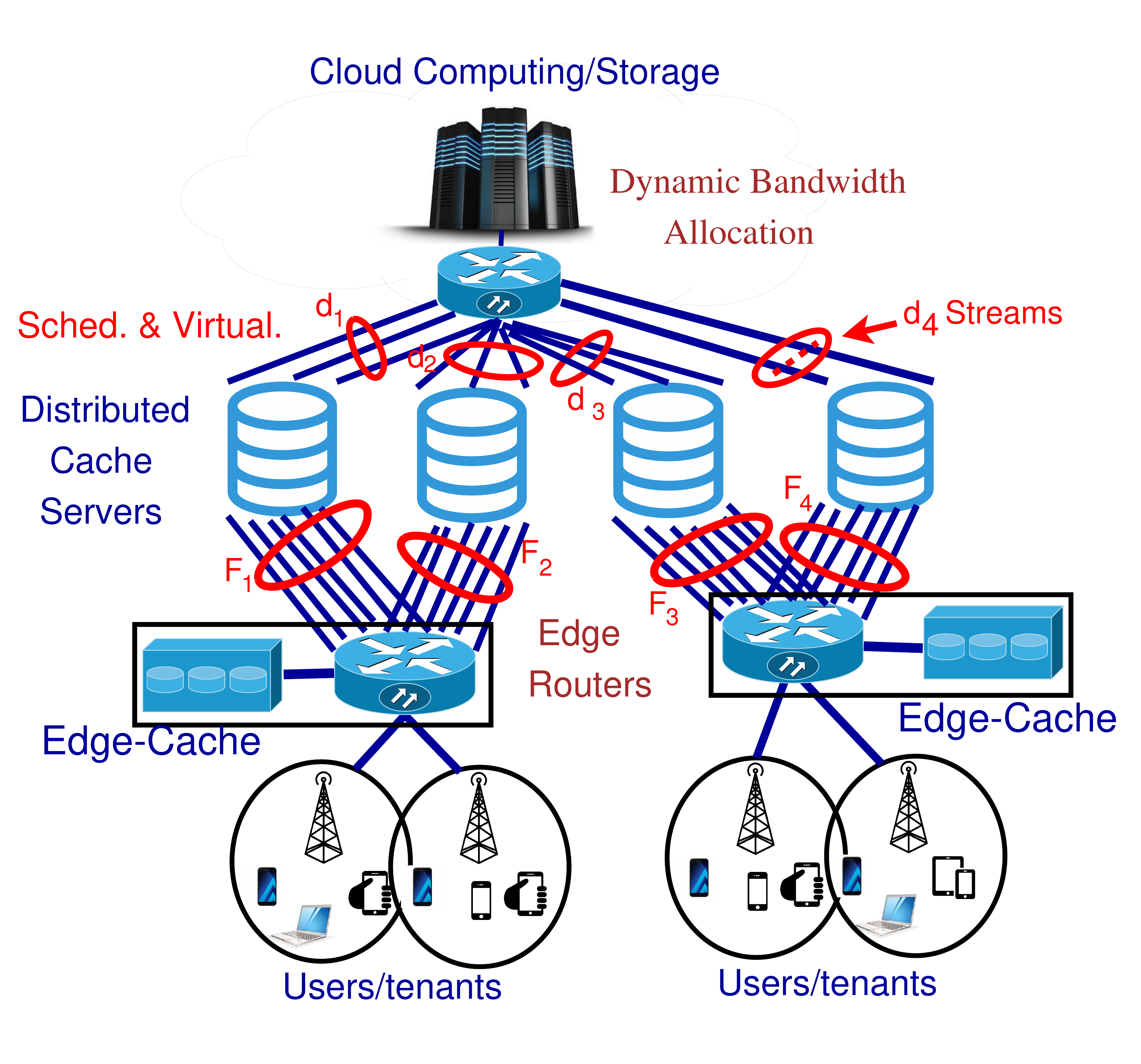}
%\vspace{-.4in}
\caption{An illustration of our system model for video content delivery, consisting
of a datacenter, four cache servers ($m=4$), and $2$ edge routers. $d_j$ and $F_j$ parallel connections are assumed between datacenter and cache server $j$, and datacenter and edge router, respectively. 
\label{fig:sysModel}}
%\vspace{-.20in}
\end{figure}

\begin{figure}[t]
	\centering\includegraphics[trim=0.035in 0.15in 0.25in 0.051in, clip,width=0.481\textwidth]{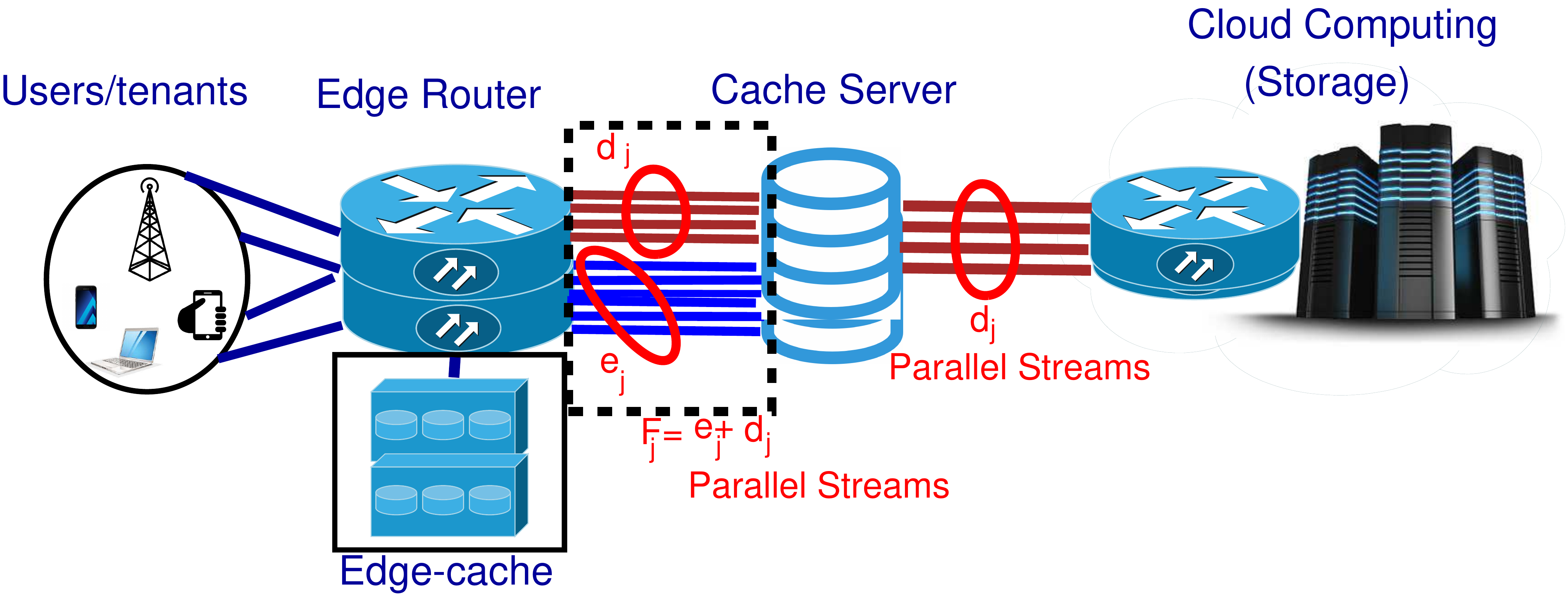}
%	\centering\includegraphics[scale=0.20]{sysModelPllstreams}
	%\vspace{-.4in}
	\caption{ A schematic illustrates the parallel streams setup between the different system model components.
		\label{PSsetup}}
	%\vspace{-.20in}
\end{figure}

We consider a content delivery network as shown in Fig. \ref{fig:sysModel},
consisting of a single datacenter that has an origin server, 
$m$ geographically-distributed cache servers denoted by $j=1,\ldots,m$,  and edge-cache storage nodes associated with the edge routers $\ell \in \{1,2,\cdots, R\}$, where $R$ is the total number of edge-routers, as depicted in Figure \ref{PSsetup}.    The compute cache servers (also called storage nodes) are located close to the edge of the network and thus provide lower access latency for end users.  We also assume that each cache server is connected to one edge router.
%Since there is no sharing of cache servers, we only consider one edge router in this work. 
Further, the connection from the edge router to the users is not considered as a bottleneck. Thus, the edge router is considered as a combination of users and is the last hop for our analysis. We also note that the link from the edge router to the end users is not controlled by the service provider and thus cannot be considered for optimized resource allocation from the network. The service provider wishes to optimize the links it controls for efficient quality of experience to the end user.

A set of $r$ video files (denoted by $i=1,\ldots,r$) are stored
 in the datacenter, where video file $i$ is divided into $L_{i}$ equal-size segments each of length $\tau$ seconds. We assume that the first $L_{j,i}$ chunks of video $i$ are stored on cache server $j$. Even though we consider a fixed cache placement, we note that $L_{j,i}$ are optimization variables and can be updated when sufficient arrival rate change is detected.

We assume that the bandwidth between the data center and the cache server $j$ (service edge router $\ell$) is split into $d_{j}$ parallel streams, where the streams are  denoted as $PS^{(d,j)}_{\beta_j,\ell}$ for $\beta_j = 1, \cdots, d_j$. Further,  the bandwidth between the cache server $j$ and the edge router $\ell$ is divided into $f_j^{(\ell)}$ parallel streams, denoted as $PS^{(f,j)}_{\zeta_j,\ell}$ for $\zeta_j^{(\ell)} = 1, \cdots, f_j^{(\ell)}$, and $\ell=1,2,\cdots,R$. Multiple parallel streams are assumed for video streaming since multiple video downloads can happen simultaneously. Since we care about stall duration, obtaining multiple videos simultaneously is helpful as the stall durations of multiple videos can be improved. We further assume that $f_j^{(\ell)}$ PSs are divided into two set of streams $d_j$ and $e_j^{(\ell)}$. This setup is captured in Figure \ref{PSsetup}. The first $d_j$ parallel streams are denoted as $PS^{(\overline{d},j)}_{\beta_j,\ell}$ for $\beta_j = 1, \cdots, d_j$, and for all $\ell$ while the remaining $e_j^{(\ell)}$ streams are denoted as $PS^{(e,j)}_{\nu_j,\ell}$ for $\nu_j = 1, \cdots, e_j^{(\ell)}$. In order to consider the splits, $PS^{(d,j)}_{\beta_j,\ell}$ gets $\{w_{j,\beta_{j},\ell}^{(d)},\,\beta_{j}=1,\ldots,d_{j}\}$ fraction of the bandwidth from the data center to the cache server $j$. Similarly, $PS^{(\overline{d},j)}_{\beta_j,\ell}$ gets $\{w_{j,\beta_{j},\ell}^{(\overline{d})},\,\beta_{j}=1,\ldots,d_{j}\}$ fraction of bandwidth from cache server $j$ to the edge router $\ell$ and $PS^{(e,j)}_{\nu_j,\ell}$ gets $\{w_{j,\nu_{j},\ell}^{(e)},\,\nu_{j}=1,\ldots,e_{j}^{(\ell)}\}$ fraction of bandwidth from cache server $j$ to the edge router $\ell$. Thus, we have 
%\begin{equation}
%\vspace{-.12in}
%\sum_{\beta_j=1}^{d_j}w_{j,\beta_{j}}^{(d)} \le 1
%\end{equation}
%\begin{equation}
%\sum_{\beta_j=1}^{d_j}w_{j,\beta_{j}}^{(c)} +  \sum_{\nu_j=1}^{e_j}w_{j,\nu_{j}}^{(e)} \le 1,
%\end{equation}
\begin{equation}
%\vspace{-.12in}
\sum_{\beta_j=1}^{d_j}w_{j,\beta_{j},\ell}^{(d)} \le 1,\,\,\,\,\,\,\,\,\,\,\,\,\,\,\,\,\,\,\,
\sum_{\beta_j=1}^{d_j}w_{j,\beta_{j},\ell}^{(\overline{d})} +  \sum_{\nu_j=1}^{e_j^{(\ell)}}w_{j,\nu_{j},\ell}^{(e)} \le 1,\label{sumWs}
\end{equation}
for all $j = 1, \cdots, m$ and $\ell=1,\cdots,R$. We note that the sum of weights may be less than $1$ and some amount of the bandwidth may be wasted. While the optimal solution will satisfy this with equality since for better utilization, we do not need to explicitly enforce the equality constraint. We note that if the cache server serves multiple edge routers, the parallel streams between cloud storage and cache server will be the sum of $d_j$ to each edge router thus making the problem separated for each edge router. For ease, we will sometime omit $\ell$ to focus on links to one edge router only and the same procedure can be used for each edge router. 

We assume that the service time of a segment for data transfer from the data center to the cache server $j$ is shifted-exponential with rate $\alpha_{j,\ell}^{(d)}$ and a shift of $\eta_{j,\ell}^{(d)}$ while that between the cache server $j$ and the edge router $\ell$ is also shifted-exponential with rate $\alpha_{j,\ell}^{(f_j)}$ and a shift of $\eta_{j,\ell}^{(f_j)}$. The shifted exponential distribution can be seen as an approximation of the realistic service time distribution in the prior works, e.g., \cite{sproutVaneet017}, and references therein.
%We validate this assumption in our  analysis via experiments on the testbed which will be explained in Section \ref{testbed}. Our measurement of real
%service time distribution shows that the service time can be well approximated by a shifted exponential distribution. 
We also note that the rate of a parallel stream is proportional to the bandwidth split. Thus, the service time distribution of $PS^{(d,j)}_{\beta_j,\ell}$, $PS^{(\overline{d},j)}_{\beta_j,\ell}$, and $PS^{(e,j)}_{\nu_j,\ell}$, denoted as $\alpha_{j,\beta_{j},\ell}^{(d)}$, $\alpha_{j,\beta_{j},\ell}^{(\overline{d})}$, and $\alpha_{j,\nu_{j},\ell}^{(\overline{d})}$, respectively, and are given as follows.
%\begin{eqnarray}
%\vspace{-.4in}
%\alpha_{j,\beta_{j}}^{(d)}&=& w_{j,\beta_{j}}^{(d)} \alpha_j^{(d)}\label{alphad}\\
%\alpha_{j,\beta_{j}}^{(c)}&=& w_{j,\beta_{j}}^{(c)} \alpha_j^{(c)}\label{alphac}\\
%\alpha_{j,\nu_{j}}^{(e)} &=& w_{j,\nu_{j}}^{(e)} \alpha_j^{(c)}\label{alphae},
%\vspace{-.4in}
%\end{eqnarray}
\begin{eqnarray}
%\vspace{-.4in}
\alpha_{j,\beta_{j},\ell}^{(d)}= w_{j,\beta_{j},\ell}^{(d)} \alpha_j^{(d)}\label{alphad},\,\,\,\, \\
\alpha_{j,\beta_{j},\ell}^{(\overline{d})}= w_{j,\beta_{j},\ell}^{(\overline{d})} \alpha_{j,\ell}^{(f_j)}\label{alphac},\,\,\,\ 
\alpha_{j,\nu_{j},\ell}^{(e)} = w_{j,\nu_{j},\ell}^{(e)} \alpha_j^{(f_j)}\label{alphae},
%\vspace{-.4in}
\end{eqnarray}
for all $\beta_j$, $\nu_j$, and  $\ell$. We further define the moment generating functions of the service times of $PS^{(d,j)}_{\beta_j,\ell}$, $PS^{(\overline{d},j)}_{\beta_j,\ell}$, and $PS^{(e,j)}_{\nu_j,\ell}$ as $M^{(d)}_{j,\beta_j,\ell}$, $M^{(\overline{d})}_{j,\beta_j,\ell}$, and $M^{(\overline{d})}_{j,\nu_j,\ell}$, which are defined as follows. 
%\begin{eqnarray}
%\vspace{-.4in}
%M_{j,\beta_{j}}^{(d)} & =\frac{\alpha_{j,\beta_{j}}^{(d)}}{\alpha_{j,\beta_{j}}^{(d)}-t}\\
%M_{j,\beta_{j}}^{(c)} & =\frac{\alpha_{j,\beta_{j}}^{(c)}}{\alpha_{j,\beta_{j}}^{(c)}-t}\\
%M_{j,\nu_{j}}^{(e)} & =\frac{\alpha_{j,\nu_{j}}^{(e)}}{\alpha_{j,\nu_{j}}^{(e)}-t}\label{Me_jnuj}
%\vspace{-.4in}
%\end{eqnarray}
\begin{align}
\vspace{-.4in}
&
M_{j,\beta_{j},\ell}^{(d)}  =\frac{\alpha_{j,\beta_{j},\ell}^{(d)}e^{\eta_{j,\beta_j,\ell}^{(d)}t}}{\alpha_{j,\beta_{j},\ell}^{(d)}-t},\,\,\,\,\\
&
M_{j,\beta_{j},\ell}^{(\overline{d})}  =\frac{\alpha_{j,\beta_{j},\ell}^{(\overline{d})}e^{\eta_{j,\beta_j,\ell}^{(\overline{d})}t}}{\alpha_{j,\beta_{j},\ell}^{(\overline{d})}-t},\,\,\,\,\\
&
M_{j,\nu_{j},\ell}^{(e)}  =\frac{\alpha_{j,\nu_{j},\ell}^{(e)}e^{\eta_{j,\nu_j,\ell}^{(e)}t}}{\alpha_{j,\nu_{j},\ell}^{(e)}-t}\label{Me_jnuj}
\vspace{-.4in}
\end{align}

We also assume that there is a start-up delay of $d_{s}$ (in seconds)
for the video which is the duration in which the content can be buffered
but not played. 
Table \ref{tab:Key-Notations-Used} (in Appendix \ref{notation}) summarizes the key used notation in this paper.

\subsection{Edge-cache Model}

Edge cache $\ell \in \{1,2,\cdots, R\}$, where $R$ is the total number of edge-routers, stores the video content closer to end users. This improves the QoE to end users. We assume a limited cache size at the edge-router (edge-cache) of a maximum capacity of $C_{\ell,e}$ seconds, at edge-router $\ell$. When a file is requested by the user, the edge cache is first checked to see if the file is there completely or partly (in case some other user is watching that content). If the file is not in the edge cache, the space for this video file is created in the edge cache, and a file or some other video files have to be evicted so as not to violate the space constraint.
%
%
%Let the maximum capacity of edge-cache $\ell$ be $C_{\ell,e}$ seconds, i.e., edge-cache can store
%up to $C_{\ell,e}$ seconds of video files. 

We consider edge cache policy as follows. The file $i$ is removed from edge cache if it has not been accessed in time $\omega_{i,\ell}$ after its last request time from edge-router $\ell$. The parameter $\omega_{i,\ell}$ is a variable that can be optimized based on the file preference and its placement in the CDN cache. This caching policy is motivated by LRU since the file is evicted if it has not been used in some time in the past. The key advantages of this approach is that (i) It is tunable, in the sense that the parameters $\omega_{i,\ell}$ can be optimized, and (ii) the performance of the policy is easier to optimize as compared to LRU. When a file $i$ is requested, and someone has already requested from edge router $\ell$ in the last $\omega_{i,\ell}$ time units, the file is obtained from the edge router. Even if the file may not be completely in the edge-router  yet (not yet finished downloading), the downloaded part is given directly to the new user and the remaining content is delivered as it becomes available to the edge cache. This is akin to multicasting the remaining part of the video to multiple users \cite{Aggarwal:2009:EIS:1631272.1631330}.

\begin{figure}[t]
	\centering
	\includegraphics[trim=00.0in 0.00in 0.00in 0.00in, clip, width=0.48\textwidth]{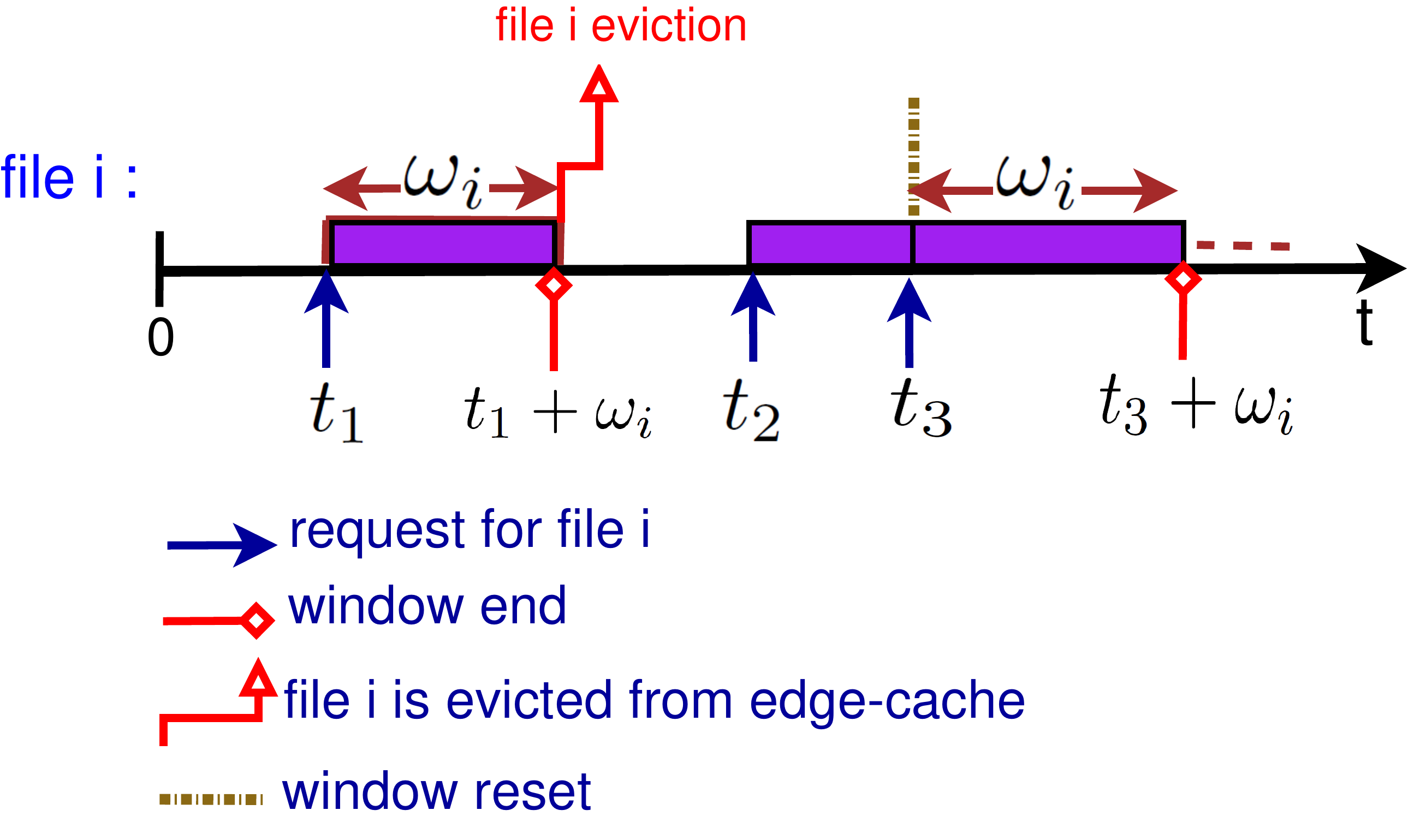}
	%	\vspace{-.45in}
	\caption{\small A schematic showing an example of how a time-line of a file $i$ can change according to the request pattern of a video file $i$.
		\label{fig:windowSize}}
	%	\vspace{-.25in}
\end{figure}

An illustration of the evolution of caching policy is illustrated in  Figure  \ref{fig:windowSize}, where the index $\ell$ is omitted since we consider the procedure at a single edge router. Video file $i$ is requested at three times $t_1$, $t_2$, and $t_3$. At $t_1$, the file enters the edge cache. Since it had not been requested in $\omega_i$ time units, it is evicted. When the file is again requested at $t_2$, the space for the file is reserved in the edge cache. The file, when requested at $t_3$ is within the $\omega_i$ duration from $t_2$ and thus will be served from the edge cache. If the file is still not in the edge cache completely, it will obtain the part already there directly while the remaining part will be streamed as it becomes available. Since the file $i$ was not requested for time $\omega_i$ after $t_3$, the file is again evicted from the edge cache.

% shows an example of how a time-line of a file $i$ can change according to the request pattern of a video file $i$. \textbf{[To add further explanation after fixing the figures]}

We note that  the arrival rate of the files is random. Thus, this file eviction policy may not satisfy the maximum edge cache constraint at all times. In order to handle this, we will first assume for the analytical optimization that the probability that the cache capacity of edge-cache $\ell$ is violated is bounded by $\epsilon_{\ell}$ which is small. That could lead us to obtain a rough estimate on the different parameters in the system.  The hard constraint on the capacity can be made in run-time, by evicting the files that are closest to be going out based on when they were requested and the corresponding $\omega_{i,\ell}$. This online adaptation will be explained in Appendix \ref{sec:adapt}.

\subsection{Queueing Model and Two-stage probabilistic scheduling} \label{quModel_PS}

If cache server $j$ is chosen for accessing video file $i$ on edge router $\ell$, the first $L_{j,i}$ chunks are obtained from one of the $e_j^{(\ell)}$ parallel streams $PS^{(e,j)}_{\nu_j,\ell}$. Further, the remaining $L_i - L_{j,i}$ chunks are obtained from the data center where a choice of $\beta_j$ is made from $1, \cdots, d_j$ and the chunks are obtained from the stream $PS^{(d,j)}_{\beta_j}$  which after being served from this queue is enqueued in the queue for the stream $PS^{(\overline{d},j)}_{\beta_j,\ell}$. However, if video file $i$ is already requested at time $t_i$ within a window of size $\omega_i$, the request will be served from the edge-cache and will not be sent to a higher level in the hierarchy, e.g., cache server.

%\begin{align}
%\alpha_{j,\nu_{j}}^{(e)} & =w_{j,\nu_{j}}^{(e)}\alpha_{j}^{(e)}\\
%\alpha_{j,\beta_{j}}^{(c)} & =w_{j,\beta_{j}}^{(c)}\alpha_{j}^{(c)}\\
%\alpha_{j,\zeta_{j}}^{(d)} & =w_{j,\zeta_{j}}^{(d)}\alpha^{(d)}
%\end{align}
%

We assume that the arrival of requests at the edge router $\ell$ for each
video $i$ form an independent Poisson process with a known rate $\lambda_{i,\ell}$. In order to serve the request for file $i$, we need to choose three things - (i) Selection of Cache server $j$, (ii) Selection of $\nu_j$ to determine one of $PS^{(e,j)}_{\nu_j,\ell}$ streams to deliver cached content, (iii) Selection of $\beta_j$ to determine one of $PS^{(d,j)}_{\beta_j}$ streams from the data-center which automatically selects the stream $PS^{(\overline{d},j)}_{\beta_j,\ell}$ from the cache server, to obtain the non-cached content from the datacenter. Thus, we will use a two-stage probabilistic scheduling to select the cache server and the parallel streams. For a file request at edge-router $\ell$, we choose server $j$ with probability $\pi_{i,j,\ell}$ for file $i$ randomly. Further, having chosen the cache server, one of the $e_j$ streams is chosen with probability $p_{i,j,\nu_j,\ell}$. Similarly, one of the $d_j$ streams is chosen with probability $q_{i,j,\beta_j,\ell}$. We note that these probabilities only have to satisfy 
%\begin{eqnarray}
%\vspace{-.3in}
%\sum_{j=1}^m \pi_{i,j} &=& 1 \forall i\\
%\sum_{\nu_j = 1}^{e_j} p_{i,j,\nu_j} &=& 1  \forall i,j
%\vspace{-.3in}
%\end{eqnarray}
%
%\begin{eqnarray}
%\vspace{-.3in}
%\sum_{\beta_j = 1}^{d_j} q_{i,j,\beta_j} &=& 1  \forall i,j\\
%\pi_{i,j}, p_{i,j,\nu_j}, q_{i,j,\beta_j} &\ge& 0  \forall i,j, \beta_j, \nu_j
%\vspace{-.3in}
%\end{eqnarray}
%\begin{eqnarray}
%\vspace{-.3in}
%\sum_{j=1}^m \pi_{i,j} &=& 1 \forall i\\
%\sum_{\nu_j = 1}^{e_j} p_{i,j,\nu_j} &=& 1  \forall i,j\\
%\sum_{\beta_j = 1}^{d_j} q_{i,j,\beta_j} &=& 1  \forall i,j\\
%\pi_{i,j}, p_{i,j,\nu_j}, q_{i,j,\beta_j} &\ge& 0  \forall i,j, \beta_j, \nu_j
%\vspace{-.3in}
%\end{eqnarray}
\begin{align}
&
\sum_{j=1}^m \pi_{i,j,\ell} = 1 \forall i, \ell;\,\,\,\,\\
&
\sum_{\nu_j = 1}^{e_j^{(\ell)}} p_{i,j,\nu_j,\ell} = 1  \forall i,j,\ell;\,\,\,\,\\
&
\sum_{\beta_j = 1}^{d_j} q_{i,j,\beta_j,\ell} = 1  \forall i,j,\ell,\\
&
\pi_{i,j,\ell}, p_{i,j,\nu_j,\ell}, q_{i,j,\beta_j,\ell} \ge 0  \forall i,j, \beta_j, \nu_j, \ell
\end{align}
%In order to serve the incoming request at edge-router whose connected
%to server $j$, we need to assign one of the $(e_{j})$ PSs
%to bring the cached segments from the cache and we need also to select
%one of the $d_{j}$ PSs to bring the remaining segments from the DC,
%if any. Let $p_{j,\nu_{j}}$ be the probability of choosing the stream
%$\nu_{j}$-out-of-$(e_{j})$ PSs associated with server $j$
%and $q_{j,\beta_{j}}$ be the probability of choosing the stream $\beta_{j}$-out-of-$d_{j}$
%PSs that are dedicated to serve the non-cached segements from the DC, and let $f_{j,\zeta_{j}}$ be the probability of choosing the stream $\zeta_j$-out-of-$d_j$ PSs that are connected to the datacenter (as shown in Figure \ref{fig:sysPS}).
%These choices are feasible if and only if 
%
%\begin{align}
%\sum_{\nu_{j}=1}^{e_{j}}p_{j,\nu_{j}} & =1\,,\forall j\\
%\sum_{\beta_{j}=1}^{d_{j}}q_{j,\beta_{j}} & =1\,,\forall j\\
%\sum_{\zeta_{j}=1}^{d_{j}}f_{j,\zeta_{j}} & =1\,,\forall j
%\end{align}
%
%\subsection{Queuing Model }

We note that since  file $i$ is removed from the edge cache after time $\omega_i$, the requests at the cache server are no longer Poisson. We note that this could be alleviated by assuming that every time the file is requested, the time $\omega_i$ is chosen using an exponential distribution. This change of distribution will make the distribution of requests at the cache server Poisson thus alleviating the issue. In the following, we will assume constant $\omega_i$, while still approximate the request pattern at cache servers as Poisson which holds when the times for which file remains in the edge cache is chosen using an exponential distribution. This approximation turns out to be quite accurate as will be shown in the evaluation
results. Further, such approximations of Poisson arrivals are widely used in the literature in similar fashions. In particular, it is used to characterize the coherence time of an LRU-based caching, e.g., see \cite{garetto2016unified} and references therein. Further, in
\cite{schwartz1987telecommunication} (Ch.9, page 470) authors approximate the arrivals of new
and retransmitted packets in CSMA protocol as Poisson even
though they are not due to the dependencies between them.

Since sampling of Poisson process is Poisson, and superposition of independent Poisson processes is also Poisson, we get the aggregate arrival rate at $PS^{(d,j)}_{\beta_j,\ell}$, $PS^{(\overline{d},j)}_{\beta_j,\ell}$, and $PS^{(e,j)}_{\nu_j,\ell}$, denoted as $\Lambda^{(d)}_{j,\beta_{j},\ell}$,  $\Lambda^{(\overline{d})}_{j,\beta_{j},\ell}$, and  $\Lambda^{(e)}_{j,\nu_{j},\ell}$, respectively are given as follows. 
%\begin{eqnarray}
%\vspace{-.12in}
%\Lambda^{(d)}_{j,\beta_{j}} &=& \sum_{i=1}^r \lambda_{i} \pi_{i,j}q_{i,j,\beta_{j}}\label{lambdad_betaj}\\
%\Lambda^{(c)}_{j,\beta_{j}} &=& \Lambda^{d}_{j,\beta_{j}} \label{lambdac_betaj}\\
%\Lambda^{(e)}_{j,\nu_{j}} &=& \sum_{i=1}^r \lambda_{i} \pi_{i,j}p_{i,j,\nu_{j}}\label{lambdae_nuj}
%\vspace{-.12in}
%\end{eqnarray}
\begin{eqnarray}
\Lambda^{(d)}_{j,\beta_{j},\ell} = \sum_{i=1}^r \lambda_{i,\ell} \pi_{i,j,\ell}q_{i,j,\beta_{j},\ell}e^{-\lambda_{i,\ell} \omega_{i,\ell}}\label{lambdad_betaj};\,\,\,\,
\Lambda^{(\overline{d})}_{j,\beta_{j},\ell} = \Lambda^{(d)}_{j,\beta_{j},\ell} \label{lambdac_betaj},\\
\Lambda^{(e)}_{j,\nu_{j},\ell} = \sum_{i=1}^r \lambda_{i,\ell} \pi_{i,j,\ell}p_{i,j,\nu_{j},\ell}e^{-\lambda_{i,\ell} \omega_i}\label{lambdae_nuj}
\end{eqnarray}
%Then, the arrival of file requests at PS $\nu_{j}$ at node $j$ forms
%a Poisson Process with rate $\Lambda_{j,\nu_{j}}=\sum_{i}\lambda_{i} \pi_{i,j}p_{j,\nu_{j}}$
%which is the superposition of $r(e_{j})$ Poisson processes
%each with rate $\lambda_{i} \pi_{i,j} p_{j,\nu_{j}}.$ 
%
%If video $i$ segments are not all stored in the cache, we
%request the non-cached segments from the DC. Let $L_{c,i}$ denotes
%the number of cached segments of video file $i$, i.e., $L_{c,i}\leq L_{i}$.
%Hence, the remaining $(L_{i}-L_{c,i})$ segments need to be brought
%(served) from the DC. In order to schedule the request for the $(L_{i}-L_{c,i})$
%segments of video file $i$ from a PS $\beta_{j}$, we need first
%to characterize the arrival rate at the PS $\zeta_{j}$. Since the
%arrival process and the service time of the cache server are memoryless
%(i.e., M/M/1), it is easy to show that the arrival of file requests
%at PS $\zeta_{j}$ at node $j$ forms a Poisson Process with rate $\Lambda_{j,\zeta_{j}}=\sum_{i}\lambda_{i}\pi_{i,j}f_{j,\zeta_{j}}$
%which is the superposition of $rd_{j}$ Poisson processes each with
%rate $\lambda_{i}\pi_{i,j}f_{j,\beta_{j}}$. Here, we also assume that the service time of a segment
%at node $j$ at PS $\zeta_{j}$, i.e., $X_{j,\zeta_{j}}^{(d)}$, follows
%also exponential distribution with rate $\alpha_{j,\zeta_{j}}^{(d)}$
%and $M_{j,\zeta_{j}}^{(d)}(t)$ can be expressed as 
%\begin{equation}
%M_{j,\zeta_{j}}^{(d)}(t)=\frac{\alpha_{j,\beta_{j}}^{(d)}}{\alpha_{j,\zeta_{j}}^{(d)}-t}
%\end{equation}
\begin{lemma}
If $\omega_{i}$ follows an exponential distribution (i.e., not fixed) with parameter $\nu_i$, the probability that the request of video file $i$ is directed to the distributed cache servers and/or central server is given by
\begin{equation}
P(\widetilde{t}_{i}>\omega_{i})=\frac{\nu_{i}}{\nu_{i}+\lambda_{i}}
\end{equation} 
\end{lemma}

\begin{proof}
Let $\widetilde{t}_i$ be the inter-arrival request for video file $i$. Hence, if  video file $i$ is requested within $\omega_{i}$ since its last request, video $i$ request will be served from the edge-cache (i.e., $\widetilde{t}_i \leq \omega_{i}$).
%The request for video file $i$ is served from edge-cache unless it is not locally stored. 
In contrast, if video file $i$ is not at the edge-cache, the request is forwarded to higher hierarchy levels (distributed storage cache and/or central cloud stroage). 
Since $\widetilde{t}_i$ and $\omega_{i}$ are exponentially distributed with paramters $\lambda_{i}$ and $\nu_i$, respectively, the probability of directing file $i$ request to the distributed storage cache and/or central cloud stroage is given by $P(\widetilde{t}_{i}>\omega_{i})$. This proves the statement of the Lemma. 

\end{proof}

\begin{lemma}
	\label{PoisArr2ndQ}
 When the service time distribution of datacenter server (first queue) is given by shifted exponential distribution, the arrivals at the cache servers (second queue) are Poisson.
\end{lemma}

\begin{proof}
	The proof is provided in  Appendix \ref{2ndQArr}.
\end{proof}

\subsection{Distribution of Edge Cache Utilization}

We will now investigate the distribution of the edge-cache utilization at any time. This will help us in bounding the probability that the edge cache is more than the capacity of the cache. In the analytic part, we will bound this probability. However, the online adaptations in Appendix \ref{sec:adapt} will provide an adaptation to maintain the maximum edge cache capacity constraint at all times. 

Let $X_{i,\ell}$ be the random variable corresponding to amount of space in the edge-cache $\ell$ for video file $i$. Since the file arrival rate is Poisson, and the file is in the edge-cache $\ell$ if it has been requested in the last $\omega_i$ seconds. Then,  $X_{i,\ell}$ is given as 
%size of edge-cache if video file $i$ is stored at cache of router $j$. Since arrival of video files is Poisson, the size of the edge-cache $j$ if file $i$ is stored, $X_j$, can be written as
\begin{equation}
X_{i,\ell}=\begin{cases}
\tau\,L_{i} & \text{with prob. \ensuremath{1-e^{-\lambda_{i,\ell}\omega_{i,\ell}}}}\\
0 & \text{with prob. }\ensuremath{e^{-\lambda_{i,\ell}\omega_{i,\ell}}}
\end{cases}
\end{equation}
where $1-e^{-\lambda_{i,\ell}\omega_{i,\ell}}$ is the probability that file $i$ is requested within a window-size of $\omega_{i,\ell}$ time units. The total utilization of the edge-cache $j$ is given as 
\begin{equation}
X_{\ell}=\sum_{i=1}^{r}X_{i,\ell}
\end{equation}

The mean and variance of $X_{\ell}$ can be found to be $\sum_{i}^{r} \tau L_i (1-e^{-\lambda_{i,\ell} \omega_{i,\ell}}$ and $\sum_{i}^{r} (\tau L_i)^2 e^{-\lambda_{i,\ell} \omega_{i,\ell}}(1-e^{-\lambda_{i,\ell} \omega_{i,\ell}})$, respectively. Since $r$ is large, we will approximate the distribution of $X_j$ by a Gaussian distribution with mean $\sum_{i}^{r} \tau L_i (1-e^{-\lambda_{i,\ell} \omega_{i,\ell}}$ and variance $\sum_{i}^{r} (\tau L_i)^2 e^{-\lambda_{i,\ell} \omega_{i,\ell}}(1-e^{-\lambda_{i,\ell} \omega_{i,\ell}})$. This distribution is then used as a constraint in the design of $\omega_{i,\ell}$, where the constraint bounds the probability that the edge cache utilization is higher than the maximum capacity of the edge cache. Since $X_{\ell}$ can be well approximated by a Gaussian distribution, the edge cache  utilization, can be probabilistically bounded as follows, 
\begin{equation}
%\varPhi_{\ell}(C_{\ell,e})\leq\epsilon_{\ell}
\intop_{C_{\ell,e}}^{\infty}\frac{1}{\sqrt{2\pi\sigma^{2}}}e^{-\frac{(x-\mu)^{2}}{2\sigma^{2}}}dx\leq\epsilon_{\ell} \label{edgeCacheConst}
\end{equation}
where $\mu = \sum_{i}^{r} \tau L_i (1-e^{-\lambda_{i,\ell} \omega_{i,\ell}})$, $\sigma^2 = \sum_{i}^{r} (\tau L_i)^2 e^{-\lambda_{i,\ell} \omega_{i,\ell}}(1-e^{-\lambda_{i,\ell} \omega_{i,\ell}})$ are the mean and variance, respectively.
% and  $\varPhi_{\ell}(C_{\ell,e})$ is the CDF of the Gaussian distribution.

%{\bf Abubakr, Add the exact equation for the $\epsilon$ constraint here using the distribution functions here, and cite that in the formulation. }

%We note that $X_j$ has a distribution, which is given by a Poisson Binomial Distribution (PBD).

%Since we know the distribution of the size of video files in the edge-cache, we can efficiently design the window-size $\omega_i$ by which the miss rate of edge-cache is minimized. We note that optimizing over $\omega_i$ is important to better utilize the role of the  edge-cache and improve the QoE of users accordingly. 

% This work will characterize the stall duration tail
%probability using the two-stage probabilistic scheduling and allocation of bandwidth
%weights. We note that the arrival rates are given in terms of the video files,
%and the service rate above is provided in terms of segment at each
%server. The analysis would require detailed consideration of the different segments in a video. 

\section{Stall Duration Tail Probability} \label{sec:StallTailProb}

This section will characterize the stall duration tail probability using the two-stage probabilistic scheduling and allocation of bandwidth weights. We note that the arrival rates are given in terms of the video files, and the service rate above is provided in terms of segment at each server. The analysis would require detailed consideration of the different segments in a video. In this section, we will assume that the edge-router for the request $\ell$ is known, and thus we omit the subscript/superscript $\ell$  to simplify notations.

In order to find the stall durations, we first consider the case where file $i$ is not in the edge cache and has to be requested from the CDN. We also assume that the cache server $j$ is used, with the streams $\beta_j$ and $\nu_j$ known. We will later consider the distribution of these choices to compute the overall metric.  In order to compute stall durations, we would first calculate the download time of each of the video segment, which accounts for the first $L_{j,i}$ segments at the cache $j$ and the later $L_i-L_{j,i}$ segments at the server. After the download times are found, the play times of the different contents are found. The detailed calculations are shown in Appendix \ref{sec:DownLoadPlay}, where the distribution of $T_{i,j,\beta_j,\nu_j}^{\left(g\right)}$, the time that segment $g$ begins to play at the client $i$ given that it is downloaded from $\beta_j$ and $\nu_j$ queues, is found.
The stall duration for the request of file $i$ from $\beta_{j}$ queue,  $\nu_{j}$ queue and server $j$, if not in the edge-cache, i.e., $\Gamma^{(i,j,\beta_{j},\nu_{j})}_{U}$ is given as
\begin{equation}
\Gamma^{(i,j,\beta_{j},\nu_{j})}_{U}=T_{i,j,\beta_{j},\nu_{j}}^{(L_{i})}-d_{s}-\left(L_{i}-1\right)\tau,
\label{stallTime}
\end{equation}
as explained in Appendix \ref{sec:DownLoadPlay}. We use this expression to derive a tight bound on the SDTP.

The stall duration tail probability of a video file $i$ is defined as the
probability that the stall duration  is greater
than a pre-defined threshold $\sigma$. Since exact evaluation of stall duration  
is hard
\cite{Joshi:13, Abubakr_TON}, we cannot evaluate 
%$\text{\text{Pr}}\left(\Gamma^{(i,j,\beta_j,\nu_j)}\geq\sigma\right)$
$\text{\text{Pr}}\left(\Gamma^{(i)}_{tot}\geq\sigma\right)$
in closed-form, where $\Gamma^{(i)}_{tot}$ is random variable indicating the overall stall duration for file $i$. In this section, we derive a tight upper bound on the SDTP through the two-stage Probabilistic Scheduling as follows. 

We first note that the  expression in equation \eqref{stallTime} (Appendix \ref{sec:DownLoadPlay})  accounts only for the stalls that would be incurred if the video segments are not accessed from the edge-cache (including stored, or multicasted). However, the user would experience lower stalls if the requested content is accessed from the edge-cache. Thus, we need an expectation over the choice of whether the file is accessed from the edge server, and the choice of $(j,\beta_j, \nu_j)$ in addition to the queue statistics.  For a video file $i$ requested at time $\widetilde{t}_i$ after the last request for file $i$,  the stall duration for the request of file $i$  can be  expressed as follows:
%Let $\Gamma_{tot}^{(i,j,\beta_{j},\nu_{j})}$ give the overall stall duration accounting for the edge-cache, and representing that $(j,\beta_j, \nu_j)$ is used only if the file is not accessed from the edge cache. Then,  for a video file $i$ requested at time $\widetilde{t}_i$ after the last request for file $i$,  the stall duration for the request of file $i$ from server $j$ can be  expressed as follows:
\begin{equation}\label{ve1}
\!\Gamma_{tot}^{(i)}\overset{d}{=}\begin{cases}
\left(\Gamma_{tot}^{(i)}-\widetilde{t}_{i}\right)^{+} & 0\leq\widetilde{t}_{i}\leq\omega_{i}\\
\Gamma_{U}^{(i,j,\beta_{j},\nu_{j})} & \widetilde{t}_{i}>\omega_{i}
\end{cases}
\end{equation}
where $\overset{d}{=}$ means equal in distribution. This is because if file $i$ is requested again within $\omega_i$ time, then the multicast or stored file can lead to the reduced stall duration based on how much time has passed since the last request. Further, if the file has not been requested in the last $\omega_i$ time units, then the file has to be obtained from the CDN, and thus the expression of random variable $\Gamma_{tot}^{(i)}$ also includes randomness over the choice of $(j,\beta_j, \nu_j)$ in this case. From  \eqref{ve1}, we can obtain the following result. 

%Since the inter-arrival time is exponential, we have  
%To avoid cumbersome notation, we drop the usual $i,j,\beta_j,\nu_j$ notation; instead we let
%$r$ to be a generic notation for request of video file $i$ served from queue $\nu_j$ from $\beta_j$ at server $j$. {\bf $r$ is number of files and not anything generic - please change all notations to be precise. } The, we can write 
%\begin{equation}
%\!\!\!\!\!e^{h\Gamma_{tot}^{(i,j,\beta_{j},\nu_{j})}}=\begin{cases}
%\text{max}\left(e^{h\left(\Gamma_{U}^{(i,j,\beta_{j},\nu_{j})}-\omega_{i}+\widetilde{t}_{i}\right)},1\right) & 0\leq\widetilde{t}_{i}\leq\omega_{i}\\
%e^{h\Gamma_{U}^{(i,j,\beta_{j},\nu_{j})}} & \widetilde{t}_{i}>\omega_{i}
%\end{cases}\!\!\!\label{ehG}
%\end{equation}

\begin{lemma}
For a given choice of $(j,\beta_j, \nu_j)$,  $\mathbb{E}\left[e^{h_{i}\Gamma_{tot}^{(i)}}\right] $  is bounded as
\begin{align}
\mathbb{E}\left[e^{h_{i}\Gamma_{tot}^{(i)}}\right] & \leq \widetilde{c}+\widetilde{a}\,{\mathbb E}\left[e^{h_{i}\Gamma_{U}^{(i,j,\beta_j,\nu_j)}}\right] \label{E_stall_tot}
\end{align}
where  $c=1-e^{-\lambda_{i}\omega_{i}}$, $a=e^{-\lambda_{i}\omega_{i}}$, $b=\Biggl[1-\frac{\lambda_{i}}{\lambda_{i}+h_{i}}\left(1-e^{-\left(\lambda_{i}+h_{i}\right)\omega_{i}}\right)\Biggr]$, $\widetilde{c}=c/b$ and $\widetilde{a}=a/b$. \label{lem:expon}
\end{lemma}
\begin{proof}
The proof is provided in Appendix \ref{proof_exp}.
\end{proof}

%the moment generating function of the service time for all video files from parallel stream $PS^{(e,j)}_{\nu_j}$   is given by   

%where $\overline{D}_{i,j,\beta_{j},\nu_{j}}^{(v,t)}=e^{(\widetilde{\sigma}-\overline{\sigma})t_{i}}\mathbb{E}\left[e^{t_{i}D_{i,j,\beta_{j},\nu_{j}}^{(v)}}\right]$. 
%where $(f)$ follows from the two-stage Probabilistic Scheduling, $\widetilde{\sigma}=(L_{i}-z+1)\tau
%$, $\overline{\sigma}=\text{\ensuremath{\sigma}}+d_{s}+\left(L_{i}-1\right)\tau$,  is given by (\ref{D_nu_j}). 
We next derive $\mathbb{E}\left[e^{h_iD_{i,j,\beta_j,\nu_j}^{(v)}}|(j,\beta_j,\nu_j)\right]$ using the following two lemmas, which will be used in the main result. The key idea is that we characterize the download and play times of each segments and use them in determining the SDTP of each video file request.  
%can be calculated as shown in the following two lemmas. 
\begin{lemma} \label{v_in_cache}
For $v\leq L_{j,i}$,   $\mathbb{E}\left[e^{h_iD_{i,j,\beta_j,\nu_j}^{(v)}}|(j,\beta_j,\nu_j)\right]$ is given by 
\begin{eqnarray}
&&\mathbb{E}\left[e^{h_iD_{i,j,\beta_j,\nu_j}^{(v)}}|(j,\beta_j,\nu_j)\right]\nonumber\\
&=&
\frac{(1-\rho_{j,\nu_{j}}^{(e)})t_i}{t_i-\Lambda_{j,\nu_{j}}^{(e)}(B_{j,\nu_{j}}^{(e)}(t_i)-1)}\left(\frac{\alpha_{j,\nu_{j}}^{(e)}e^{\eta_{j,\nu_j}^{(e)}t}}{\alpha_{j,\nu_{j}}^{(e)}-t_i}\right)^{v}
\end{eqnarray}
\end{lemma}
\begin{proof}
The proof follows from \eqref{D_nu_j} in Appendix \ref{sec:DownLoadPlay} by replacing $g$ by $v$ and rearranging the terms in the result.
\end{proof}

\begin{lemma}
	\label{lemmaDLfromDC}
	For $v\geq(L_{j,i}+1)$,   $\mathbb{E}\left[e^{h_iD_{r}^{(v)}}|(j,\beta_j,\nu_j)\right]$ is given by 
\begin{eqnarray}
&&\mathbb{E}\left[e^{h_{i}D_{r}^{(v)}}|(j,\beta_j,\nu_j)\right] \nonumber\\
& \leq& {\mathbb E}\left[e^{h_{i}U_{i,j,\beta_{j},v,L_{j,i}}}|(j,\beta_j,\nu_j)\right]+\nonumber \\
&& \sum_{w=L_{j,i}+1}^{v}\frac{(1-\rho_{j,\beta_{j}}^{(d)})t_{i}}{h_{i}-\Lambda_{j,\beta_{j}}^{(d)}(B_{j,\beta_{j}}^{(d)}(h_{i})-1)}\times\nonumber \\
&& \left(\frac{\alpha_{j,\beta_{j}}^{(d)}e^{\eta_{j,\beta_{j}}^{(d)}h_{i}}}{\alpha_{j,\beta_{j}}^{(d)}-h_{i}}\right)^{w-L_{j,i}-1}\left(\frac{\alpha_{j,\beta_{j}}^{(\overline{d})}e^{\eta_{j,\beta_{j}}^{(\overline{d})}h_{i}}}{\alpha_{j,\beta_{j}}^{(\overline{d})}-h_{i}}\right)^{v-w+1},
\end{eqnarray}
where ${\mathbb E}\left[e^{h_{i}U_{i,j,\beta_{j},v,L_{j,i}}}|(j,\beta_j,\nu_j)\right]$
and $B_{j,\beta_{j}}^{(d)}$ are given in Appendix \ref{sec:DownLoadPlay}, equations \eqref{U_MGF_Lji}, and \eqref{Bc}, respectively.
\end{lemma}
\begin{proof}
The proof is provided in Appendix \ref{LemmaDLfromDC}.
\end{proof}

\begin{corollary}
	\label{ttfc}
The (expected) time to the first chunk (TTFC) can be obtained from equation \eqref{eq:T_s_main_stall} (Lemma \ref{meanthm}) by setting $d_s=0$ and $g_i=1$.
\end{corollary}

%\begin{proof}
%	The detailed steps are provided in Appendix \refeq{lemmaSDTPtail}.
%\end{proof}

Using these expressions, the following theorem summarizes the stall duration tail probability for file $i$. We include the edge router index $\ell$ in all the expressions in the result for the ease of using it in the following section. 
\begin{theorem}
	\label{tailLemma}
The stall distribution tail probability for video
file $i$ requested through edge router $\ell$ is bounded by
\begin{align}
 & \text{\text{Pr}}\left(\Gamma_{tot}^{(i,\ell)}\geq\sigma\right)\leq\nonumber\\
& \sum_{j=1}^{m}\pi_{i,j,\ell}\times\Biggl[\overline{c}_{\ell}+\widetilde{a}_{\ell}\,e^{-h_{i}\sigma}+\overline{a}_{\ell}\sum_{\nu_{j}=1}^{e_{j}^{(\ell)}}p_{i,j,\nu_{j},\ell}\times\nonumber\nonumber\\
& \sum_{\beta_{j}=1}^{d_{j}}q_{i,j,\beta_{j},\ell}e^{h_{i}L_{i}\tau}\times\left(\delta^{(e,\ell)}+\delta^{(\overline{d},\ell)}+\delta^{(d,\overline{d},\ell)}\right)\label{deltaAll}
\end{align}
for  $\rho_{j,\beta_{j}}^{(d)}<1$, $\rho_{j,\beta_{j},\ell }^{(\overline{d})}<1$, $\rho_{j,\nu_{j},\ell}^{(e)}<1$, 
where the auxiliary variables in the statement of the Theorem are defined as 
\begin{eqnarray}
\!\ensuremath{\!\!\!\!\!\!\!\!\!\!\!\!\delta^{(e,\ell)}} & \!\!\!\!\!\!\!\!\!\!\!= & \!\!\!\!\!\!\!\!\!\!\!\frac{\widetilde{M}_{j,\nu_{j}}^{(e,\ell)}(h_{i})(1-\rho_{j,\beta_{j},\ell}^{(e)})t_{i}((\widetilde{M}_{j,\nu_{j}}^{(e,\ell)}(h_{i}))^{L_{j,i}}\!-1\!)}{(h_{i}-\Lambda_{j,\beta_{j},\ell}^{(e)}(B_{j,\beta_{j}}^{(e,\ell)}(h_{i})-1))(\widetilde{M}_{j,\nu_{j}}^{(e,\ell)}(h_{i}))-1)}\,\,\,\,\,\,\,\label{deltae}\\
\!\delta^{(\overline{d},\ell)} & =\!\!\!\!\! & \ensuremath{\frac{(1-\rho_{j,\beta_{j},\ell}^{(\overline{d})})t_{i}(\widetilde{M}_{j,\nu_{j}}^{(\overline{d},\ell)}(h_{i}))^{L_{j,i}-L_{i}}}{h_{i}-\Lambda_{j,\beta_{j},\ell}^{(\overline{d})}(B_{j,\beta_{j}}^{(\overline{d},\ell)}(h_{i})-1)}}\\
\!\!\!\!\!\!\!\!\!\!\!\!\!\delta^{(d,\overline{d},\ell)}\!\!\!\!\!\!\! & \!\!\!\!\!\!\!=\!\!\!\!\!\!\!\!\!\! & \!\!\gamma^{(d,\ell)}\!\!\left(\!\!\!\frac{(\widetilde{M}_{j,\beta_{j}}^{(d,\overline{d},\ell)}(h_{i}))^{L_{i}-L_{j,i}}\!\!-(L_{i}\!\!-L_{j,i}\!\!)}{(\widetilde{M}_{j,\beta_{j}}^{(d,\overline{d},\ell)}(h_{i}))-1}\!\!+\!\!\xi_{i,j,\beta_{j},}^{(d,\overline{d},\ell)}\!\!\right)\,\!\!\\
\!\xi_{i,j,\beta_{j},}^{(d,\overline{d},\ell)} & =\!\!\!\!\! & \frac{\widetilde{M}_{j,\beta_{j}}^{(d,\overline{d},\ell)}(h_{i})\left((\widetilde{M}_{j,\beta_{j}}^{(d,\overline{d},\ell)}(h_{i}))^{L_{i}-L_{j,i}-1}-1\right)}{(\widetilde{M}_{j,\beta_{j}}^{(d,\overline{d},\ell)}(h_{i}))-1}\\
\!\gamma^{(d,\ell)} & \!\!\!\!\!\!=\!\!\!\!\!\!\!\!\!\! & \ensuremath{\frac{(1-\rho_{j,\beta_{j}}^{(d)})t_{i}(\widetilde{M}_{j,\beta_{j}}^{(\overline{d},\ell)}(h_{i}))^{L_{i}+1}}{\left[h_{i}-\Lambda_{j,\beta_{j}}^{(d)}(B_{j,\beta_{j}}^{(d)}(h_{i})-1)\right](\widetilde{M}_{j,\beta_{j}}^{(d)}(h_{i}))^{L_{j,i}+1}}}
\end{eqnarray}

\begin{eqnarray}
\ensuremath{\widetilde{M}_{j,\beta_{j}}^{(d)}(h_{i})=\!\frac{\alpha_{j,\beta_{j}}^{(d)}e^{\eta_{j,\beta_{j}}-h_{i}\tau}}{\alpha_{j,\beta_{j}}^{(d)}-h_{i}},} & \ensuremath{}\\
\widetilde{M}_{j,\beta_{j}}^{(\overline{d},\ell)}(h_{i})=\!\frac{\alpha_{j,\beta_{j},\ell}^{(\overline{d})}e^{\eta_{j,\beta_{j},\ell}-h_{i}\tau}}{\alpha_{j,\beta_{j},\ell}^{(\overline{d})}-h_{i}}\\
\ensuremath{\widetilde{M}_{j,\nu_{j}}^{(e,\ell)}(h_{i})=\!\frac{\alpha_{j,\nu_{j},\ell}^{(e)}e^{\eta_{j,\nu_{j},\ell}-h_{i}\tau}}{\alpha_{j,\nu_{j},\ell}^{(e)}-h_{i}}},\!\!\!\!\!\! & \ensuremath{\ensuremath{}}
\end{eqnarray}
\begin{eqnarray}
\widetilde{M}_{j,\beta_{j},\ell}^{(d,\overline{d})}(h)=\frac{\alpha_{j,\beta_{j}}^{(d)}(\alpha_{j,\beta_{j},\ell}^{(\overline{d})}-h)e^{\eta_{j,\beta_{j}}^{(d)}h}}{\alpha_{j,\beta_{j},\ell}^{(\overline{d})}(\alpha_{j,\beta_{j}}^{(d)}-h)e^{\eta_{j,\beta_{j},\ell}^{(\overline{d})}h}},\,\,\,\forall j,\beta_{j}\label{tailEqn}\\
\overline{c}_{\ell}=\frac{(1-e^{-\lambda_{i,\ell}\omega_{i,\ell}})e^{-h_{i}\sigma}}{1-\frac{\lambda_{i}}{\lambda_{i}+h_{i}}\left(1-e^{-\left(\lambda_{i}+h_{i}\right)\omega_{i}}\right)}\label{eq:cbar}\\
\widetilde{a}_{\ell}=\frac{e^{-\lambda_{i,\ell}\omega_{i,\ell}}}{1-\frac{\lambda_{i}}{\lambda_{i}+h_{i}}\left(1-e^{-\left(\lambda_{i}+h_{i}\right)\omega_{i}}\right)}\label{eq:awidetilde}\\
\overline{a}_{\ell}=\widetilde{a}_{\ell}e^{-h_{i}(\sigma+d_{s}+(L_{i}-1)\tau)}\label{eq:abar}
\end{eqnarray}

%Further, the moment generating function terms exist by satisfying of (\ref{alpha_j_nuj}--\ref{mgf_gz_e_queue}), (\ref{alpha_betaj_d})--(\ref{mgf_beta_j_c}) for $t=t_i$.

%\end{equation}
\label{lemmatailEqn}
\end{theorem}
\begin{proof}
The detailed steps are provided in Appendix \refeq{lemmaSDTPtail}. 

\end{proof}

We note that  $\delta^{(e,\ell)}=\delta^{(\overline{d},\ell)}=0$, if the storage server nodes  are not hosting the requested video files and $\delta^{(\overline{d},d,\ell)}$ has nonzero value only if some sWe can also derive the mean stall duration for video file $i$ in a similar fashion. The interested reader is referred to Appendix \ref{sec:mean} for detailed treatment of this metric.

\section{Optimization Problem Formulation and Proposed Algorithm }
\label{sec:SDTP}
\subsection{Problem Formulation}
We define 
%$\boldsymbol{\pi}=(\pi_{i,j}\forall i=1,\cdots,r\text{ and }j=1,\cdots,m)$,
%$\boldsymbol{p}=(p_{i,,j,\nu_{j}}\forall i=1,\cdots,r\,,j=1,\cdots,m\,,\,\nu_{j}=1,\ldots e_{j}),$
%$\boldsymbol{q}=(q_{i,,j,\beta_{j}}\forall i=1,\cdots,r\,,j=1,\cdots,m\,,\,\beta_{j}=1,\ldots d_{j}),$ $\boldsymbol{t}=\left(t_{1},t_{2},\ldots,t_{r}\right)$, 
%
%$\boldsymbol{w}_{j,\beta_{j}}^{(c)}=(w_{j,1}^{(c)},w_{j,2}^{(c)},\ldots,w_{j,d_{j}}^{(c)},\text{ and }j=1,\cdots,m)$,
%$\boldsymbol{w}_{j,\beta_{j}}^{(e)}=(w_{j,1}^{(e)},w_{j,2}^{(e)},\ldots,w_{j,d_{j}}^{(e)},\text{ and }j=1,\cdots,m)$,
%$\boldsymbol{w}_{j,\beta_{j}}^{(d)}=(w_{j,1}^{(d)},w_{j,2}^{(d)},\ldots,w_{j,d_{j}}^{(d)},\text{ and }j=1,\cdots,m)$,
%$\boldsymbol{L}_{j,i}=(L_{j,i},\,\forall i,\,\forall j)$
$\ensuremath{\ensuremath{\boldsymbol{\pi}=(\pi_{i,j,\ell}\forall i=1,\cdots,r\text{ and }j=1,\cdots,m,\ell=1,\cdots,R)},}\ensuremath{\boldsymbol{p}=(p_{i,,j,\nu_{j},\ell},(\nu_{j},\ell)\in\{(\nu_{j},\ell):\nu_{j}\in\{1,\ldots,e_{j}^{(\ell)}\},\ell\in\{1,\dots,R\}\},\forall i=1,\cdots,r\,,j=1,\cdots,m\,,\,\nu_{j}=1,\ldots e_{j}^{(\ell)},\ell=1,\cdots,R)},\ensuremath{\ensuremath{\boldsymbol{q}=(q_{i,,j,\beta_{j},\ell}\forall i=1,\cdots,r\,,j=1,\cdots,m\,,\,\beta_{j}=1,\ldots d_{j},\ell=1,\cdots,R)}},\ensuremath{\boldsymbol{h}=\left(h_{1},t_{2},\ldots,h_{r}\right)\,},\ensuremath{\ensuremath{\boldsymbol{w}^{(\overline{d})}=(w_{j,1,\ell}^{(\overline{d})},\ldots,w_{j,d_{j},\ell}^{(d)}\ensuremath{,}\text{ and }j=1,\cdots,m,\ell=1,\cdots,R)}\,,}\ensuremath{\boldsymbol{w}^{(e)}=(w_{j,1,\ell}^{(e)},w_{j,2,\ell}^{(e)},\ldots,w_{j,e_{j}^{(\ell)},\ell}^{(e)},\text{ and }j=1,\cdots,m,\ell=1,\ldots,R),\,}\ensuremath{\ensuremath{\boldsymbol{w}^{(d)}=(w_{j,1,\ell}^{(d)},\ldots,w_{j,d_{j},\ell}^{(d)},\text{ and }j=1,\cdots,m)}},\ensuremath{\boldsymbol{L}=(L_{j,i},\,\forall i=1,\,\text{ and }j=1,\cdots,m)}$
  and $\boldsymbol{\omega}=(\omega_{i,\ell}, \forall i,\ell)$.
Our goal is to minimize the SDTP  over the
choice of cache and datacenter access decisions, bandwidth allocation
weights, portion (number) of cached segments, time window over which we maintain the video files at edge-cache and auxiliary bound parameters.

To incorporate for weighted fairness and differentiated services,
we assign a positive weight $\kappa_{i,\ell}$ for each file $i$. Without
loss of generality, each file $i$ is weighted by the arrival rate
$\lambda_{i,\ell}$ in the objective (so larger arrival rates are weighted
higher). However, any other weights can be incorporated to accommodate
for weighted fairness or differentiated services. Let $\overline{\lambda}=\sum_{i,\ell}\lambda_{i,\ell}$
be the total arrival rate. Hence, $\kappa_{i,\ell}=\lambda_{i,\ell}/\overline{\lambda}$
is the ratio of file $i$ requests. Hence, the objective is the minimization
of stall duration tail probability, averaged over all the file requests,
and is given as $\sum_{i,\ell}\frac{\lambda_{i,\ell}}{\overline{\lambda}}\,\text{\text{Pr}}\left(\Gamma^{(i)}\geq\sigma\right)$.
By using the expression for SDTP in Section \ref{sec:StallTailProb},
the optimization problem can be formulated as follows.
\vspace{-.1in}
\begin{align}
&
\sum_{\ell=1}^{R}\sum_{i=1}^{r}\frac{\lambda_{i,\ell}}{\overline{\lambda}_{i,\ell}}\sum_{j=1}^{m}\pi_{i,j,\ell}\times\Biggl[\overline{c}_{\ell}+\widetilde{a}_{\ell}\,e^{-h_{i}\sigma}+\overline{a}_{\ell}\sum_{\nu_{j}=1}^{e_{j}}p_{i,j,\nu_{j},\ell}\times\nonumber\nonumber \\
& \sum_{\beta_{j}=1}^{d_{j}}q_{i,j,\beta_{j},\ell}e^{h_{i}L_{i}\tau}\times\left(\delta^{(e)}+\delta^{(\overline{d})}+\delta^{(d,\overline{d})}\right)\label{eq:optfun}\\
& \boldsymbol{\text{s.t.}}\nonumber\\
&
\eqref{sumWs}-\eqref{lambdae_nuj},\eqref{edgeCacheConst}, \eqref{deltae}-\eqref{eq:abar}
%\eqref{alphad}-
%\eqref{Me_jnuj},\,
%\eqref{lambdac_betaj},\,
%\eqref{lambdae_nuj},\,
%\eqref{edgeCacheConst},\,
%%\eqref{Be},
%%\eqref{rhod},
%%\eqref{Bd},
%%\eqref{rhoc},
%\eqref{eq:delta_e}-\eqref{eq:mgf_dpar}
%%\eqref{Bc},
%%\eqref{deltaAll}
\label{formulas}
\\ 
%&
%\widetilde{M}_{j,\beta_{j},\ell}^{(d,\overline{d})}(h)=\frac{\alpha_{j,\beta_{j},\ell}^{(d)}(\alpha_{j,\beta_{j},\ell}^{(\overline{d})}-h)e^{\eta_{j,\beta_j,\ell}^{(d)}h}}{\alpha_{j,\beta_{j},\ell}^{(\overline{d})}(\alpha_{j,\beta_{j},\ell}^{(d)}-h)e^{\eta_{j,\beta_j,\ell}^{(\overline{d})}h}},\,\,\,\forall j,\beta_{j}\\
& \rho_{j,\beta_{j},\ell}^{(d)}=\sum_{i=1}^{r}\lambda_{i,\ell}\pi_{i,j}q_{i,j,\beta_{j},\ell}e^{\lambda_{i,\ell} \omega_{i,\ell}}\frac{L_{i}-L_{j,i}}{\alpha_{j,\beta_{j},\ell}^{(d)}}<1,\,\,\,\forall j,\beta_{j}\label{eq:rho_beta_d}\\
& \rho_{j,\beta_{j},\ell}^{(\overline{d})}=\sum_{i=1}^{r}\lambda_{i,\ell}\pi_{i,j,\ell}q_{i,j,\beta_{j},\ell}e^{\lambda_{i,\ell} \omega_{i,\ell}}\frac{L_{i}-L_{j,i}}{\alpha_{j,\beta_{j},\ell}^{(\overline{d})}}<1\,\forall j,\beta_{j},\ell\label{eq:rho_beta_c}\\
& \rho_{j,\nu_{j},\ell}^{(e)}=\sum_{i=1}^{r}\lambda_{i,\ell}\pi_{i,j,\ell}p_{i,j,\nu_{j},\ell}e^{\lambda_{i,\ell} \omega_{i,\ell}}\frac{L_{j,i}}{\alpha_{j,\nu_{j},\ell}^{(e)}}<1,\,\,\,\forall j,\nu_{j},\ell\label{eq:rho_nu_e}
\end{align}

\begin{align}
%& \sum_{j=1}^{m}\pi_{i,j,\ell}=1,\,\,\,\forall i,\ell \label{eq:sum_pi_ij}\\
%& \sum_{\nu_{j}=1}^{e_{j}}p_{i,,j,\nu_{j},\ell}=1,\,\,\,\forall i,j,\ell \label{eq:sum_piij_nu}\\
%& \sum_{\beta_{j}=1}^{d_{j}}q_{i,j,\beta_{j},\ell}=1,\,\,\,\forall i,j,\ell\label{eq:sum_q_ij_beta}\\
%& \pi_{i,j,\ell},\,p_{i,,j,\nu_{j},\ell},\,q_{i,,j,\beta_{j},\ell}\geq0,\,\,\,\forall i,j,\beta_{j},\nu_{j},\ell\label{eq:pi_q_p_gz}\\
%& w_{j,\beta_{j},\ell}^{(\overline{d})},\,w_{j,\beta_{j}}^{(d)},\,w_{j,\nu_{j},\ell}^{(e)}\geq0,\,\,\,\forall j,\beta_{j},\nu_{j},\ell\label{eq:w_cde_gz}\\
%& \sum_{\beta_{j}=1}^{d_{j}}w_{j,\beta_{j}}^{(d)}\leq1,\,\,\,\forall j
%\label{eq:sum_w_d}\\
%& \sum_{\beta_{j}=1}^{d_{j}}w_{j,\beta_{j},\ell}^{(\overline{d})}+\sum_{\nu_{j}=1}^{e_{j}}w_{j,\nu_{j},\ell}^{(e)}\leq1,\,\,\,\forall j,\ell\label{eq:sum_w_de}\\
& 
\sum_{i}L_{j,i}\leq C_{j},\,L_{j,i}\geq0,\,\,\,\forall i,j\label{eq:capConst}\\
%& 
%\mathbb{P}(N_{\ell}>N_t)\leq \epsilon_{\ell}\label{P_N_j}\\
& h_{i}<\alpha_{j,\beta_{j}}^{(d)},
%
%\,\,\,\forall i,j,\beta_{j}
%\label{eq:t_leq_alpha_beta_d}
%&
h_{i}<\alpha_{j,\beta_{j},\ell}^{(\overline{d})},
%\,\,\,\forall i,j,\beta_{j},\ell\
%label{eq:t_leq_alpha_beta_c}
%&
h_{i}<\alpha_{j,\nu_{j},\ell}^{(e)},\,\,\,\forall i,j,\nu_{j},\ell \label{eq:t_leq_alpha_nu_e}\\
& 
0<h_{i}-\Lambda_{j,\beta_{j}}^{(d)}(B_{j,\beta_{j},\ell}^{(d)}(h_{i})-1),\,\,\,\forall i,j,\beta_{j},\ell\label{eq:mgf_beta_d}\\
& 0<h_{i}-\Lambda_{j,\beta_{j},\ell}^{(\overline{d})}(B_{j,\beta_{j},\ell}^{(\overline{d})}(h_{i})-1),\,\,\,\forall i,j,\beta_{j}\label{h_const}\\
& 0<h_{i}-\Lambda_{j,\nu_{j},\ell}^{(e)}(B_{j,\nu_{j},\ell}^{(e)}(h_{i})-1),\,\,\,\forall i,j,\nu_{j},\ell\label{eq:mgf_beta_e}\\
&L_{j,i}\in {\mathbb Z}\label{eq:Lij}\\
\boldsymbol{\text{var}} & \text{\,\,\,\,\,}\boldsymbol{\pi,}\,\boldsymbol{q},\,\boldsymbol{p},\,\boldsymbol{h},\,\boldsymbol{w}^{(c)},\,\boldsymbol{w}^{(d)},\,\boldsymbol{w}^{(e)},\,\boldsymbol{L}, \boldsymbol{\omega} \nonumber
\end{align}

%{\bf Abubakr, there are ??, and some of the equations still may be cited from above, like the edge-cache utilization. Further, this needs index for the edge cache. }

Here, in (\ref{formulas}), equations $(1)-(3)$ give the feasibility constraints on the bandwidth allocation, while equations $(4)-(6)$ define the MGFs of the service time distributions, equations $(7)-(10)$ give the feasibility of the two-stage probabilistic scheduling and $(11)-(12)$ define the arrival rates at the different queues. Constraints 
(\ref{eq:rho_beta_d})--(\ref{eq:rho_nu_e}) ensure the stability of the systems queue (do not blow up to infinity).
% while Constraints (\ref{eq:sum_pi_ij})--(\ref{eq:pi_q_p_gz}) guarantee that the two-stage probabilistic scheduling exists. 
% 
% Further, constraints (\ref{eq:w_cde_gz})--(\ref{eq:sum_w_de}) define the bandwidth split and ensure the feasibility of bandwidth allocation weights.
%Constraint (\ref{eq:capConst}) limits the cached segments to the storage server cache capacity. Constraint (\ref{P_N_j}) limits (with  $z^{th}$ percentile) the files in the edge-cache to not exceed $N_t$. The exact expression for this constraint is given in \eqref{edgeCacheConst}. 
Constraints (\ref{eq:t_leq_alpha_nu_e})--(\ref{eq:mgf_beta_e}) ensure that the moment generating functions exist. We note that some optimization variables can be combined to form a single optimization variable which results in having only five independent and separable variables as shown below. In the next subsection, we will describe the proposed algorithm for this optimization problem. 
\subsection{Proposed Algorithm}
\label{sec:algo}

We first note that the two-stage probabilistic scheduling variables are independent and separable, thus we can combine them and define a single variable $\widetilde{\boldsymbol{\pi}}$ such that  
$\widetilde{\boldsymbol{\pi}}=\left(\boldsymbol{\pi}\,,\boldsymbol{\,p}\,\,,\boldsymbol{q}\right)$. Similarly, since the bandwidth allocation weights are independent and separable, we concatenate them in a single optimization variable $\boldsymbol{w}$, where  
$\boldsymbol{w}=\left(\boldsymbol{w}^{(e)}\,,\boldsymbol{\,w}^{(\overline{d})}\,\,,\boldsymbol{w}^{(d)}\right)$. Hence, the weighted SDTP optimization problem given in \eqref{eq:optfun}-\eqref{eq:Lij}  is optimized over five set of variables:  server and PSs scheduling probabilities $\widetilde{\boldsymbol{\pi}}$ (two-stage scheduling probabilities), auxiliary parameters $\boldsymbol{h}$, bandwidth allocation weights $\boldsymbol{w}$, cache placement $\boldsymbol{L}$, and edge cache window size optimization $\boldsymbol{\omega}$.

Clearly, the problem is non-convex in all the parameters jointly, which can be easily seen in the terms which are product of the different variables. Since the problem is non-convex, we propose an iterative algorithm to solve the problem. The proposed algorithm divides the problem into five sub-problems that optimize one variable while fixing the remaining four. These sub-problems are labeled as   (i) Server and PSs Access Optimization: optimizes  $\widetilde{\boldsymbol{\pi}}$, for given  $\boldsymbol{h}$, $\boldsymbol{w}$, $\boldsymbol{\omega}$, and $\boldsymbol{L}$,  
(ii) Auxiliary Variables Optimization: optimizes  $\boldsymbol{h}$ for given $\widetilde{\boldsymbol{\pi}}$, $\boldsymbol{w}$, $\boldsymbol{\omega}$, and $\boldsymbol{L}$,  (iii) Bandwidth Allocation Optimization: optimizes  $\boldsymbol{w}$ for given $\widetilde{\boldsymbol{\pi}}$, $\boldsymbol{h}$, $\boldsymbol{\omega}$, and  $\boldsymbol{L}$. 
(iv) Cache Placement Optimization: optimizes  $\boldsymbol{L}$ for given $\widetilde{\boldsymbol{\pi}}$, $\boldsymbol{h}$, $\boldsymbol{\omega}$, and $\boldsymbol{w}$, (v) Edge-cache Window Size Optimization: optimizes  $\boldsymbol{\omega}$ for given $\widetilde{\boldsymbol{\pi}}$, $\boldsymbol{h}$,  $\boldsymbol{L}$, and $\boldsymbol{w}$. The algorithm is summarized as follows. 

%, where $\boldsymbol{\mathcal{S}}=\left(\mathcal{S}_{1},\mathcal{S}_{2},\ldots,\mathcal{S}_{r}\right)$,
% (ii) access-scheduling probabilities  ($\boldsymbol{\pi}$- Optimization ), and finally (iii) auxiliary variables optimization ($\boldsymbol{t}$- Optimization). These sub-problems have an easier-to-handle structure and can be solved efficiently. Next, we develop an efficient algorithm solution for our optimization problem.
%
%
%The joint mean-tail stall duration optimization problem given in (\ref{eq:joint_otp_prob}) is carried out over three set of variables: chunk placement $\mathcal{S}_{i}$, $\forall$ $i$, scheduling probabilities $\pi_{i,j}$, $\forall$ $i,j$, and auxiliary parameter $t_{i}$, $\forall$ $i$. In order to solve the non-convex problem, we develop an alternating minimization algorithm which is shown to converge to a stationary solution. The proposed algorithm can be written as follows.

\begin{enumerate}[leftmargin=0cm,itemindent=.5cm,labelwidth=\itemindent,labelsep=0cm,align=left]
	\item \textbf{Initialization}: Initialize $\boldsymbol{h}$,\,$\widetilde{\boldsymbol{\pi}}$, $\boldsymbol{w}$, $\boldsymbol{\omega}$
	and $\boldsymbol{L}$ in the feasible set. 
	\item \textbf{While Objective Converges}
	\begin{enumerate}
		
		\item  Run Server Access Optimization using current values of $\boldsymbol{h}$, $\boldsymbol{w}$, $\boldsymbol{\omega}$, 
		and $\boldsymbol{L}$ to get new values of $\widetilde{\boldsymbol{\pi}}$

		\item  Run Auxiliary Variables Optimization using current values of $\widetilde{\boldsymbol{\pi}}$, $\boldsymbol{w}$, $\boldsymbol{\omega}$, 
		and $\boldsymbol{L}$ to get new values of $\boldsymbol{h}$

		\item  Run Bandwidth Allocation Optimization using current values
		of $\widetilde{\boldsymbol{\pi}}$, $\boldsymbol{h}$, $\boldsymbol{L}$, and $\boldsymbol{\omega}$, to get new values of $\boldsymbol{w}$.

		\item  Run Cache Placement Optimization using current values
		of $\widetilde{\boldsymbol{\pi}}$, $\boldsymbol{h}$,   $\boldsymbol{w}$, and $\boldsymbol{\omega}$ to get new values of $\boldsymbol{L}$.

		\item  Run Edge-cache Window Size Optimization using current values
		of $\widetilde{\boldsymbol{\pi}}$, $\boldsymbol{h}$,   $\boldsymbol{w}$, and $\boldsymbol{L}$ to get new values of $\boldsymbol{\omega}$.
		
	\end{enumerate}
\end{enumerate}

The proposed algorithm performs an alternating optimization over the different aforementioned dimensions, such that each sub-problem is shown to have convex constraints and thus can be efficiently solved using the iNner cOnVex Approximation (NOVA)  algorithm proposed in \cite{scutNOVA}. The subproblems are explained in detailed in Appendix \ref{opt_mention}. 

%\subsubsection*{Convergence of the Proposed Algorithm}

We first initialize 
$\widetilde{\boldsymbol{\pi}}$, $\boldsymbol{w}$, $\boldsymbol{h}$, $\boldsymbol{\omega}$, and $\boldsymbol{L}$ $\forall i, j, \nu_j, \beta_j$ such that the choice is feasible for the problem. Then, we do alternating minimization over the five sub-problems defined above. Since each sub-problem can only decrease the objective (properties of convergence of subproblems to a stationary point is given in Appendix \ref{opt_mention}) and the overall problem is bounded from below, we have the following result.

\begin{theorem} 
	The proposed algorithm  converges to a  stationary solution.
\end{theorem}

%We next describe the five sub-problems along with the proposed solutions for the sub-problems.
Appendix \ref{sec:adapt}  describes how our algorithm can be used in an online fashion to keep track of the systems dynamics at the edge-cache. 

%Next, we present our evaluation results for our proposed scheme considering both offline and online versions of our proposed algorithm. 

%\textbf{Example:} add example to further illustrate. xxxx
%\newpage
\section{Implementation and Evaluation}\label{sec:num}

In this section, we evaluate our proposed  algorithm for weighted stall duration tail probability. 

{\scriptsize{}}
\begin{table}[b]
{\scriptsize{}\caption{{\small{}
The value of $\alpha_j$  used in the evaluation results with units of 1/ms. We set $\eta_{j,\beta_j}^{(d)}=\eta_{j,\beta_j}^{(\overline{d})}=\eta_{j,\nu_j}^{(e)}=14$ ms.
%\vspace{-.2in}
%For $1<\ell\leq10$, both $\beta_{j}^{(\ell)}$ and $\alpha_{j}^{(\ell)}$are scaled by $DR^{(\ell)}/DR^{(1)}$.
%\vspace{-.2in}
\label{tab:Storage-Nodes-Parameters}}}
}{\scriptsize \par}

{\scriptsize{}}%
\begin{tabular}{|c|>{\centering}p{2.53cc}|>{\centering}p{2.53cc}|c|c|c|}
\multicolumn{1}{>{\centering}p{2.53cc}}{{\scriptsize{}Node 1}} & \multicolumn{1}{>{\centering}p{2.53cc}}{{\scriptsize{}Node 2}} & \multicolumn{1}{>{\centering}p{2.53cc}}{{\scriptsize{}Node 3}} & \multicolumn{1}{c}{{\scriptsize{}Node 4}} & \multicolumn{1}{c}{{\scriptsize{}Node 5}} & \multicolumn{1}{c}{{\scriptsize{}Node 6}}\tabularnewline
\hline 
{\scriptsize{}$82.00$} & {\scriptsize{}$76.53$} & {\scriptsize{}$71.06$} & {\scriptsize{}$65.6$} & {\scriptsize{}$60.13$} & {\scriptsize{}$54.66$}\tabularnewline
\hline 
\end{tabular}{\scriptsize \par}

{\scriptsize{}}%
\begin{tabular}{|c|>{\centering}p{2.53cc}|>{\centering}p{2.53cc}|c|c|c|}
 \multicolumn{1}{>{\centering}p{2.53cc}}{{\scriptsize{}Node 7}} & \multicolumn{1}{>{\centering}p{2.53cc}}{{\scriptsize{}Node 8}} & \multicolumn{1}{>{\centering}p{2.53cc}}{{\scriptsize{}Node 9}} & \multicolumn{1}{c}{{\scriptsize{}Node 10}} & \multicolumn{1}{c}{{\scriptsize{}Node 11}} & \multicolumn{1}{c}{{\scriptsize{}Node 12}}\tabularnewline
\hline 
{\scriptsize{}$49.20$} & {\scriptsize{}$44.28$} & {\scriptsize{}$39.36$} & {\scriptsize{}$34.44$} & {\scriptsize{}$29.52$} & {\scriptsize{}$24.60$}\tabularnewline
\hline 
\end{tabular}{\scriptsize \par}
%\vspace{-.2in}
\end{table}
{\scriptsize \par}

\begin{figure}[t]
	\centering
	\includegraphics[trim=0.0in 0in 1.25in 0.45in, clip, width=0.47\textwidth]{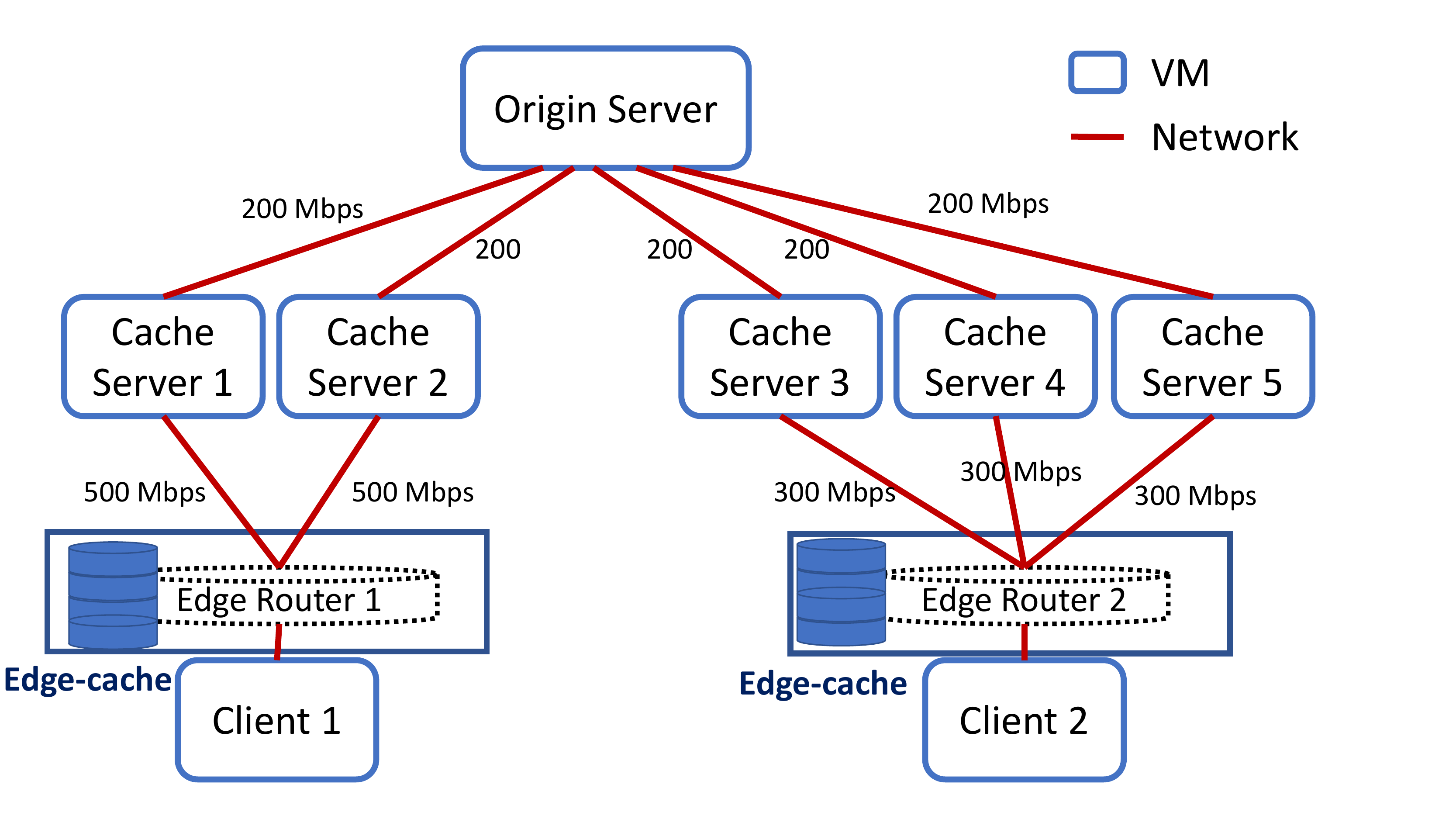}
	%	\vspace{-0.18in}
	\captionof{figure}{\small Testbed in the cloud.}
	\label{fig:alphaScaling}
	%	\vspace{-0.4in}
\end{figure}

%\hspace{.05in}

\begin{figure*}[htbp]
\begin{minipage}{.48\textwidth}
	\centering
	\includegraphics[trim=0.1in 0.01in 4.10in 0.01in, clip, width=\textwidth]{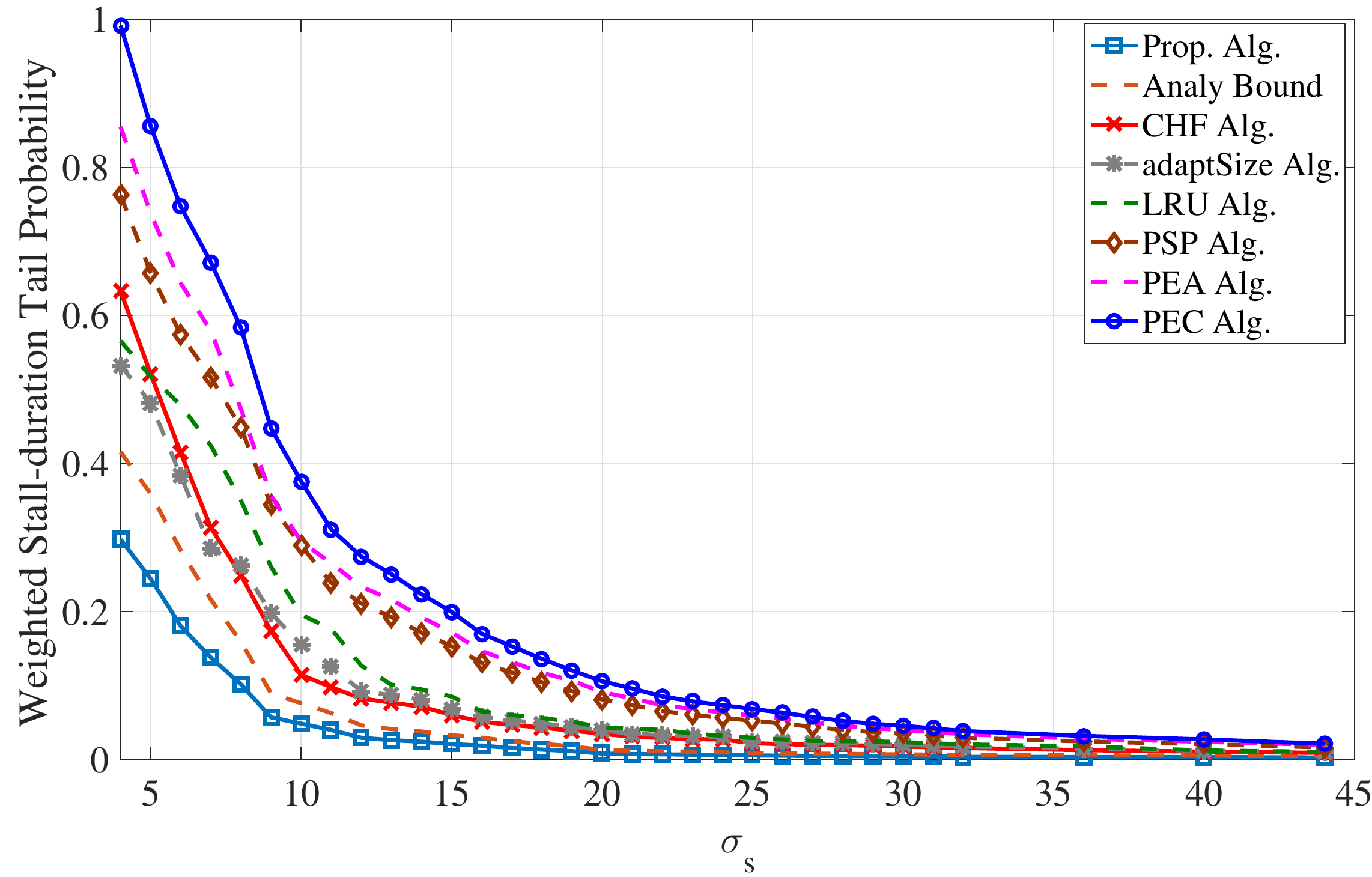}
%\vspace{-.2in}
\captionof{figure}{ \small Weighted SDTP versus $\sigma_s$.}
\label{fig:SDTP_vs_sigma}
\end{minipage}
\hspace{0.25in}
\begin{minipage}{.48\textwidth}
	\centering
\includegraphics[trim=0.05in 0in 3.9in 0.0in, clip, width=\textwidth]{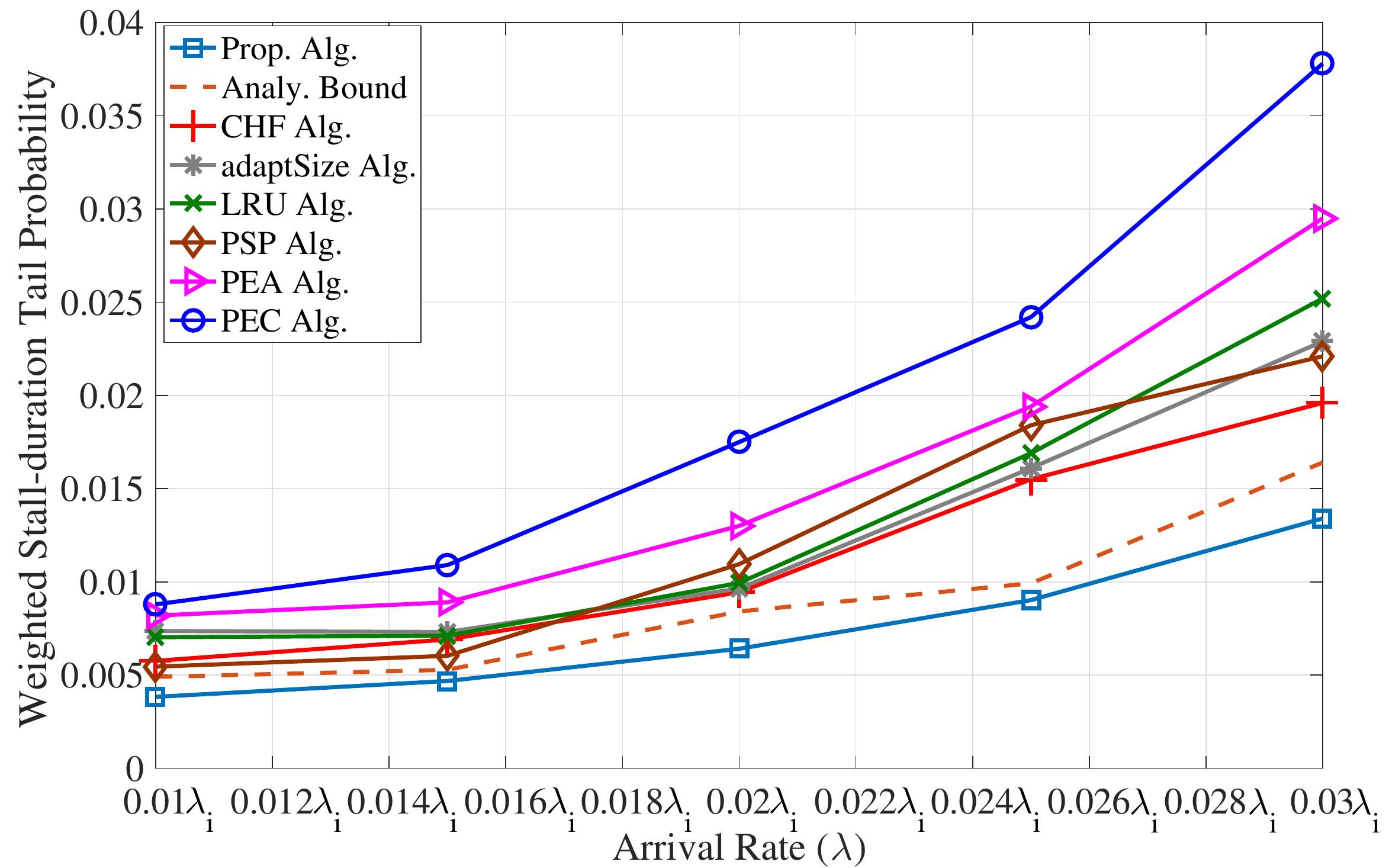}
\vspace{-.2in}
\captionof{figure}{ \small Weighted stall-duration tail probability versus arrival rate of video files.  We vary the arrival rate of the files from $0.01\lambda_i$ to $0.03\lambda_i$ with an increment step of $0.002$, where $\lambda_i$ is the base arrival rate. }%Here, we compare our proposed approach with the four considered baselines.}
\label{fig:arrRate}
\end{minipage}
\end{figure*}

\subsection{Testbed Configuration and Parameter Setup}
\label{testbed}
% Testbed configuration
\begin{table}[t]
	   \caption{\small Testbed Configuration}
  \small
   \centering
   %\topcaption{Table captions are better up top} % requires the topcapt package
  \resizebox{.48\textwidth}{!}{
%\begin{tabular*}{.5\textwidth}{@{} llr @{}}
   \begin{tabular}{@{} llr @{}} % Column formatting, @{} suppresses leading/trailing space
      \hline
      \multicolumn{2}{c}{Cluster Information}         \\
      %\cmidrule(r){1-2} % Partial rule. (r) trims the line a little bit on the right; (l) & (lr) also possible
      \hline
      Control Plane     &       Openstack Kilo    \\ 
      VM flavor                 &     1 VCPU, 2GB RAM, 20G storage (HDD)  \\ 
      \hline
      \hline
      \multicolumn{2}{c}{Software Configuration} \\ 
      \hline
      Operating System       &  Ubuntu Server 16.04 LTS \\ 
      Origin Server(s)          &  Apache Web Server~\cite{apacheweb}: Apache/2.4.18 (Ubuntu)     \\ 
      Cache Server(s)           & Apace Traffic Server~\cite{trafficserv} 6.2.0 (build \# 100621)    \\ 
      Client                & Apache JMeter~\cite{jmeter} with HLS plugin~\cite{hlsplugin} \\
      \hline
   \end{tabular}
}
%   \caption{\small Testbed Configuration}
   \label{tbl:testbed}
%   \vspace{-0.3in}
\end{table}

We construct an experimental environment in a virtualized cloud environment 
managed by Openstack~\cite{openstack} to investigate our proposed SDTP framework.  We allocated one VM for an origin server and 5 VMs for cache servers intended
to simulate two locations (e.g., different states). We implement the proposed online caching mechanism in the edge cache that takes the inputs of $\omega_{i,\ell}$ at each edge router. When a video file is requested, it is stored in the edge-cache for a window size of $\omega_{i,\ell}$ time units (unless requested again in this window). For the future requests within $\omega_i$ or concurrent user requests, the requests for the video chunks are served from the edge-cache, and thus future/concurrent users would experience lower stall duration. If the file can be accessed from the edge router,  higher caching level is not used for this request which consequently  reduces the traffic at the core backbone servers. If the file cannot be accessed from the edge router, it goes to the distributed cache. We assume some segments, i.e., $L_{j,i}$, of video file $i$ are stored in the distributed cache node $j$, and are  served from the cache nodes. The non-cached segments are served from the data-center.  The schematic of our testbed is illustrated in Figure~\ref{fig:alphaScaling}. Since the two edge-routers are likely in different states, they may not share the cache servers which is the setup we study in the experiments. We note that the theoretical approach proposed earlier is general and can work with shared cache servers across multiple edge routers.

One VM per location is used for generating client workloads. 
Table~\ref{tbl:testbed} summarizes a detailed configuration used for the experiments.  
For client workload, we exploit a popular HTTP-traffic generator, Apache JMeter, with a plug-in 
that can generate traffic using HTTP Streaming protocol. 
We assume the amount of available bandwidth between origin server and each cache server is 200 Mbps, 
500 Mbps between cache server 1/2 and edge router 1, and 300 Mbps between cache server 3/4/5 and edge router 2.
In this experiment, to allocate bandwidth to the clients, we throttle the client (i.e., JMeter) traffic according to the plan generated by our algorithm.  We consider $1000$ threads (i.e., users) and set $e_j^{(\ell)}=40$ for all $\ell= 1, 2$, $d_j=20$. Segment size $\tau$ is set to be equal to $8$ seconds. Each edge cache is assumed to have a capacity, equivalent to $15\%$ of the total size of the video files. Further, distributed cache servers can store up to $35\%$ out of the total number of video file segments. The values of $\alpha_{j}$ and $\eta_{j}$ are summarized in Table I.

% and the distributed cache servers are assumed to store only $35\%$ out of the total number of video file segments. Unless otherwise explicitly stated, the edge cache capacity is limited to $15\%$ of the total size of the video files. 

Video files are generated based on Pareto distribution \cite{arnold2015pareto} (as it is a commonly used distribution for file sizes \cite{Vaphase}) with shape factor of $2$ and scale of $300$, respectively. While we stick in the experiment to these parameters, our analysis and results remain applicable for any setting given that the system maintains stable conditions under the chosen parameters. Since we assume that the video file sizes are not heavy-tailed, the first $500$ file-sizes that are less than 60 minutes are chosen. When generating video files, the size of each video file is rounded up to the multiple of $8$ seconds. For the arrival rates, we use the data from our production system for  500 hot files from two edge routers, and use those arrival rates. The aggregate arrival rates at edge router 1 and router 2 are $\Lambda_1 = 0.01455s^{-1}$, $\Lambda_2 = 0.02155s^{-1}$, respectively. 

%, as explained earlier, i.e., Figure \ref{fig:flowchart} and Figure \ref{fig:windowSize}. 

In order to generate the policy for the implementation, we assume uniform scheduling,  $\pi_{i,j}=k/n$, $p_{j,\nu_j} = 1/e_j$, $q_{j,\beta_j} = 1/d_j$. Further, we choose  $t_{i}=0.01$, $w_{j, \nu_j}^{(e)}=1/e_j $, $w_{j, \beta_j}^{(\overline{d})}=1/d_j$ and $w_{j, \beta_j}^{(d)}=1/d_j$. However, these choices of the initial parameters may not be feasible. Thus, we modify the parameter initialization to be closest norm feasible solutions. Using the initialization, the proposed algorithm is used to obtain the parameters. These parameters are then used to control the bandwidth allocation, distributed cache content placement, the probabilistic scheduling parameters, and the edge caching window sizes. Based on these parameters, the proposed online algorithm is implemented. Since we assume the arrivals of video files are Poisson (and hence inter-arrival time is exponential with $\lambda_i$ for file $i$), we generate a sequence of $10000$ video file arrivals/requests corresponding to the different files at each edge router. Upon an arrival of a video file at edge-cache, we apply our proposed online mechanism. For each segment, we used JMeter built-in reports to estimate the downloaded time of each segment and then plug these times into our model to obtain the stall duration which will be used for evaluation of the proposed method.

%Based on one week trace from our production system, we estimate the aggregate arrival rates at edge router 1 and router 2 to be $\Lambda_1 = 0.01455s^{-1}$, $\Lambda_2 = 0.02155s^{-1}$, respectively. Then, HLS sampler (i.e., request) is sent every $3$s. We assume $40\%$ of the segments are stored in the cache and hence the remaining segments are served from the origin server. The per file arrival rate is also estimated from our production system and the video files with the highest arrival rates (most popular ones) are stored in the storage cache servers. The video files are $300$s of length and the segment length is set to be $8$s. 

%For each segment, we used JMeter built-in reports to estimate the downloaded time of each segment and then plug these times into our model to get the SDTP. We first run the optimization problem online to get the optimized system variables. Then, we use those optimized parameters to track the online updates in the system and change the edge-cache content according to our proposed strategy. 

%In the following subsection, we provide a list of competitive strategies to compare with. 

\subsection{Baselines}
We  compare our proposed approach with multiple strategies, which are described as follows. 
\begin{enumerate}[leftmargin=0cm,itemindent=.5cm,labelwidth=\itemindent,labelsep=0cm,align=left]
	%1
	\item {\em Projected Equal Server-PSs Scheduling, Optimized Auxiliary variables, Cache Placement, Edge-cache Window-Size, and Bandwidth Wights (PEA):} Starting with the initial solution mentioned above, the problem in \eqref{eq:optfun} is optimized over the choice of $\boldsymbol{h}$, $\boldsymbol{w}$,  $\boldsymbol{L}$, and $\boldsymbol{\omega}$ (using Algorithms \ref{alg:NOVA_Alg1}, \ref{alg:NOVA_Alg3},  
	\ref{alg:NOVA_Alg5}, and \ref{alg:NOVA_Alg6} 
	respectively)  using alternating minimization. Thus, the values of $\pi_{i,j}$, $p_{i,j,\nu_j}$, and $q_{i,j,\beta_j}$ will be approximately close to $k/n$, $1/e_j$, and $1/d_j$, respectively,  for all $i,j,\nu_j,\beta_j$.

%	%2
%	\item {\em Projected Equal Bandwidth, Optimized Access Servers and PS scheduling Probabilities, Auxiliary variables, Edge-cache Window-Size, and cache placement (PEB):} Starting with the initial solution mentioned above, the problem in \eqref{eq:optfun} is optimized over the choice of $\widetilde{\boldsymbol{\pi}}$
%	, $\boldsymbol{h}$,  $\boldsymbol{L}$, and $\boldsymbol{\omega}$ (using Algorithms \ref{alg:NOVA_Alg1Pi}, \ref{alg:NOVA_Alg1},   \ref{alg:NOVA_Alg5}, and \ref{alg:NOVA_Alg6}, respectively) using alternating minimization. Thus, the bandwidth allocation weights, $w_{j,\nu_j}^{(e)}$, $w_{j,\beta_j}^{(\overline{d})}$, $w_{j,\beta_j}^{(d)}$ will be approximately $1/e_j$, $1/d_j$, and $1/d_j$, respectively. 

	%% 3
	\item {\em Projected Proportional Service-Rate, Optimized Auxiliary variables, Bandwidth Wights, Edge-cache Window-Size, and Cache Placement (PSP):} In the initialization, the access probabilities among the servers, are given as  $\ensuremath{\pi_{i,j}=\frac{\mu_{j}}{\sum_{j}\mu_{j}},\,\forall i,j}$.
	This policy assigns servers proportional to their service rates. The choice of all parameters are then modified to the closest norm feasible solution.  Using this initialization, the problem in \eqref{eq:optfun} is optimized over the choice of $\boldsymbol{h}$, $\boldsymbol{w}$,  $\boldsymbol{L}$, and $\boldsymbol{\omega}$, (using Algorithms \ref{alg:NOVA_Alg1},  \ref{alg:NOVA_Alg3},  \ref{alg:NOVA_Alg5}, and
	\ref{alg:NOVA_Alg6}, respectively)  using alternating minimization.

	%% 4
	\item {\em Projected Equal Caching, Optimized Scheduling Probabilities, Auxiliary variables and Bandwidth Allocation Weights (PEC):}
	In this strategy, we divide the cache size equally among the video files. Thus, the size of each file in the cache is the same (unless file is smaller than the cache size divided by the number of files). Using this initialization, the problem in \eqref{eq:optfun} is optimized over the choice of $\ensuremath{\widetilde{\boldsymbol{\pi}}}$, $\boldsymbol{h}$, $\boldsymbol{w}$, and $ \boldsymbol{\omega}$  (using Algorithms  \ref{alg:NOVA_Alg1Pi}, \ref{alg:NOVA_Alg1},  \ref{alg:NOVA_Alg3}, and \ref{alg:NOVA_Alg6}, respectively)  using alternating minimization. 
	
		%% 5
	\item {\em Caching Hot Files, Optimized Scheduling Probabilities, Auxiliary variables, Edge-cache Window-Size, and Bandwidth Allocation Weights (CHF):}
	In this strategy, we cache entirely the files that have the largest arrival rates in the storage cache server. Such hot file caching policies have been studied in the literature, see  \cite{cacheChes02} and references therein.  Using this initialization, the problem in \eqref{eq:optfun} is optimized over the choice of $\ensuremath{\widetilde{\boldsymbol{\pi}}}$, $\boldsymbol{h}$, $\boldsymbol{w}$, and $ \boldsymbol{\omega}$  (using Algorithms  \ref{alg:NOVA_Alg1Pi}, \ref{alg:NOVA_Alg1},  \ref{alg:NOVA_Alg3}, and \ref{alg:NOVA_Alg6}, respectively)  using alternating minimization. 

		%% 6
\item {\em Caching based on Least-Recently-Used bases at edge-cache and Caching-Hottest files at storage nodes, Optimized Scheduling Probabilities, Auxiliary variables, Storage Cache Placement, and Bandwidth Allocation Weights (LRU):}
In this strategy, a file is entirely cached in the edge-cache servers upon request if space permits; otherwise, the least-recently used file(s) is removed first to evacuate the needed space for the new file. Further, the hottest files are partially cached in the distributed storage cache servers. Such hot file caching policies have been studied in the literature, e.g., \cite{cacheChes02} and references therein.  Using this initialization, the problem in \eqref{eq:optfun} is optimized over the choice of $\ensuremath{\widetilde{\boldsymbol{\pi}}}$, $\boldsymbol{h}$, and $\boldsymbol{w}$,  (using Algorithms  \ref{alg:NOVA_Alg1Pi}, \ref{alg:NOVA_Alg1},  and \ref{alg:NOVA_Alg3}, respectively)  using alternating minimization. 	

\item { \em Caching  at edge-cache based on adaptSize policy \cite{berger2017adaptsize} and Caching-Hottest files at storage nodes, Optimized Scheduling Probabilities, Auxiliary variables, Storage Cache Placement, and Bandwidth Allocation Weights (adaptSize)}: This policy is a probabilistic admission policy in which a video file is admitted into the cache with
probability $e^{-size/c}$ so as larger objects are admitted with lower probability and the parameter $c$ is tuned to
maximize the object hit rate (OHR), defined as the probability that a requested file is found in the cache. In particular, given a $c$ and an estimate on the arrival rate for the requests for each video file, one can estimate the
probability that a given file will be served from the edge-cache. One can then use these probabilities to compute the OHR as a function of $c$ and then optimize. The value of  $c$ is recomputed after a certain number of file requests, using a sliding window approach. We refer the reader to \cite{berger2017adaptsize} for a more in-depth description. 

\item {\em Caching  at edge-cache based on variant of LRU policy \cite{garetto2016unified}, Caching-Hottest files at storage nodes, Optimized Scheduling Probabilities, Auxiliary variables, Storage Cache Placement, and Bandwidth Allocation Weights ($xLRU$)}: 
We denote by $xLRU$ one of the these policies: $qLRU$, $kLRU$, and $k$Random. 
A $qLRU$ policy is the same as LRU except that files are only added with probability $q$.
In $kLRU$, 
requested files must traverse $k-1$ additional virtual LRU caches before it is added to the actual cache. kRandom is the same as $kLRU$ except files are evicted from the cache at random. The other optimization parameters are optimized the same way as in the adaptSize policy.

\end{enumerate} 

\subsection{Experimental Results}

%\textit{Service Time Distribution:} We first run  experiments to measure the actual service time distribution in our cloud environment. Figure \ref{fig:serTimeFig} depicts the cumulative distribution function (CDF) of the chunk service
%time for different bandwidths. Using these results, we show that the service time of the chunk can be well approximated by a shifted-exponential distribution with a rate of 24.60s, 29.75s for a bandwidth of 25 Mbps and 30 Mbps, respectively. These results also verify that actual service time does not follow an exponential distribution. This observation has also been made earlier in \cite{Yu-TON16}. Further, the parameter for the exponential is almost proportional to the bandwidth while the shift is nearly constant, which validates the model.

%This observation is also evident because a typical real service distribution is unlikely to have positive probabilities for even very small service times.

\textit{SDTP performance for different $\sigma$:} Figure \ref{fig:SDTP_vs_sigma} shows the decay of weighted SDTP $\sum_{i=1}^{r}\frac{\lambda_i}{\overline{\lambda}_i}\mathbb{P}(\Gamma^{(i)}>\sigma)$ with $\sigma$ (in seconds) for the considered policies. Notice
that SDTP Policy solves the optimal weighted stall tail probability via proposed alternating optimization algorithm. Also, this figure
represents the complementary cumulative distribution function (ccdf) of the proposed algorithm as well as the selected baselines. We further observe that uniformly accessing servers and simple service-rate-based scheduling are unable to optimize the
request scheduler based on factors like chunk placement, request arrival rates, different stall weights, thus leading
to much higher SDTP. Moreover, the figure shows that an entire video file
does not have to be present in the edge-cache. That's because when the user requests a cached video, it is served by first sending the portion of the video locally present at edge-cache
while obtaining the remainder from the distributed cache servers and/or the origin server, and transparently passing it on to the client. In addition, we see that the analytical (offline) SDTP is very close to the actual (online) SDTP measurement on our testbed. Further, since adaptSize policy does not intelligently incorporate the arrival rates in adding/evicting the video files, it fails to significantly reduce the SDTP.
To the best of our knowledge, this is the first work to jointly consider all key design degrees of freedom, including bandwidth allocation among different parallel streams, cache content placement,  the request scheduling, window-size of the edge-cache and the modeling variables associated with the SDTP bound. 

\textit{Arrival Rates Comparisons:} Figure \ref{fig:arrRate} shows the effect of increasing system workload, obtained by varying the arrival rates of the video files from $0.01s^{-1}$ to $0.03s^{-1}$ with an increment step of $0.002s^{-1}$ on the SDTP. We notice a significant improvement of the QoE metric with the proposed strategy as compared to the baselines. Further, the gap between the analytical offline bound and actual online SDTP is small which validates the tightness of our proposed SDTP bound. Further, while our algorithm optimizes the system parameters offline, this figure shows that an online version of our algorithm can be used to keep track of the systems dynamics and thus achieve an improved performance.

\begin{figure}[t]
%\centering\includegraphics[scale=0.50]{figs_up/cdf_actual_impl_v3}
\centering\includegraphics[trim=0.01in 0.05in 4.250in 0.0in, clip, width=0.41\textwidth]{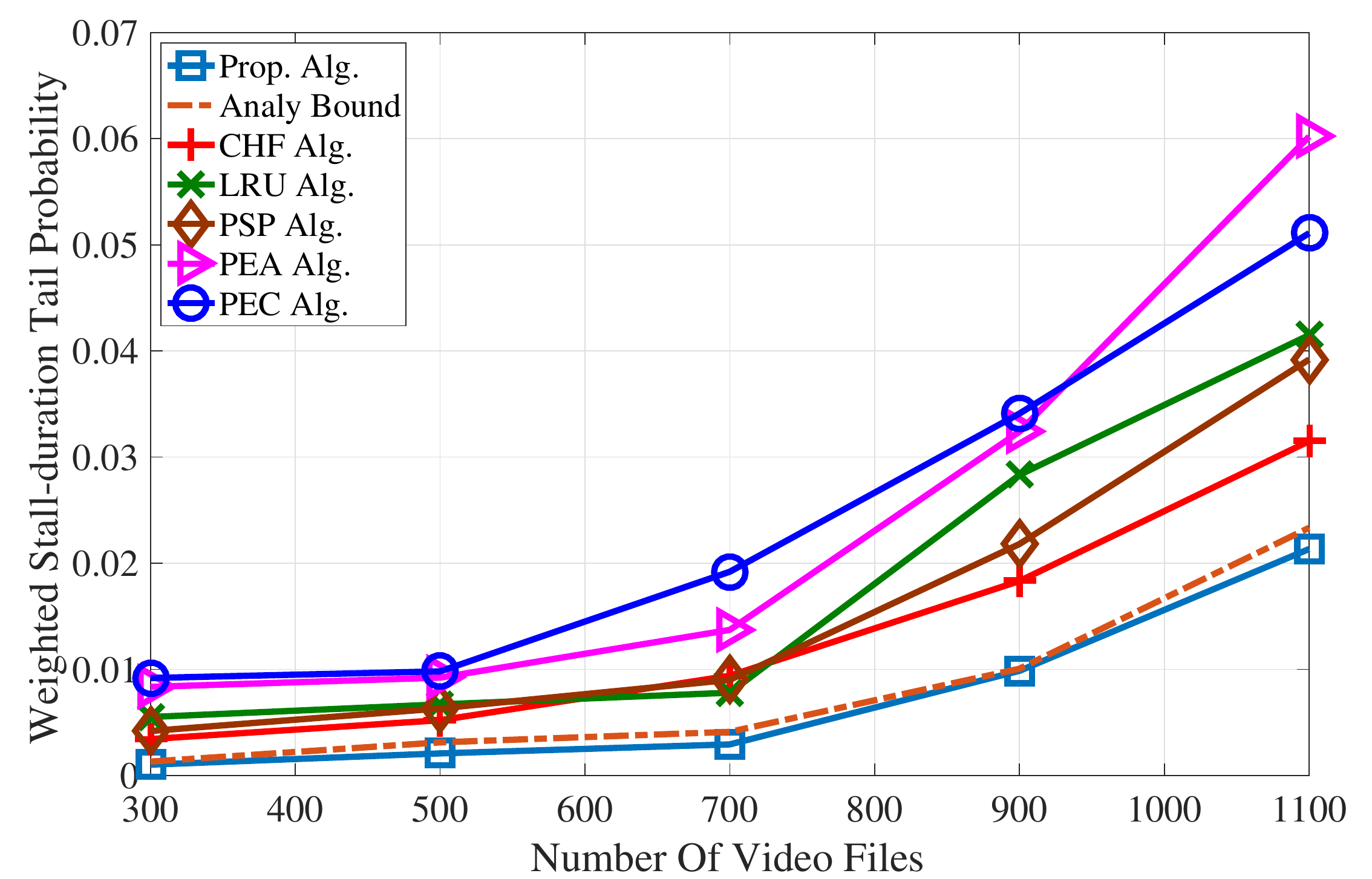}
%\vspace{-.45in}
\caption{\small Comparison of implementation results of our SDTP Algorithm to Analytical SDTP and PEA-based SDTP.
\label{fig:noFiles}}
%\vspace{-.25in}
\end{figure}

\textit{Effect of Number of files:} Figure \ref{fig:noFiles} shows the impact of varying the number of files from $150$ to $550$
on the weighted SDTP for the online algorithm. Clearly, weighted SDTP increases with the number of files, which brings in more workload (i.e., higher arrival rates). However, our
optimization algorithm optimizes new files along with existing
ones to keep overall weighted SDTP
at a low level. We note that the proposed optimization strategy
effectively reduces the tail probability and outperforms
the considered baseline strategies. Thus, joint optimization
over all optimization parameters help reduce the tail
probability significantly. Also, the gap between online and offline performance is almost negligible which reflects the robustness of our algorithm.

%\textit{Effect of increasing the edge cache:} Figure \ref{fig:missRate} plots the weighted SDTP for our online SDTP algorithms. We increase the edge-cache size from $0.10 C_{tot}$ to $0.40 C_{tot}$ with an increment step of $0.05 C_{tot}$. The parameters (e.g., $\boldsymbol{\omega}$, $\widetilde{\boldsymbol{\pi}}$, $\boldsymbol{h}$, $\boldsymbol{w}$, and $\boldsymbol{L}$) of the SDTP algorithm are first optimized offline and then we run the online algorithm using these optimized parameters. We set  the probability that the cache capacity is exceeded to  $0.05$.  As expected, the weighted SDTP decreases as the edge-cache capacity increases since more video files can be served from the edge-cache without getting the video content from the higher hierarchical storage levels which reduces the stall accordingly.
 
 Additional performance evaluation is provided in Appendix  \ref{edgeCacheEval} and Appendix \ref{tradeoff_sec}.

% 
% 
% 
% We see that the SDTP slightly increases when $\epsilon$ increases because the system can tolerate more violations when $\epsilon$ is high. However, in all given points, we observe that the SDTP does not increase significantly since our proposed algorithm adapts to the systems dynamics to maintain the SDTP at low levels. Thus, the proposed algorithm optimizes the systems parameters efficiently and thus reduce the the SDTP.     

%\begin{figure}[t]
%	\centering\includegraphics[trim=0.01in 0.0in 0.50in 0.0in, clip, width=0.35\textwidth]{res/TCPBWQ1_50Mbps_fitting}
%%	\vspace{-.45in}
%	\caption{\small
%		Comparison of actual chunk service time distribution and shifted-exponential distribution with the corresponding mean and shift. It verifies that the actual service time of a chunk can be well approximated by a shifted exponential distribution.
%		\label{fig:serTimeFig}}
%%	\vspace{-.25in}
%\end{figure}

%\begin{figure}[t]
%	\centering\includegraphics[trim=0.01in 0.05in 0.150in 0.0in, clip, width=0.45\textwidth]{res/SDTP_vs_edgeCacheCapacity_v2}
%	%	\vspace{-.45in}
%	\caption{\small Weighted SDTP versus edge-cache capacity. The edge-cache capacity varies from $0.10\times C_{tot}$ to $0.40\times C_{tot}$, where $C_{tot}$ is the total size of all files, i.e., $\sum_{i=1}^{N}\tau L_i$. \label{fig:missRate}}
%	%	\vspace{-.25in}
%\end{figure}

%\begin{figure}[ht]
%	\centering\includegraphics[scale=0.30]{newFigs/NoVideos_expm}
%%	\vspace{-.45in}
%	\caption{\small Comparison of implementation results of our SDTP Algorithm to Analytical SDTP and PEA-based SDTP.
%		\label{fig:impFig}}
%%	\vspace{-.25in}
%\end{figure}
\section{Conclusion}\label{sec:conc}

This paper proposes a  CDN-based edge-cache-aided over-the-top multicast video streaming system, where the video content is partially stored on distributed cache servers and access-dependent online edge caching strategy is used at the edge-cache. Further, this paper optimizes the weighted stall duration tail probability by considering two-stage probabilistic scheduling for the choice of servers and the parallel streams between the server and the edge router. Using the two-stage probabilistic scheduling and the  edge caching mechanism, upper bound on the stall duration tail probability is characterized.  Further, an optimization problem that minimizes the weighted stall duration tail probability is formulated, over the choice of two-stage probabilistic scheduling, bandwidth allocation, cache placement, edge-cache parameters, and the auxiliary variables in the bound.  An efficient algorithm is proposed to solve the optimization problem and the experimental results on a virtualized cloud system managed by Openstack depict the improved performance of our proposed algorithm as compared to the considered baselines. Possible extensions to accommodate multiple quality levels and different chunk sizes are discussed in Appendix \ref{qltyext}.  However, a complete treatment of adaptive bit-rate video streaming is left as a future work. 

\section{Acknowledgment}
We would like to thank Prof. Tian Lan from GWU for helpful discussions related to this work. 

%\newpage
%\clearpage
%\bibliographystyle{plain}
%\bibliographystyle{ACM-Reference-Format}
\bibliographystyle{IEEEtran}
%\pagebreak{}
\bibliography{vidStallRef,allstorage,Tian,ref_Tian2,ref_Tian3,Vaneet_cloud,Tian_rest}
\newpage
\clearpage
\appendices
%\section{Appendices}
%\section{Proof of Lemma \label{tailLemma}}

%\input{dltime}
%\input{dltimelast}
%\input{pltime}
\section{Proof of Lemma \ref{PoisArr2ndQ} }
\label{2ndQArr}

We note  that the arrivals at the second queue in  two $M/M/1$ tandem networks are Poisson is well known in the queueing theory literature \cite{gross011}. The service distribution of the first queue in this paper is a shifted exponential distribution. We note that the deterministic shift also does not change the distribution since the number of arrivals in any time window in the steady state will be the same. Thus, the arrival distribution in the second queue will still be Poisson. 
\section{Download and Play Times of a Segment not Requested in $\omega_i$}

\label{sec:DownLoadPlay}

Since we assume the edge-router index, we will omit $\ell$ in this section. In order to characterize the stall duration tail probability, we need
to find the download time and the play time of different video segments, for any server $j$ and streams with the choice of $\beta_j$ and $\nu_j$, assuming that they are not requested in $\omega_i$. The optimization over these decision variables will be considered in this paper. 
%In the following, we will first estimate the download time of segments from the cache servers and from datacenter, respectively. Then, based on the download time of each segment, we derive the play time of different segments and then give the stall duration expression for each segment request.
\subsection{Download Times of first $L_{j,i}$ Segments}

We consider a queueing model, where $W_{j,\nu_{j}}^{(e)}$ denotes the random  waiting time of all video files in the queue of $PS^{(e,j)}_{\nu_j}$ before file $i$ request is served, 
and $Y_{i,j,\nu_{j}}^{(e,g)}$ be the (random service time of a coded chunk
$g$ for file $i$ from server $j$ and queue $\nu_{j}$. Then, for $g\le L_{j,i}$, the random download time of the first $L_{j,i}$ segments $g\in\{1,\ldots,L_{j,i}\}$ if file $i$ from stream $PS^{(e,j)}_{\nu_j}$ is given as 
 \begin{equation}
D_{i,j,\beta_j,\nu_{j}}^{(g)}=W_{j,\nu_{j}}^{(e)}+\sum_{v=1}^{g}Y_{i,j,\nu_{j}}^{(e,v)}
\end{equation}
Since video file $i$ consists of $L_{j,i}$ segments stored at cache
server $j$, the total service time for video file $i$ request at queue
$PS^{(e,j)}_{\nu_j}$, denoted by $ST_{i,j,\nu_{j}}$, is given as 
\begin{equation}
ST_{i,j,\nu_{j}}=\sum_{v=1}^{L_{j,i}}Y_{i,j,\nu_j}^{(e,v)}
\end{equation}
Hence, the service time of the video files at the parallel stream $PS^{(e,j)}_{\nu_j}$ is given as  
\begin{equation}
R_{j,\nu_{j}}^{(e)}=\begin{cases}
ST_{i,j,\nu_{j}}^{(e,L_{j,i})} \ \   \text { with prob. } \frac{\lambda_{i}\pi_{i,j}p_{i,j,\nu_j}e^{\lambda_i \omega_i}}{\Lambda_{j,\nu_{j}}^{(e)}}& \forall i\end{cases}
\end{equation}

We can show that the moment generating function of the service time for all video files from parallel stream $PS^{(e,j)}_{\nu_j}$   is given by 
\begin{equation}
B_{j,\nu_{j}}^{(e)}(t)={\mathbb E}[e^{tR_{j,\nu_{j}}^{(e)}}] =\sum_{i=1}^{r}\frac{\lambda_{i}\pi_{i,j}p_{i,j,\nu_j}e^{\lambda_i \omega_i}}{\Lambda_{j,\nu_{j}}^{(e)}}\left(\frac{\alpha_{j,\nu_{j}}^{(e)}e^{\eta_{j,\nu_j}^{(e)}t}}{\alpha_{j,\nu_{j}}^{(e)}-t}\right)^{L_{j,i}}\label{Be}
\end{equation}
Further, based on our 2-stage scheduling policy, the load intensity at $PS^{(e,j)}_{\nu_j}$  is as follows 
\begin{align}
\rho_{j,\nu_{j}}^{(e)} & =\Lambda_{j,\nu_{j}}^{(e)}B_{j,\nu_{j}}^{(e)'}(0)\\
& =\sum_{i=1}^{r}\lambda_{i}\pi_{i,j}p_{i,j,\nu_{j}}e^{\lambda_{i}\omega_{i}}L_{j,i}\left(\eta_{j,\nu_{j}}^{(e)}+\frac{1}{\alpha_{j,\nu_{j}}^{(e)}}\right)\label{rhoe}
\end{align}

Since the arrival is Poisson and the service time is shifted-exponentially distributed, the moment generating function (MGF) of the waiting time at queue $PS^{(e,j)}_{\nu_j}$ can be calculated usingthe  Pollaczek-Khinchine formula, i.e.,
%(since M/M/1 is a special case of M/G/1) 
%and is given by
\begin{align}
\mathbb{E}\left[e^{tW_{j,\nu_{j}}^{(e)}}\right] & =\frac{(1-\rho_{j,\nu_{j}}^{(e)})t}{t-\Lambda_{j,\nu_{j}}^{(e)}(B_{j,\nu_{j}}^{(e)}(t)-1)}
\end{align}
From the MGF of $W_{j,\nu_{j}}^{(e)}$ and the service
time, the MGF of the download time of segment $g$ from the queue $PS^{(e,j)}_{\nu_j}$ for file $i$ is then
\begin{equation}
\mathbb{E}\left[e^{tD_{i,j,\beta_j,\nu_{j}}^{(g)}}\right]=\frac{(1-\rho_{j,\nu_{j}}^{(e)})t}{t-\Lambda_{j,\nu_{j}}^{(e)}(B_{j,\nu_{j}}^{(e)}(t)-1)}\left(\frac{\alpha_{j,\nu_{j}}^{(e)}e^{\eta_{j,\nu_j}^{(e)}t}}{\alpha_{j,\nu_{j}}^{(e)}-t}\right)^{g}
\label{D_nu_j}.
\end{equation}
We note that the above is defined only when MGFs exist, i.e., 
\begin{eqnarray}
t&<& \alpha_{j,\nu_{j}}^{(e)}\label{alpha_j_nuj}\\
0 &<& t-\Lambda_{j,\nu_{j}}^{(e)}(B_{j,\nu_{j}}^{(e)}(t)-1)
\label{mgf_gz_e_queue}
\end{eqnarray}

%Finally, the overall download time of all the $L_{j,i}$ segments of video $i$ at the client,
%$D_{i}^{(g)}$ , is given by
%
%
%\begin{equation}
%D_{i}^{(g)}=\sum_{j=1}^{m}\pi_{i,j}\sum_{\nu_{j}=1}^{e_{j}}p_{j,\nu_{j}}D_{i,j,\nu_{j}}^{(g)},\,\,\,\,\,g\leq L_{j,i}
%\end{equation}

\subsection{Download Times of last $(L_i - L_{j,i})$ Segments }

Since the later video segments $(L_i - L_{j,i})$ are downloaded from the data center, we need to schedule them to the $\beta_j$ streams using the proposed probabilistic scheduling policy.
%and thus the choice is among the 
We first determine the time it takes for chunk $g$ to depart the first queue (i.e., $\beta_j$ queue at datacenter). For that, we define the time of chunk $g$ to depart the first queue as
\begin{equation}
E_{i,j,\beta_{j}}^{(g)}=W_{j,\beta_{j}}^{(d)}+\sum_{v=L_{j,i}+1}^{L_i}Y^{(d,v)}_{j,\beta_{j}},
\end{equation}
where $W_{j,\beta_{j}}^{(d)}$ is the waiting time from $PS^{(d,j)}_{\beta_j}$ for the earlier video segments, and $Y^{(d,v)}_{j,\beta_{j}}$ is the service time for obtaining segment $v$ from the queue of $PS^{(d,j)}_{\beta_j}$. Using similar analysis for that of deriving the MGF of download time of chunk $g$  as in the last section, we obtain
 \begin{equation}
 \mathbb{E}\left[e^{tE_{i,j,\beta_{j}}^{(g)}}\right]=\frac{(1-\rho_{j,\beta_{j}}^{(d)})t}{t-\Lambda^{(d)}_{j,\beta_{j}}(B_{j,\beta_{j}}^{(d)}(t)-1)}\left(\frac{\alpha_{j,\beta_{j}}^{(d)}e^{\eta_{j,\beta_j}^{(d)}t}}{\alpha_{j,\beta_{j}}^{(d)}-t}\right)^{g-L_{j,i}-1},
 \end{equation}
where the load intensity at queue $\beta_j$ at datacenter, $\rho_{j,\beta_{j}}^{(d)}$
 \begin{align}
\rho_{j,\beta_{j}}^{(d)} & =\sum_{i=1}^{r}\lambda_{i}\pi_{i,j}q_{i,j,\beta_j}e^{\lambda_i \omega_i}\left( L_i-L_{j,i}\right)\left( \eta_{j,\beta_j}^{(d)}+ \frac{1}{\alpha_{j,\beta_{j}}^{(d)}}\right) \label{rhod}\\
B_{j,\beta_{j}}^{(d)}(t) & =\sum_{i=1}^{r}\frac{\lambda_{i}\pi_{i,j}q_{i,j,\beta_j}e^{\lambda_i \omega_i}}{\Lambda_{j,\beta_{j}}^{(d)}}\left(\frac{\alpha_{j,\beta_{j}}^{(d)}e^{\eta_{j,\beta_j}^{(d)}t}}{\alpha_{j,\beta_{j}}^{(d)}-t}\right)^{L_{i}-L_{j,i}}\label{Bd}
\end{align}
% $\rho_{j,\beta_{j}}^{(d)} = ??$ and $B_{j,\beta_{j}}^{(d)}(t) = ??$.
%{\color{blue} Add a couple of sentences here to explain why it can be derived recursively.} 

To find the download time of video segments from the second queue (at cache server $j$), we notice that the download time for segment $g$ includes the waiting to download all previous segments and the idle time if the segment $g$ is not yet downloaded from the first queue ($PS^{(d,j)}_{\beta_j}$), as well as the service time of the segment from $PS^{(\overline{d},j)}_{\beta_j}$. Then, the download time of the video segments from the second queue (i.e.,$PS^{(\overline{d},j)}_{\beta_j}$) can be derived by a set of recursive equations, with download time of the first (initial) segment $(L_{j,i}+1)$ as 
 \begin{equation}
D_{i,j,\beta_{j},\nu_j}^{(L_{j,i}+1)} = \max(W_{j,\beta_{j}}^{(\overline{d})},  E_{j,\beta_{j}}^{(L_{j,i}+1)}) + Y^{(\overline{d},L_{j,i}+1)}_{j,\beta_{j}},
 \end{equation}
where $W_{j,\beta_{j}}^{(\overline{d})}$ is the waiting time from queue $PS^{(\overline{d},j)}_{\beta_j}$ for the previous video segments, and $Y^{(\overline{d},v)}_{j,\beta_{j}}$ is the required service time for obtaining segment $v$ from the queue of $PS^{(\overline{d},j)}_{\beta_j}$. The download time of the following segments ($g>L_{j,i}+1$) is given by the following recursive equation 
\begin{equation}
D_{i,j,\beta_{j},\nu_j}^{(g)} = \max(D_{i,j,\beta_{j},\nu_j}^{(g-1)},  E_{i,j,\beta_{j}}^{(g)}) + Y^{(\overline{d},g)}_{j,\beta_{j}}.
\end{equation}
%This is because the download time for segment $g$ has to 

With the above recursive equations from $y=L_{j,i}$ to $y=g$, we can obtain that 
\begin{equation}
D_{i,j,\beta_{j},\nu_j}^{(g)} = \max_{y=L_{j,i}}^g U_{i,j,\beta_j,g,y}, \label{DUP}
\end{equation}
where
\begin{equation}
U_{i,j,\beta_j,g,L_{j,i}} = W_{j,\beta_{j}}^{(\overline{d})} + \sum_{h= L_{j,i}+1}^gY^{(\overline{d},h)}_{j,\beta_{j}}
\end{equation}
Similarly, for $y>L_{j,i}$, we have
\begin{equation}
U_{i,j,\beta_j,g,y} = E_{i,j,\beta_{j}}^{(y)} + \sum_{h= y}^gY^{(\overline{d},h)}_{j,\beta_{j}}.
\end{equation}

It is easy to see that the moment generating function of $U_{i,j,\beta_j,g,y}$ for $y=L_{j,i}$ is given by
\begin{equation}
\mathbb{E}[e^{tU_{i,j,\beta_{j},g,L_{j,i}}}]=  \!\!\frac{(1-\rho_{j,\beta_{j}}^{(\overline{d})})t}{t-\Lambda_{j,\beta_{j}}^{(c)}(B_{j,\beta_{j}}^{(\overline{d})}(t)-1)}\!\!\left(\!\!\!\frac{\alpha_{j,\beta_{j}}^{(\overline{d})}e^{\eta_{j,\beta_{j}}^{(\overline{d})}t}}{\alpha_{j,\beta_{j}}^{(\overline{d})}-t}\!\!\right)^{g-L_{j,i}},
\label{U_MGF_Lji}
\end{equation}
where the load intensity at queue $\beta_j$ at cache server $j$, $\rho_{j,\beta_{j}}^{(\overline{d})}$ is given by 
\begin{align}
\rho_{j,\beta_{j}}^{(\overline{d})} & =\sum_{i=1}^{r}\lambda_{i}\pi_{i,j}q_{i,j,\beta_j}e^{\lambda_i \omega_i}\left( L_i-L_{j,i}\right)\left( \eta_{j,\beta_j}^{(\overline{d})}+ \frac{1}{\alpha_{j,\beta_{j}}^{(\overline{d})}}\right)\label{rhoc}\\
B_{j,\beta_{j}}^{(\overline{d})}(t) & =\sum_{i=1}^{r}\frac{\lambda_{i}\pi_{i,j}q_{i,j,\beta_j}e^{\lambda_i \omega_i}}{\Lambda_{j,\beta_{j}}^{(c)}}\left(\frac{\alpha_{j,\beta_{j}}^{(\overline{d})}e^{\eta_{j,\beta_j}^{(\overline{d})}t}}{\alpha_{j,\beta_{j}}^{(\overline{d})}-t}\right)^{L_{i}-L_{j,i}}\label{Bc}
\end{align}
Similarly, the moment generating function of $U_{i,j,\beta_j,g,y}$ for $y>L_{j,i}$ is given as 
\begin{align}
\mathbb{E}[e^{tU_{i,j,\beta_{j},g,y}}] & =\overline{W}_{j,\beta_{j}}^{(d)}\times \nonumber\\
& \left(\frac{\alpha_{j,\beta_{j}}^{(d)}e^{\eta_{j,\beta_{j}}^{(d)}t}}{\alpha_{j,\beta_{j}}^{(d)}-t}\right)^{y-L_{j,i}-1}\left(\frac{\alpha_{j,\beta_{j}}^{(\overline{d})}e^{\eta_{j,\beta_{j}}^{(\overline{d})}t}}{\alpha_{j,\beta_{j}}^{(\overline{d})}-t}\right)^{g-y+1}
\label{U_MGF_all}
\end{align}
%{\mathbb E}[e^{tU_{i,j,\beta_j,g,y}}] = \frac{(1-\rho_{j,\beta_{j}}^{(d)})tB_{j,\beta_{j}}^{(d)}(t)}{t-\Lambda^{(d)}_{j,\beta_{j}}(B_{j,\beta_{j}}^{(d)}(t)-1)}\left(\frac{\alpha_{j,\beta_{j}}^{(d)}}{\alpha_{j,\beta_{j}}^{(d)}-t}\right)^{y-L_{j,i}-1}\left(\frac{\alpha_{j,\beta_{j}}^{(c)}}{\alpha_{j,\beta_{j}}^{(c)}-t}\right)^{g-y+1}.
where $\overline{W}_{j,\beta_{j}}^{(d)}$
$\overline{W}_{j,\beta_{j}}^{(d)}=\frac{(1-\rho_{j,\beta_{j}}^{(d)})tB_{j,\beta_{j}}^{(d)}(t)}{t-\Lambda_{j,\beta_{j}}^{(d)}(B_{j,\beta_{j}}^{(d)}(t)-1)}$.
We further note that these moment generating functions are only defined when the MGF functions exist, i.e., 
%\begin{eqnarray}
%t&<& \alpha_{j,\beta_{j}}^{(d)}\label{alpha_betaj_d}\\
%0 &<& t-\Lambda_{j,\beta_{j}}^{(d)}(B_{j,\beta_{j}}^{(d)}(t)-1)\\
%t&<& \alpha_{j,\beta_{j}}^{(\overline{d})}\\
%0 &<& t-\Lambda_{j,\beta_{j}}^{(\overline{d})}(B_{j,\beta_{j}}^{(\overline{d})}(t)-1)
%\label{mgf_beta_j_c}
%\end{eqnarray}
\begin{eqnarray}
t<\alpha_{j,\beta_{j}}^{(d)}\label{alpha_betaj_d};\,\,\,\,
0 < t-\Lambda_{j,\beta_{j}}^{(d)}(B_{j,\beta_{j}}^{(d)}(t)-1)\\
t< \alpha_{j,\beta_{j}}^{(\overline{d})};\,\,\,\,
0 < t-\Lambda_{j,\beta_{j}}^{(\overline{d})}(B_{j,\beta_{j}}^{(\overline{d})}(t)-1)
\label{mgf_beta_j_c}
\end{eqnarray}

\subsection{Play Times of different Segments}

Next, we find the play time of different video segments. Recall that $D_{i,j,\beta_{j},\nu_{j}}^{(g)}$ is the download time of segment
$g$ from $\nu_{j}$ and $\beta_{j}$ queues at client $i$. We further define $T_{i,j,\beta_j,\nu_j}^{\left(g\right)}$ as the time that segment $g$ begins to play at the client $i$, given that it is downloaded from $\beta_j$ and $\nu_j$ queues. This start-up delay of the video is denoted by $d_{s}$. Then, the first segment is ready for play at the maximum
of the startup delay and the time that the first segment can be downloaded. This means
\begin{equation}
T_{i,j,\beta_j,\nu_j}^{(1)}=\mbox{max }\left(d_{s},\,D_{i,j,\beta_j,\nu_j}^{(1)}\right).
\end{equation}

For $1<g\le L_{i}$, the play time of segment $g$ of video file $i$ is
given by the maximum of (i) the time to download the segment
and (ii) the time to play all previous segment plus the time
to play segment $g$ (i.e., $\tau$ seconds). Thus, the play time of segment
$g$ of video file $i$, when requested from server $j$ and from $\nu_j$ and $\beta_j$ queues, can be expressed as 
%$T_{i,j,\beta_j,\nu_j}^{(q)}$
\begin{equation}
T_{i,j,\beta_j,\nu_j}^{(q)}=\mbox{max }\left(T_{i,j,\beta_j,\nu_j}^{(q-1)}+\tau,\,D_{i,j,\beta_j,\nu_j}^{(q)}\right).\label{eq:T_ij_general}
\end{equation}
%Equation (\ref{eq:T_ij_general}) 
This results in a set of recursive equations, which further yield by
\begin{align}
T_{i,j,\beta_{j},\nu_{j}}^{(L_{i})} & =\mbox{max }\left(T_{i,j,\beta_{j},\nu_{j}}^{(L_{i}-1)}+\tau,\,D_{i,j,\beta_{j},\nu_{j}}^{(L_{i})}\right)\nonumber \\
 & =\mbox{max }\left(T_{i,j,\beta_{j},\nu_{j}}^{(L_{i}-2)}+2\tau,\,D_{i,j,\beta_{j},\nu_{j}}^{(L_{i}-1)}+\tau,\,D_{i,j,\beta_{j},\nu_{j}}^{(L_{i})}\right)\nonumber \\
 & =\!\mbox{max}\left(d_{s}+(L_{i}-1)\tau,\,\max_{z=2}^{L_{i}+1}D_{i,j,\beta_{j},\nu_{j}}^{(z-1)}+(L_{i}-z+1)\tau\right)\nonumber \\
 & =\overset{L_{i}+1}{\underset{z=1}{\text{max}}}\,\mathcal{F}_{i,j,\beta_{j},\nu_{j},z}
 \label{T_ijbetanuF}
\end{align}
where $\mathcal{F}_{i,j,\beta_{j},\nu_{j},z}$ is expressed as
\begin{equation}
\mathcal{F}_{i,j,\beta_{j},\nu_{j},z}=\begin{cases}
d_{s}+(L_{i}-1)\tau & \,,z=1\\
D_{i,j,\beta_{j},\nu_{j}}^{(z-1)}+(L_{i}-z+1)\tau & 1<z\leq L_{i}
\end{cases} \label{F_1}
\end{equation}

We now get the MGFs of the 
$\mathcal{F}_{i,j,\nu_{j},\beta_{j},z}$ to use in characterizing the play time of the different segments. Towards this goal, we plug Equation (\ref{F_1}) into $\mathbb{E}\left[e^{t\mathcal{F}_{i,j,\nu_{j},\beta_{j},z}}\right]$ and obtain
%can be calculated as follows
\begin{align}
& \mathbb{E}\left[e^{t\mathcal{F}_{i,j,\nu_{j},\beta_{j},z}}\right]=\\
& \begin{cases}
e^{(d_{s}+(L_{i}-1)\tau)t} & \,,z=1\nonumber\\
e^{(L_{i}-z+1)\tau t}\mathbb{E}\left[e^{tD_{i,j,\beta_{j},\nu_{j}}^{(z-1)}}\right] & 1<z\leq(L_{i}+1)
\end{cases}\label{MGF_{F}}
\end{align}
where $\mathbb{E}\left[e^{tD_{i,j,\beta_{j},\nu_{j}}^{(z-1)}}\right]$ can be calculated using equation (\ref{D_nu_j}) when $1<z\leq (L_{j,i}+1)$ and using equation (\ref{U_MGF_all}) when $z>L_{j,i}+1$ .
%
%where $Z_{D_{i,j,\nu_{j}}^{(z-1)}}$ is given by (\ref{D_nu_j}) and $Z_{D_{i,j,\beta_{j},\nu_j}^{(z-1)}}$ is given as follows.
%
%\begin{align}
%Z_{D_{i,j,\beta_{j},\nu_{j}}^{(z-1)}} & =\mathbb{E}\left[e^{tD_{i,j,\beta_{j},\nu_{j}}^{(z-1)}}\right]=\mathbb{E}\left[e^{t(\underset{y}{\text{\ensuremath{\max}}}U_{i,j,\beta_{j},g,y})}\right]\\
% & =\mathbb{E}\left[\underset{y}{\text{\ensuremath{\max}}}\,\,e^{tU_{i,j,\beta_{j},g,y}}\right]\\
% & \leq\sum_{y}e^{tU_{i,j,\beta_{j},g,y}}\\
% & =\left[e^{tU_{i,j,\beta_{j},g,L_{j,i}}}\right]+\sum_{w=L_{j,i}+1}^{g}\frac{(1-\rho_{j,\beta_{j}}^{(d)})tB_{j,\beta_{j}}^{(d)}(t)}{t-\Lambda_{j,\beta_{j}}^{(d)}(B_{j,\beta_{j}}^{(d)}(t)-1)}\left(\frac{\alpha_{j,\beta_{j}}^{(d)}}{\alpha_{j,\beta_{j}}^{(d)}-t}\right)^{w-L_{j,i}-1}\left(\frac{\alpha_{j,\beta_{j}}^{(c)}}{\alpha_{j,\beta_{j}}^{(c)}-t}\right)^{g-w+1}.
% \label{MGF_Z_D_last}
%\end{align}
%where $\left[e^{tU_{i,j,\beta_{j},g,L_{j,i}}}\right]$ is given in (\ref{U_MGF_Lji}).

The last segment should be completed by time $d_{s}+L_{i}\tau$ (which is the time at which the playing of the $L_i-1$ segment finishes). Thus, 
the difference between the play time of the last segment $T_{i,j,\beta_{j},\nu_{j}}^{(L_{i})}$ and
$d_{s}+\left(L_{i}-1\right)\tau$ gives the stall duration. We note that
the stalls may occur before any segment and hence this difference will give
the sum of durations of all the stall periods before any segment. Thus, the stall duration for the request of file $i$ from $\beta_{j}$ queue,  $\nu_{j}$ queue and server $j$, i.e., $\Gamma^{(i,j,\beta_{j},\nu_{j})}_{U}$ is given as
\begin{equation}
\Gamma^{(i,j,\beta_{j},\nu_{j})}_{U}=T_{i,j,\beta_{j},\nu_{j}}^{(L_{i})}-d_{s}-\left(L_{i}-1\right)\tau 
\label{stallTime}
\end{equation}

Next, we use this expression to derive a tight bound on the SDTP. 
\section{Proof of Lemma \ref{lem:expon}}\label{proof_exp}

From \eqref{ve1}, we can get
\begin{equation}
e^{h_{i}\Gamma_{tot}^{(i)}}\overset{d}{=}\begin{cases}
\text{max}\left(e^{h_{i}\left(\Gamma_{tot}^{(i)}-\widetilde{t}_{i}\right)},1\right) & 0\leq\widetilde{t}_{i}\leq\omega_{i}\\
e^{h_{i}\Gamma_{U}^{(i,j,\beta_j,\nu_j)}} & \widetilde{t}_{i}>\omega_{i}
\end{cases} \label{ehG}
\end{equation}

By taking the expectation of both sides in \eqref{ehG}, we can write
%\begin{equation}
%\mathbb{E}\left[e^{h\Gamma}|t_{i}\right]\leq\begin{cases}
%1+\mathbb{E}\left[e^{h\left(\Gamma_{U}^{(i,j,\beta_{j},\nu_{j})}-\omega_{i}+\widetilde{t}_{i}\right)}\left|t_{i}\right.\right] & 0\leq\widetilde{t}_{i}\leq\omega_{i}\\
%e^{h\Gamma_{U}^{(i,j,\beta_{j},\nu_{j})}} & \widetilde{t}_{i}>\omega_{i}
%\end{cases}
%\end{equation}
\begin{equation}
\!\!\!\!\mathbb{E}\left[e^{h_i\Gamma_{tot}^{(i)}}|\widetilde{t}_{i}\right]\leq\begin{cases}
1+\mathbb{E}\left[e^{h_i\left(\Gamma_{tot}^{(i)}-\widetilde{t}_{i}\right)}\right] & 0\leq\widetilde{t}_{i}\leq\omega_{i}\\
{\mathbb E}[e^{h_i\Gamma_{U}^{(i,j,\beta_j,\nu_j)}}] & \widetilde{t}_{i}>\omega_{i}
\end{cases},
\end{equation}
where the expectation in the second case is over the choice of $(j,\beta_j,\nu_j)$ in addition to the queue statistics with arrival and departure rates. Since the arrivals at edge-cache of video files are Poisson, the time till  first request for file $i$, i.e., $\widetilde{t}_i$, is exponentially distributed with rate $\lambda_i$. By averaging over $\widetilde{t}_i$, we have
%\begin{align}
%\mathbb{E}\left[e^{h\Gamma}\right] & \leq\intop_{0}^{\omega_{i}}\left(1+\mathbb{E}\left[e^{h\left(\Gamma_{U}^{(i,j,\beta_{j},\nu_{j})}-\omega_{i}+\widetilde{t}_{i}\right)}\right]\right)\lambda_{i}e^{-\lambda_{i}\widetilde{t}_{i}}d\widetilde{t}_{i}\\
%& +\intop_{\omega_{i}}^{\infty}\lambda_{i}e^{h\Gamma_{U}^{(i,j,\beta_{j},\nu_{j})}}e^{-\lambda_{i}\widetilde{t}_{i}}d\widetilde{t}_{i}\nonumber 
%\end{align}
\begin{align}
\mathbb{E}\left[e^{h_i\Gamma_{tot}^{(i)}}\right] & \leq \intop_{0}^{\omega_{i}}\left(1+\mathbb{E}\left[e^{h_i\left(\Gamma_{tot}^{(i)}-\widetilde{t}_{i}\right)}\right]\right)\lambda_{i}e^{-\lambda_{i}\widetilde{t}_{i}}d\widetilde{t}_{i}\\
& +\intop_{\omega_{i}}^{\infty}\lambda_{i} {\mathbb E}[e^{h_i\Gamma_{U}^{(i,j,\beta_j,\nu_j)}}]e^{-\lambda_{i}\widetilde{t}_{i}}d\widetilde{t}_{i}\nonumber 
\end{align}

Performing the integration  and simplifying the expressions, we get 
%\begin{align}
%\mathbb{E}\left[e^{h_{i}\Gamma}\right] & =\left(1-e^{-\lambda_{i}\omega_{i}}\right)+e^{h_{i}\Gamma_{U}^{(i,j,\beta_{j},\nu_{j})}}e^{-\lambda_{i}\omega_{i}}\nonumber \\
%& +\frac{\lambda_{i}e^{-h_{i}\omega_{i}}e^{h\Gamma_{U}^{(i,j,\beta_{j},\nu_{j})}}}{\lambda_{i}-h_{i}}\left(1-e^{-(\lambda_{i}-h_{i})\omega_{i}}\right)
%\end{align}
\begin{align}
\mathbb{E}\left[e^{h_{i}\Gamma_{tot}^{(i)}}\right] & \leq\left(1-e^{-\lambda_{i}\omega_{i}}\right)+{\mathbb{E}}\left[e^{h_{i}\Gamma_{U}^{(i,j,\beta_{j},\nu_{j})}}\right]e^{-\lambda_{i}\omega_{i}}\nonumber\\
& +\frac{\lambda_{i}\mathbb{E}\left[e^{h_{i}\Gamma_{tot}^{(i)}}\right]}{\lambda_{i}+h_{i}}\left(1-e^{-(\lambda_{i}+h_{i})\omega_{i}}\right)
\end{align}

This can be further  simplified as follows. 
%\begin{align}
%\mathbb{E}\left[e^{h_{i}\Gamma}\right] & =\left(1-e^{-\lambda_{i}\omega_{i}}\right)+e^{h_{i}\Gamma_{U}^{(i,j,\beta_{j},\nu_{j})}}e^{-\lambda_{i}\omega_{i}}\nonumber \\
%& +\frac{\lambda_{i}e^{h\Gamma_{U}^{(i,j,\beta_{j},\nu_{j})}}}{\lambda_{i}-h_{i}}\left(e^{-h_{i}\omega_{i}}-e^{-\lambda_{i}\omega_{i}}\right)
%\end{align}
\begin{align}
\mathbb{E}\left[e^{h_{i}\Gamma_{tot}^{(i)}}\right] & \leq\frac{\left(1-e^{-\lambda_{i}\omega_{i}}\right)+e^{-\lambda_{i}\omega_{i}}{\mathbb{E}}\left[e^{h_{i}\Gamma_{U}^{(i,j,\beta_{j},\nu_{j})}}\right]}{1-\frac{\lambda_{i}}{\lambda_{i}+h_{i}}\left(1-e^{-\left(\lambda_{i}+h_{i}\right)\omega_{i}}\right)}\nonumber\\
& \overset{(a)}{=}\frac{c}{b}+\frac{a}{b}{\mathbb{E}}\left[e^{h_{i}\Gamma_{U}^{(i,j,\beta_{j},\nu_{j})}}\right]\nonumber\\
& \overset{(b)}{=}\widetilde{c}+\widetilde{a}\,{\mathbb{E}}\left[e^{h_{i}\Gamma_{U}^{(i,j,\beta_{j},\nu_{j})}}\right]
 \label{E_stall_tot_p}
\end{align}
where $(a)$ and $(b)$ follow by setting $c=1-e^{-\lambda_{i}\omega_{i}}$, $a=e^{-\lambda_{i}\omega_{i}}$, $b=\Biggl[1-\frac{\lambda_{i}}{\lambda_{i}+h_{i}}\left(1-e^{-\left(\lambda_{i}+h_{i}\right)\omega_{i}}\right)
\Biggr]$, $\widetilde{c}=c/b$ and $\widetilde{a}=a/b$. We also recall that the expectation in ${\mathbb E}\left[e^{h_{i}\Gamma_{U}^{(i,j,\beta_j,\nu_j)}}\right]$ is over the choice of $(j,\beta_j, \nu_j)$ and the queue arrival/departure statistics. 

%\section{A schematic of Our TestBed }

%
%%Here, we need to show that the arrivals still Poisson when the service time is shifted exponential. 
%
%
%%Let $X(t)$ denote the number of customers in the system at time $t$. 
%
%
%%Also, let $SE$ denote the shifted-exponential distribution. 
%
%%Since the $M/SE/1$ process is a birth and death process, it follows that ${X(t), t \geq 0}$ is time reversible. Going forward in time, the time epochs at which $X(t)$ increases by $1$ constitute a Poisson process since these are just the arrival times of customers of an exponential with shifted mean \cite{gross011}. Hence, by time reversibility the time points at which
%$%X(t)$ increases by $1$ when we go backward in time also constitute a Poisson
%%process. But these latter epochs are exactly the points of time when customers
%%depart. Hence, the departure times constitute a Poisson process with rate $\lambda$.
%

\section{Proof of Lemma \ref{lemmaDLfromDC} } \label{LemmaDLfromDC}

\label{lemmaStall}

We have 
%\begin{align}
%\mathbb{E}\left[e^{t_iD_{i,j,\beta_{j},\nu_{j}}^{(v)}}\right] & =\mathbb{E}\left[e^{t_i(\max_{y=L_{j,i}}^v U_{i,j,\beta_{j},g,y})}\right]\nonumber \\
% & =\mathbb{E}\left[\max_{y=L_{j,i}}^v\,\,e^{t_iU_{i,j,\beta_{j},g,y}}\right]\nonumber \\
% & \leq\sum_{y=L_{j,i}}^ve^{t_iU_{i,j,\beta_{j},g,y}}\nonumber \\
% & =\left[e^{t_iU_{i,j,\beta_{j},g,L_{j,i}}}\right]+\sum_{w=L_{j,i}+1}^{v}\frac{(1-\rho_{j,\beta_{j}}^{(d)})t_iB_{j,\beta_{j}}^{(d)}(t_i)}{t_i-\Lambda_{j,\beta_{j}}^{(d)}(B_{j,\beta_{j}}^{(d)}(t_i)-1)}\left(\frac{\alpha_{j,\beta_{j}}^{(d)}}{\alpha_{j,\beta_{j}}^{(d)}-t_i}\right)^{w-L_{j,i}-1}\left(\frac{\alpha_{j,\beta_{j}}^{(c)}}{\alpha_{j,\beta_{j}}^{(c)}-t_i}\right)^{v-w+1}
%\end{align}
\begin{eqnarray}
&&\mathbb{E}\left[e^{t_{i}D_{i,j,\beta_{j},\nu_{j}}^{(v)}}|(j,\beta_j,\nu_j)\right] \nonumber\\
& \overset{(a)}{=} &\mathbb{E}\left[e^{t_{i}(\max_{y=L_{j,i}}^{v}U_{i,j,\beta_{j},g,y})}|(j,\beta_j,\nu_j)\right]\nonumber\\
& =&\mathbb{E}\left[\max_{y=L_{j,i}}^{v}\,\,e^{t_{i}U_{i,j,\beta_{j},g,y}}|(j,\beta_j,\nu_j)\right]\nonumber\\
& \leq& \sum_{y=L_{j,i}}^{v}{\mathbb E}\left[e^{t_{i}U_{i,j,\beta_{j},g,y}}|(j,\beta_j,\nu_j)\right]\nonumber\\
& =&{\mathbb E}\left[e^{t_{i}U_{i,j,\beta_{j},g,L_{j,i}}}|(j,\beta_j,\nu_j)\right]+\nonumber\\
&& \sum_{w=L_{j,i}+1}^{v}\frac{(1-\rho_{j,\beta_{j}}^{(d)})t_{i}}{t_{i}-\Lambda_{j,\beta_{j}}^{(d)}(B_{j,\beta_{j}}^{(d)}(t_{i})-1)}\times\nonumber\\
&& \left(\frac{\alpha_{j,\beta_{j}}^{(d)}e^{\eta_{j,\beta_{j}}^{(d)}}}{\alpha_{j,\beta_{j}}^{(d)}-t_{i}}\right)^{w-L_{j,i}-1}\left(\frac{\alpha_{j,\beta_{j},\ell}^{(\overline{d})}e^{\eta_{j,\beta_{j},\ell}^{(\overline{d})}}}{\alpha_{j,\beta_{j},\ell}^{(\overline{d})}-t_{i}}\right)^{v-w+1}, 
 \label{dt_v_apend}
\end{eqnarray}
where $(a)$ follows from \eqref{DUP}, the inequality above follows by replacing the $\text{max}_y(.)$ by $\sum_{y}(.)$. Moreover, the last step follows from \eqref{U_MGF_all}. Hence, we can write 
\begin{align}
& \mathbb{E}\left[e^{t_{i}D_{i,j,\beta_{j},\nu_{j}}^{(v)}}|(j,\beta_{j},\nu_{j})\right]\nonumber\\
& \leq\!\!\frac{(1-\rho_{j,\beta_{j}}^{(\overline{d})})t}{t-\Lambda_{j,\beta_{j}}^{(c)}(B_{j,\beta_{j}}^{(\overline{d})}(t)-1)}\!\!\left(\!\!\!\frac{\alpha_{j,\beta_{j}}^{(\overline{d})}e^{\eta_{j,\beta_{j}}^{(\overline{d})}t}}{\alpha_{j,\beta_{j}}^{(\overline{d})}-t}\!\!\right)^{g-L_{j,i}}+\nonumber\\
& \sum_{w=L_{j,i}+1}^{v}\frac{(1-\rho_{j,\beta_{j}}^{(d)})t_{i}}{t_{i}-\Lambda_{j,\beta_{j}}^{(d)}(B_{j,\beta_{j}}^{(d)}(t_{i})-1)}\times\nonumber\\
& \left(\frac{\alpha_{j,\beta_{j}}^{(d)}e^{\eta_{j,\beta_{j}}^{(d)}}}{\alpha_{j,\beta_{j}}^{(d)}-t_{i}}\right)^{w-L_{j,i}-1}\left(\frac{\alpha_{j,\beta_{j}}^{(\overline{d})}e^{\eta_{j,\beta_{j}}^{(\overline{d})}}}{\alpha_{j,\beta_{j}}^{(\overline{d})}-t_{i}}\right)^{v-w+1}, \label{dlwTimeAbove}
\end{align}
this proves the statement of the Lemma.

\section{Proof of Theorem \ref{tailLemma}}\label{lemmaSDTPtail}
%\begin{proof}

The SDTP for the request of file $i$  can be bounded using Markov Lemma as follows
\begin{equation}
\mathbb{P}\left(\Gamma_{tot}^{(i)}\geq\sigma\right)\leq\frac{\mathbb{E}\left[e^{h_{i}\Gamma_{tot}^{(i)}}\right]}{e^{h_{i}\sigma}} \label{Markineq}
\end{equation}

This can be further simplified as follows

\begin{align}
& \mathbb{P}\left(\Gamma_{tot}^{(i)}\geq\sigma\right)\nonumber \\
& \overset{(c)}{\leq}\widetilde{c}\,e^{-h_{i}\sigma}+\widetilde{a}\,e^{-h_{i}\sigma}{\mathbb E}\left[e^{h_{i}\Gamma_{U}^{(i,j,\beta_j,\nu_j)}}\right]\nonumber \\
& =\widetilde{c}\,e^{-h_{i}\sigma}+\widetilde{a}\,e^{-h_{i}\sigma}{\mathbb E}\left[e^{h_{i}(T_{i,j,\beta_j,\nu_j}^{(L_i)}-(d_{s}+(L_{i}-1)\tau))}\right]\nonumber \\
& =\widetilde{c}\,e^{-h_{i}\sigma}+\widetilde{a}\,e^{-h_{i}\sigma}e^{-h_{i}(d_{s}+(L_{i}-1)\tau)}{\mathbb E}\left[e^{h_{i}T_{i,j,\beta_j,\nu_j}^{(L_i)}}\right]\nonumber \\
& \overset{(d)}{=}\overline{c}+\overline{a}\,{\mathbb E}\left[e^{h_{i}\underset{z}{\text{max}}\left(\mathcal{F}_{i,j,\beta_j,\nu_j,z}\right)}\right]\nonumber \\
& =\overline{c}+\overline{a}\,{\mathbb E}\left[\underset{z}{\text{max}}\left(e^{h_{i}\mathcal{F}_{i,j,\beta_j,\nu_j,z}}\right)\right]\nonumber \\
& \overset{(e)}{\leq}\overline{c}+\overline{a}\,\sum_{z=1}^{L_{i}+1}{\mathbb E}\left[e^{h_{i}\mathcal{F}_{i,j,\beta_j,\nu_j,z}}\right]\nonumber \\
& \overset{(f)}{=}\overline{c}+\overline{a}\,\Biggl(e^{h_{i}(d_{s}+(L_{i}-1)\tau)}+\nonumber \\
& +\sum_{v=1}^{L_{i}}e^{h_{i}(L_{i}-v)\tau}\mathbb{E}\left[e^{h_{i}D_{i,j,\beta_j,\nu_j}^{(v)}}\right]\Biggr)
\end{align}
%\begin{align}
%&\text{\text{Pr}}\left(T_{i,j,\beta_{j},\nu_{j},z}^{(L_{i})}\geq\overline{\sigma}\right) \\
%& \leq\sum_{z=1}^{L_{i}+1}\mathbb{P}\left(\text{\ensuremath{\mathcal{F}_{i,j,\beta_{j},\nu_{j},z}}}\geq\overline{\sigma}\right)\\
% & \overset{(e)}{=}e^{t_{i}(d_{s}+(L_{i}-1)\tau-\overline{\sigma})}\nonumber\\
% & +\sum_{z=2}^{L_{i}+1}e^{(L_{i}-z+1)\tau t_{i}-\overline{\sigma}t_{i}}\mathbb{E}\left[e^{t_{i}D_{i,j,\beta_{j},\nu_{j}}^{(z-1)}}\right]\\
% & =e^{-t_{i}\sigma}+\sum_{z=2}^{L_{i}+1}e^{(\widetilde{\sigma}-\overline{\sigma})t_{i}}\mathbb{E}\left[e^{t_{i}D_{i,j,\beta_{j},\nu_{j}}^{(z-1)}}\right]\\
% & =\left[e^{-t_{i}\sigma}+\sum_{v=1}^{L_{i}}e^{(\widetilde{\sigma}-\overline{\sigma})t_{i}}\mathbb{E}\left[e^{t_{i}D_{i,j,\beta_{j},\nu_{j}}^{(v)}}\right]\right],
%\label{eq:stall_final}
%\end{align}
where $\mathcal{F}_{i,j,\beta_j,\nu_j,z}$ and $D_{i,j,\beta_j,\nu_j}^{(v)}$ are given in Appendix \ref{sec:DownLoadPlay} in Equations \eqref{F_1} and \eqref{D_nu_j}, respectively. Further,  $(c)$ follows from (\ref{E_stall_tot}), $(d)$ follows from \eqref{T_ijbetanuF}  and by setting $\overline{c}=\widetilde{c}\,e^{-h_{i}\sigma}$,  $\ensuremath{\overline{a}=\widetilde{a}\,e^{-h_{i}(\sigma+d_{s}+(L_{i}-1)\tau)}}$, $(e)$ follows by upper bounding the maximum by the sum, and (f) follows  from \eqref{F_1}.
%
%
%
% and  $\widetilde{\sigma}=(L_{i}-z+1)\tau
%$. 
Using the two-stage probabilistic scheduling, the SDTP for video file $i$ is further bounded by 
%\text{\text{Pr}}\left(\Gamma^{(i)}\geq\sigma\right) & \le \sum_{j=1}^{m}\pi_{i,j}\left[e^{-t_{i}\sigma}+\sum_{\nu_{j}=1}^{e_{j}}p_{i,j,\nu_{j}}\sum_{\beta_{j}=1}^{d_{j}}q_{i,j,\beta_{j}}\sum_{v=1}^{L_{i}}e^{(\widetilde{\sigma}-\overline{\sigma})t_{i}}\mathbb{E}\left[e^{t_iD_{i,j,\beta_{j},\nu_{j}}^{(v)}}\right]\right].
%\begin{align}
%\text{\text{Pr}}\left(\Gamma^{(i)}\geq\sigma\right) & \le\sum_{j=1}^{m}\pi_{i,j}\times \nonumber\\
%& \left[e^{-t_{i}\sigma}+\sum_{\nu_{j}=1}^{e_{j}}p_{i,j,\nu_{j}}\sum_{\beta_{j}=1}^{d_{j}}q_{i,j,\beta_{j}}\sum_{v=1}^{L_{i}}\overline{D}_{i,j,\beta_{j},\nu_{j}}^{(v,t)}\right].
% \label{Pr_Tijbetanuj1}
%\end{align}
\begin{align}
& \text{\text{Pr}}\left(\Gamma_{tot}^{(i)}\geq\sigma\right)\leq\nonumber \\
& \overline{c}+\overline{a}\,e^{h_{i}(d_{s}+(L_{i}-1)\tau)}+\overline{a}\,\sum_{j=1}^{m}\pi_{i,j}\sum_{\nu_{j}=1}^{e_{j}}p_{i,j,\nu_{j}}\nonumber\\
& \sum_{\beta_{j}=1}^{d_{j}}q_{i,j,\beta_{j}}\sum_{v=1}^{L_{i}}e^{h_{i}(L_{i}-v)\tau}\mathbb{E}\left[e^{h_{i}D_{i,j,\beta_j,\nu_j}^{(v)}}|(j,\beta_j,\nu_j)\right].
\end{align}

Using Lemmas \ref{v_in_cache} and \ref{lemmaDLfromDC} for $\mathbb{E}\left[e^{h_iD_{r}^{(v)}}|(j,\beta_j,\nu_j)\right]$, we obtain the following. 
\begin{align*}
& \text{\text{Pr}}\left(\Gamma_{tot}^{(i)}\geq\sigma\right)\leq\\
& \sum_{j=1}^{m}\pi_{i,j}\times\Biggl[\overline{c}+\overline{a}\,e^{h_{i}(d_{s}+(L_{i}-1)\tau)}+\overline{a}\,\times\nonumber\\
& \sum_{\nu_{j}=1}^{e_{j}}p_{i,j,\nu_{j}}\sum_{\beta_{j}=1}^{d_{j}}q_{i,j,\beta_{j}}\Biggl(\sum_{v=1}^{L_{j,i}}e^{t_{i}(L_{i}-v)\tau}\mathbb{E}\left[e^{t_{i}D_{r}^{(v)}}\right]+\\
& \sum_{v=1}^{L_{j,i}}e^{t_{i}(L_{i}-v)\tau}\mathbb{E}\left[e^{t_{i}D_{r}^{(v)}}\right]\Biggr)\Biggl]\\
& \overset{(f)}{\leq}\sum_{j=1}^{m}\pi_{i,j}\times\Biggl[\overline{c}+\overline{a}\,e^{h_{i}(d_{s}+(L_{i}-1)\tau)}+\overline{a}\,\times\nonumber\\
& \sum_{\nu_{j}=1}^{e_{j}}p_{i,j,\nu_{j}}\sum_{\beta_{j}=1}^{d_{j}}q_{i,j,\beta_{j}}\Biggl(\sum_{v=1}^{L_{j,i}}e^{h_{i}(L_{i}-v)\tau}\times\\
& \frac{(1-\rho_{j,\beta_{j}}^{(e)})t_{i}(M_{j,\nu_{j}}^{(e)}(h_{i}))^{v}}{h_{i}-\Lambda_{j,\beta_{j}}^{(e)}(B_{j,\beta_{j}}^{(e)}(h_{i})-1)}+\\
& +\frac{e^{h_{i}(L_{i}-v)\tau}(1-\rho_{j,\beta_{j}}^{(\overline{d})})t_{i}(M_{j,\nu_{j}}^{(\overline{d})}(h_{i}))^{L_{i}-L_{j,i}}}{h_{i}-\Lambda_{j,\beta_{j}}^{(\overline{d})}(B_{j,\beta_{j}}^{(\overline{d})}(h_{i})-1)}+\\
& +\sum_{v=L_{j,i}+1}^{L_{i}}\sum_{w=L_{j,i}+1}^{v}e^{h_{i}(L_{i}-v)\tau}\times\\
& \frac{(1-\rho_{j,\beta_{j}}^{(d)})t_{i}(M_{j,\beta_{j}}^{(d)}(h_{i}))^{w-L_{j,i}-1}}{\left[h_{i}-\Lambda_{j,\beta_{j}}^{(d)}(B_{j,\beta_{j}}^{(d)}(h_{i})-1)\right](M_{j,\beta_{j}}^{(\overline{d})}(h_{i})^{w-L_{i}-1}}\Biggr)\Biggl]\\
& =\sum_{j=1}^{m}\pi_{i,j}\times\Biggl[\overline{c}+\overline{a}\,e^{h_{i}(d_{s}+(L_{i}-1)\tau)}+\overline{a}\,\times\nonumber\\
& \sum_{\nu_{j}=1}^{e_{j}}p_{i,j,\nu_{j}}\sum_{\beta_{j}=1}^{d_{j}}q_{i,j,\beta_{j}}e^{h_{i}L_{i}\tau}\times\\
& \Biggl(\frac{\widetilde{M}_{j,\nu_{j}}^{(e)}(h_{i})(1-\rho_{j,\beta_{j}}^{(e)})t_{i}((\widetilde{M}_{j,\nu_{j}}^{(e)}(h_{i}))^{L_{j,i}}-1)}{(h_{i}-\Lambda_{j,\beta_{j}}^{(e)}(B_{j,\beta_{j}}^{(e)}(h_{i})-1))(\widetilde{M}_{j,\nu_{j}}^{(e)}(h_{i}))-1)}\\
& +\frac{(1-\rho_{j,\beta_{j}}^{(\overline{d})})t_{i}(\widetilde{M}_{j,\nu_{j}}^{(\overline{d})}(h_{i}))^{L_{j,i}-L_{i}}}{h_{i}-\Lambda_{j,\beta_{j}}^{(\overline{d})}(B_{j,\beta_{j}}^{(\overline{d})}(h_{i})-1)}+\sum_{v=L_{j,i}+1}^{L_{i}}
\end{align*}

\begin{align}
& \!\!\!\sum_{w=L_{j,i}+1}^{v}\!\!\frac{(1-\rho_{j,\beta_{j}}^{(d)})t_{i}(\widetilde{M}_{j,\beta_{j}}^{(\overline{d})}(h_{i}))^{L_{i}+1}(\widetilde{M}_{j,\beta_{j}}^{(d,\overline{d})}(h_{i}))^{w}}{\left[h_{i}-\Lambda_{j,\beta_{j}}^{(d)}(B_{j,\beta_{j}}^{(d)}(h_{i})-1)\right](\widetilde{M}_{j,\beta_{j}}^{(d)}(h_{i}))^{L_{j,i}+1}}\Biggr)\Biggl]\\
%\end{align*}
%
%\begin{align}
& =\sum_{j=1}^{m}\pi_{i,j}\times\Biggl[\overline{c}+\overline{a}\,e^{h_{i}(d_{s}+(L_{i}-1)\tau)}+\overline{a}\,\times\nonumber\nonumber \\
& \sum_{\nu_{j}=1}^{e_{j}}p_{i,j,\nu_{j}}\sum_{\beta_{j}=1}^{d_{j}}q_{i,j,\beta_{j}}e^{h_{i}L_{i}\tau}\times\nonumber \\
& \Biggl(\frac{\widetilde{M}_{j,\nu_{j}}^{(e)}(h_{i})(1-\rho_{j,\beta_{j}}^{(e)})t_{i}((\widetilde{M}_{j,\nu_{j}}^{(e)}(h_{i}))^{L_{j,i}}-1)}{(h_{i}-\Lambda_{j,\beta_{j}}^{(e)}(B_{j,\beta_{j}}^{(e)}(h_{i})-1))(\widetilde{M}_{j,\nu_{j}}^{(e)}(h_{i}))-1)}\nonumber \\
& +\frac{(1-\rho_{j,\beta_{j}}^{(\overline{d})})t_{i}(\widetilde{M}_{j,\nu_{j}}^{(\overline{d})}(h_{i}))^{L_{j,i}-L_{i}}}{h_{i}-\Lambda_{j,\beta_{j}}^{(\overline{d})}(B_{j,\beta_{j}}^{(\overline{d})}(h_{i})-1)}+\nonumber \\
& \frac{(1-\rho_{j,\beta_{j}}^{(d)})t_{i}(\widetilde{M}_{j,\beta_{j}}^{(\overline{d})}(h_{i}))^{L_{i}+1}}{\left[h_{i}-\Lambda_{j,\beta_{j}}^{(d)}(B_{j,\beta_{j}}^{(d)}(h_{i})-1)\right](\widetilde{M}_{j,\beta_{j}}^{(d)}(h_{i}))^{L_{j,i}+1}}\times\nonumber \\
& \Biggl(\frac{(\widetilde{M}_{j,\beta_{j}}^{(d,\overline{d})}(h_{i}))^{L_{i}-L_{j,i}}-(L_{i}-L_{j,i})}{(\widetilde{M}_{j,\beta_{j}}^{(d,\overline{d})}(h_{i}))-1}+\nonumber \\
& +\frac{\widetilde{M}_{j,\beta_{j}}^{(d,\overline{d})}(h_{i})\left((\widetilde{M}_{j,\beta_{j}}^{(d,\overline{d})}(h_{i}))^{L_{i}-L_{j,i}-1}-1\right)}{(\widetilde{M}_{j,\beta_{j}}^{(d,\overline{d})}(h_{i}))-1}\Biggr)\Biggl)\Biggl] \label{eqnSDTP}
\end{align}
where step $(f)$ follows by substitution of the moment generating functions, and the remaining of the steps use the sum of geometric and Arithmetico-geometric sequences. Note that the subscript $\ell$ is omitted in the above derivation for simplicity. Further,
$\delta^{(e)}=\frac{\widetilde{M}_{j,\nu_{j}}^{(e)}(h_{i})(1-\rho_{j,\beta_{j}}^{(e)})t_{i}((\widetilde{M}_{j,\nu_{j}}^{(e)}(h_{i}))^{L_{j,i}}-1)}{(h_{i}-\Lambda_{j,\beta_{j}}^{(e)}(B_{j,\beta_{j}}^{(e)}(h_{i})-1))(\widetilde{M}_{j,\nu_{j}}^{(e)}(h_{i}))-1)}$,
$\delta^{(\overline{d})}=\frac{(1-\rho_{j,\beta_{j}}^{(\overline{d})})t_{i}(\widetilde{M}_{j,\nu_{j}}^{(\overline{d})}(h_{i}))^{L_{j,i}-L_{i}}}{h_{i}-\Lambda_{j,\beta_{j}}^{(\overline{d})}(B_{j,\beta_{j}}^{(\overline{d})}(h_{i})-1)}$,
$\!\delta^{(d,\overline{d})}\!=\!\!\!\!\!\gamma^{(d)}\!\!\left(\!\!\!\frac{(\widetilde{M}_{j,\beta_{j}}^{(d,\overline{d})}(h_{i}))^{L_{i}-L_{j,i}}\!\!-(L_{i}\!\!-L_{j,i}\!\!)}{(\widetilde{M}_{j,\beta_{j}}^{(d,\overline{d})}(h_{i}))-1}\!\!+\!\xi_{i,j,\beta_{j}}^{(d,\overline{d})}\!\!\right)$,
$\!\xi_{i,j,\beta_{j}}^{(d,\overline{d})}=\!\!\!\!\!\frac{\widetilde{M}_{j,\beta_{j}}^{(d,\overline{d})}(h_{i})\left((\widetilde{M}_{j,\beta_{j}}^{(d,\overline{d})}(h_{i}))^{L_{i}-L_{j,i}-1}-1\right)}{(\widetilde{M}_{j,\beta_{j}}^{(d,\overline{d})}(h_{i}))-1}$
and $\gamma^{(d)}=\frac{(1-\rho_{j,\beta_{j}}^{(d)})t_{i}(\widetilde{M}_{j,\beta_{j}}^{(\overline{d})}(h_{i}))^{L_{i}+1}}{\left[h_{i}-\Lambda_{j,\beta_{j}}^{(d)}(B_{j,\beta_{j}}^{(d)}(h_{i})-1)\right](\widetilde{M}_{j,\beta_{j}}^{(d)}(h_{i}))^{L_{j,i}+1}}$. Further, 
$\widetilde{M}_{j,\beta_{j}}^{(d)}(h_{i})=\frac{\alpha_{j,\beta_{j}}^{(d)}e^{\eta_{j,\beta_{j}}-h_{i}\tau}}{\alpha_{j,\beta_{j}}^{(d)}-h_{i}}$,
$\widetilde{M}_{j,\beta_{j}}^{(\overline{d})}(h_{i})=\frac{\alpha_{j,\beta_{j}}^{(\overline{d})}e^{\eta_{j,\beta_{j}}-h_{i}\tau}}{\alpha_{j,\beta_{j}}^{(\overline{d})}-h_{i}}$,
$\widetilde{M}_{j,\nu_{j}}^{(e)}(h_{i})=\frac{\alpha_{j,\nu_{j}}^{(e)}e^{\eta_{j,\nu_{j}}-h_{i}\tau}}{\alpha_{j,\nu_{j}}^{(e)}-h_{i}}$.  This proves the statement of the Theorem. 
%\end{proof}

\section{Sub-problems Optimization}\label{opt_mention}

In this section, we explain how each sub-optimization problem is solved.

\subsubsection{Server-PSs Access Optimization}

Given the bandwidth allocation weights, the cache placement, edge-cache window size, and the auxiliary variables, this sub-problem can be written as follows. 

\textbf{Input: $\boldsymbol{h}$, $\boldsymbol{w}$, $\boldsymbol{\omega}$,  and  $\boldsymbol{L}$}

\textbf{Objective:} $\qquad\quad\;\;$min $\left(\ref{eq:optfun}\right)$

$\hphantom{\boldsymbol{\text{Objective:}\,}}\qquad\qquad$s.t. \eqref{formulas}-- 
\eqref{eq:rho_nu_e}, --\eqref{eq:t_leq_alpha_nu_e}--
\eqref{eq:mgf_beta_e}\\
$\hphantom{\boldsymbol{\text{Objective:}\,}}\qquad\qquad$var. $\widetilde{\boldsymbol{\pi}}$

%Now, we explain our approach in solving $\boldsymbol{\pi}$- Optimization sub-problem. We optimize over $\pi_{ij}$, $\forall i,j$ for given placement $\boldsymbol{\mathcal{S}}$ and fixed $\boldsymbol{t}$. This problem is non-convex as shown in Section \ref{sec:probForm}. Hence, 

In order to solve this problem, we use iNner cOnVex
Approximation (NOVA)  algorithm proposed in \cite{scutNOVA}. The key idea for this algorithm  is that the non-convex objective function is replaced by suitable convex approximations at which convergence to a stationary solution of the original non-convex optimization is established. NOVA solves the approximated function efficiently and maintains feasibility in each iteration. The objective function can be approximated by a convex one (e.g., proximal gradient-like approximation) such that the first order properties are preserved \cite{scutNOVA},  and this convex approximation can be used in NOVA algorithm.

Let $\widetilde{U_{q}}\left(\widetilde{\boldsymbol{\pi}},\boldsymbol{\widetilde{\pi}^{\nu}}\right)$ be the
convex approximation at iterate $\boldsymbol{{\pi}^\nu}$ to the original non-convex problem $U\left(\widetilde{\boldsymbol{\pi}}\right)$, where $U\left(\widetilde{\boldsymbol{\pi}}\right)$ is given by (\ref{eq:optfun}). Then, a valid choice of $U\left(\widetilde{\boldsymbol{\pi}};\boldsymbol{{\pi}^\nu}\right)$ is the first order approximation of  $U\left(\widetilde{\boldsymbol{\pi}}\right)$, e.g., (proximal) gradient-like approximation, i.e.,

\begin{equation}
\widetilde{U_{q}}\left(\widetilde{\boldsymbol{\pi}},\boldsymbol{\widetilde{\pi}^{\nu}}\right)=\nabla_{\widetilde{\boldsymbol{\pi}}}U\left(\widetilde{\boldsymbol{\pi}}^{\nu}\right)^{T}\left(\widetilde{\boldsymbol{\pi}}-\widetilde{\boldsymbol{\pi}}^{\nu}\right)+\frac{\tau_{u}}{2}\left\Vert \widetilde{\boldsymbol{\pi}}-\widetilde{\boldsymbol{\pi}}^{\nu}\right\Vert ^{2}
,\label{eq:U_x_u_bar}
\end{equation}
where $\tau_u$ is a regularization parameter. Note that all the constraints \eqref{formulas}-- 
\eqref{eq:rho_nu_e} are separable and
linear in $\widetilde{\pi}_{i,j,k}$.  
The NOVA Algorithm for optimizing $\widetilde{\boldsymbol{\pi}}$ is described in  Algorithm \ref{alg:NOVA_Alg1Pi}. Using the convex approximation $\widetilde{U_{\pi}}\left(\boldsymbol{\pi};\boldsymbol{{\pi}^\nu}\right)$, the minimization steps in Algorithm \ref{alg:NOVA_Alg1Pi} are convex, with linear constraints and thus can be solved using a projected gradient descent algorithm. A step-size ($\gamma
$)  is also used in the update of the iterate $\widetilde{\boldsymbol{\pi}}^{\nu}$. Note that the iterates $\left\{ \boldsymbol{\pi}^{\nu}\right\} $ generated by the algorithm are all feasible for the original problem and, further, convergence is guaranteed, as shown in \cite{scutNOVA} and described in   lemma \ref{lem_pi}. 

In order to use NOVA, there are some assumptions (given in \cite{scutNOVA}) that have to be satisfied in both original function and its approximation. These assumptions can be classified into two categories. The first category is the set of conditions that ensure that the original problem and its constraints are continuously differentiable on the domain of the function, which are satisfied in our problem. The second category is the set of conditions that ensures that the approximation of the original problem is uniformly strongly convex on the domain of the function. The latter set of conditions are also satisfied as the chosen function is strongly convex and its domain is also convex. To see this, we need to show that the constraints \eqref{eq:rho_beta_d}--\eqref{eq:t_leq_alpha_nu_e} form a convex domain in $\widetilde{\boldsymbol{\pi}}$
which is easy to see from the linearity of the constraints. Further details on the assumptions and function approximation can be found in \cite{scutNOVA}. Thus, the following result holds.

\begin{lemma}  \label{lem_pi}
	For fixed $\boldsymbol{h}$, $\boldsymbol{w}$, $\boldsymbol{\omega}$,  and  $\boldsymbol{L}$, the optimization of our problem over 
	$\widetilde{\boldsymbol{\pi}}$
	generates a sequence of  decreasing
	objective values and therefore is guaranteed to converge to a stationary point.
\end{lemma}

\begin{algorithm}[t]
	\caption{NOVA Algorithm to solve Server Access and PSs selection Optimization sub-problem\label{alg:NOVA_Alg1Pi}}
	
	\begin{enumerate}
		\item \textbf{Initialize} $\nu=0$, $k=0$,$\gamma^{\nu}\in\left(0,1\right]$,
		$\epsilon>0$,$\widetilde{\boldsymbol{\pi}}^{0}$ such that $\widetilde{\boldsymbol{\pi}}^{0}$
		is feasible ,
		\item \textbf{while} $\mbox{obj}\left(k\right)-\mbox{obj}\left(k-1\right)\geq\epsilon$
		\item $\quad$//\textit{\small{}Solve for $\widetilde{\boldsymbol{\pi}}^{\nu+1}$ with
			given $\widetilde{\boldsymbol{\pi}}^{\nu}$}{\small \par}
		\item $\quad$\textbf{Step 1}: Compute $\boldsymbol{\widehat{\pi}}\left(\widetilde{\boldsymbol{\pi}}^{\nu}\right),$
		the solution of $\boldsymbol{\widehat{\pi}}\left(\widetilde{\boldsymbol{\pi}}^{\nu}\right)=$$\underset{\widetilde{\boldsymbol{\pi}}}{\text{argmin}}$
		$\boldsymbol{\widetilde{U}}\left(\widetilde{\boldsymbol{\pi}},\widetilde{\boldsymbol{\pi}}^{\nu}\right)\,\, $ s.t.
		\eqref{formulas}--\eqref{eq:rho_nu_e}, \eqref{eq:t_leq_alpha_nu_e}--\eqref{eq:mgf_beta_e} solved using  projected gradient descent

		\item $\quad$\textbf{Step 2}: $\ensuremath{\widetilde{\boldsymbol{\pi}}^{\nu+1}=\widetilde{\boldsymbol{\pi}}^{\nu}+\gamma^{\nu}\left(\widehat{\boldsymbol{\pi}}\left(\widetilde{\boldsymbol{\pi}}^{\nu}\right)-\widetilde{\boldsymbol{\pi}}^{\nu}\right)}$.
		\item $\quad$//\textit{\small{}update index}{\small \par}
		\item \textbf{Set} $\ensuremath{\nu\leftarrow\nu+1}$
		\item \textbf{end while}
		\item \textbf{output: }$\ensuremath{\widehat{\boldsymbol{\pi}}\left(\widetilde{\boldsymbol{\pi}}^{\nu}\right)}$
	\end{enumerate}
\end{algorithm}

\begin{algorithm}[t]
	\caption{NOVA Algorithm to solve Auxiliary Variables  Optimization sub-problem\label{alg:NOVA_Alg1}}
	
	\begin{enumerate}
		\item \textbf{Initialize} $\nu=0$, $\gamma^{\nu}\in\left(0,1\right]$, $\epsilon>0$, $\boldsymbol{h}^{0}$ 
		such that $\boldsymbol{h}^{0}$  is feasible,
		\item \textbf{while} $\mbox{obj}\left(\nu\right)-\mbox{obj}\left(\nu-1\right)\geq\epsilon$
		\item $\quad$//\textit{\small{}Solve for  $\boldsymbol{h}^{\nu+1}$ with
			given $\boldsymbol{h}^{\nu}$}{\small \par}
		\item $\quad$\textbf{Step 1}: Compute $\boldsymbol{\widehat{h}}\left(\boldsymbol{h}^{\nu}\right),$
		the solution of  $\boldsymbol{\widehat{h}}\left(\boldsymbol{t}^{\nu}\right)=$$\underset{\boldsymbol{h}}{\text{argmin}}$
		$\boldsymbol{\overline{U}}\left(\boldsymbol{h},\boldsymbol{h}^{\nu}\right)$, s.t. \eqref{formulas}, \eqref{eq:t_leq_alpha_nu_e}--\eqref{eq:mgf_beta_e}, using projected gradient descent
		\item $\quad$\textbf{Step 2}: $\ensuremath{\boldsymbol{h}^{\nu+1}=\boldsymbol{h}^{\nu}+\gamma^{\nu}\left(\widehat{\boldsymbol{h}}\left(\boldsymbol{h}^{\nu}\right)-\boldsymbol{h}^{\nu}\right)}$.
		\item $\quad$//\textit{\small{}update index}{\small \par}
		\item \textbf{Set} $\ensuremath{\nu\leftarrow\nu+1}$
		\item \textbf{end while}
		\item \textbf{output: }$\ensuremath{\widehat{\boldsymbol{h}}\left(\boldsymbol{h}^{\nu}\right)}$
	\end{enumerate}
\end{algorithm}

\subsubsection{Auxiliary Variables Optimization}
Given the probability distribution of the server-PSs scheduling probabilities, the bandwidth allocation weights, edge-cache window size, and the cache placement, this subproblem can be written as follows. 

\textbf{Input: $\widetilde{\boldsymbol{\pi}}$, $\boldsymbol{w}$, $\boldsymbol{\omega}$,  and $\boldsymbol{L}$ }

\textbf{Objective:} $\qquad\quad\;\;$min $\left(\ref{eq:optfun}\right)$

$\hphantom{\boldsymbol{\text{Objective:}\,}}\qquad\qquad$s.t.  \eqref{formulas}, \eqref{eq:t_leq_alpha_nu_e}--\eqref{eq:mgf_beta_e}, \\
$\hphantom{\boldsymbol{\text{Objective:}\,}}\qquad\qquad$var. $\boldsymbol{h}$

Similar to Server-PSs Access Optimization, this optimization can be solved using NOVA algorithm. The constraint  \eqref{eq:t_leq_alpha_nu_e} is linear in $\boldsymbol{h}$. Further, the next  Lemma show that the constraints  \eqref{eq:mgf_beta_d}-- \eqref{eq:mgf_beta_e} are convex in $\boldsymbol{h}$, respectively.

\begin{lemma}
	%	\label{const68_70}
	The  constraints \eqref{eq:mgf_beta_d}--\eqref{eq:mgf_beta_e} are convex with respect to  $\boldsymbol{{h}}$.  \label{cnv_t}
\end{lemma}
\begin{proof}
	The proof is given in Appendix \ref{apd_ww}. 
\end{proof}

Algorithm \ref{alg:NOVA_Alg1}  shows the used procedure to solve for $\boldsymbol{h}$. Let $\overline{U}\left(\boldsymbol{h};\boldsymbol{h^\nu}\right)$ be the
convex approximation at iterate $\boldsymbol{h^\nu}$ to the original non-convex problem $U\left(\boldsymbol{h}\right)$, where $U\left(\boldsymbol{h}\right)$ is given by (\ref{eq:optfun}), assuming other parameters constant. Then, a valid choice of  $\overline{U}\left(\boldsymbol{h};\boldsymbol{h^\nu}\right)$ is the first order approximation of  $U\left(\boldsymbol{h}\right)$, i.e.,  
\begin{equation}
\overline{U}\left(\boldsymbol{h},\boldsymbol{h^\nu}\right)=\nabla_{\boldsymbol{h}}U\left(\boldsymbol{h^\nu}\right)^{T}\left(\boldsymbol{h}-\boldsymbol{h^\nu}\right)+\frac{\tau_{h}}{2}\left\Vert \boldsymbol{h}-\boldsymbol{h^\nu}\right\Vert ^{2}.\label{eq:U_t_u_bar}
\end{equation}
where $\tau_h$ is a regularization parameter. The detailed steps can be seen in Algorithm \ref{alg:NOVA_Alg1}. Since all the constraints  \eqref{eq:t_leq_alpha_nu_e}-- \eqref{eq:mgf_beta_e} have been shown to be convex in $\boldsymbol{h}$, the optimization problem in Step 1 of  Algorithm \ref{alg:NOVA_Alg1} can be solved by the standard projected gradient descent algorithm. 

%Similar to Access  Optimization problem, we get the minimizer for (\ref{eq:U_t_u_bar}), step 1 in Algorithm \ref{alg:NOVA_Alg1}), and then continue until convergence. 

%In this sub-problem, we solve for all $t_i$, for fixed placement and given accessing probabilities. Similar to $\boldsymbol{\pi}$- Optimization problem, we used NOVA algorithm to solve for $t$.  Algorithm \ref{alg:NOVA_Alg1}, shows the used procedure to solve for $\boldsymbol{t}$. After initialization with feasible choice, NOVA keeps generating  a sequence of monotonically decreasing $\left\{ \boldsymbol{t}^{\nu+1}\right\} $ until  convergence.

\begin{lemma} \label{lem_t}
	
	For fixed $\widetilde{\boldsymbol{\pi}}$, $\boldsymbol{w}$, $\boldsymbol{\omega}$, and  $\boldsymbol{L}$, the optimization of our problem over $\boldsymbol{h}$ generates a sequence of monotonically decreasing objective values and therefore is guaranteed to converge to a stationary point.
\end{lemma}

\subsubsection{ Bandwidth Allocation Weights Optimization}
Given the auxiliary variables,  the server access and PSs selection probabilities, edge-cache window size, and cache placement, this subproblem can be written as follows. 

\textbf{Input: $\widetilde{\boldsymbol{\pi}}$, $\boldsymbol{L}$, $\boldsymbol{\omega}$,  and $\boldsymbol{h}$}

\textbf{Objective:} $\qquad\quad\;\;$min $\left(\ref{eq:optfun}\right)$

$\hphantom{\boldsymbol{\text{Objective:}\,}}\qquad\qquad$s.t. 
\eqref{formulas}--\eqref{eq:rho_beta_c},\,
\eqref{eq:t_leq_alpha_nu_e}--\eqref{eq:mgf_beta_e},
%
%\eqref{eq:t_leq_alpha_beta_d}--\eqref{eq:mgf_beta_e},
 \\
$\hphantom{\boldsymbol{\text{Objective:}\,}}\qquad\qquad$var. $\boldsymbol{w}$

This optimization problem can be solved using NOVA algorithm. It is easy to notice that the constraints that exist in    \eqref{formulas}-- \eqref{eq:rho_beta_c}  are linear and thus convex with respect to $\boldsymbol{w}$. Further, the next two Lemmas show that the constraints  \eqref{eq:t_leq_alpha_nu_e}--\eqref{eq:mgf_beta_e}, are convex in $\boldsymbol{w}$, respectively.

\begin{lemma}
The  constraints \eqref{eq:t_leq_alpha_nu_e}--\eqref{eq:mgf_beta_e} are convex with respect to $\boldsymbol{{w}}$.  \label{cnv_ww}
\end{lemma}

\begin{proof}
The proof is given in Appendix \ref{apd_ww}. 
\end{proof}

Algorithm \ref{alg:NOVA_Alg3}  shows the used procedure to solve for $\boldsymbol{w}$. Let ${U_w}\left(\boldsymbol{w};\boldsymbol{w^\nu}\right)$ be the
convex approximation at iterate $\boldsymbol{w^\nu}$ to the original non-convex problem $U\left(\boldsymbol{w}\right)$, where $U\left(\boldsymbol{w}\right)$ is given by (\ref{eq:optfun}), assuming other parameters constant. Then, a valid choice of  ${U_w}\left(\boldsymbol{w};\boldsymbol{w^\nu}\right)$ is the first order approximation of  $U\left(\boldsymbol{w}\right)$, i.e.,  
\begin{equation}
{U_w}\left(\boldsymbol{w},\boldsymbol{w^\nu}\right)=\nabla_{\boldsymbol{w}}U\left(\boldsymbol{w^\nu}\right)^{T}\left(\boldsymbol{w}-\boldsymbol{w^\nu}\right)+\frac{\tau_{w}}{2}\left\Vert \boldsymbol{w}-\boldsymbol{w^\nu}\right\Vert ^{2}.\label{eq:U_b}
\end{equation}
where $\tau_t$ is a regularization parameter. The detailed steps can be seen in Algorithm \ref{alg:NOVA_Alg3}. Since all the constraints  have been shown to be convex, the optimization problem in Step 1 of  Algorithm \ref{alg:NOVA_Alg3} can be solved by the standard projected gradient descent algorithm. 

\begin{lemma} \label{lem_t}

For fixed $\widetilde{\boldsymbol{\pi}}$
, $\boldsymbol{h}$, $\boldsymbol{\omega}$, and  $\boldsymbol{L}$, the optimization of our problem over 
$\boldsymbol{w}$ generates a sequence of  decreasing objective values and therefore is guaranteed to converge to a stationary point.
\end{lemma}

%\input{apdx_algs}

%\subsubsection{Convergence of the Proposed Algorithm}
%
%We first initialize 
%$\widetilde{\boldsymbol{\pi}}$, $\boldsymbol{w}$, $\boldsymbol{h}$, $\boldsymbol{\omega}$, and $\boldsymbol{L}$ $\forall i, j, \nu_j, \beta_j$ such that the choice is feasible for the problem. Then, we do alternating minimization over the four sub-problems defined above. Since each sub-problem can only decrease the objective and the overall problem is bounded from below, we have the following result.
%
%\begin{theorem} 
%	The proposed algorithm  converges to a  local optimal solution.
%\end{theorem}

%\section{Algorithm Pseudo-codes for the Sub-problems}\label{apdx_alg}

\begin{algorithm}[t]
	\caption{NOVA Algorithm to solve Bandwidth Allocation Optimization sub-problem\label{alg:NOVA_Alg3}}	
	\begin{enumerate}
		\item \textbf{Initialize} $\nu=0$, $\gamma^{\nu}\in\left(0,1\right]$, $\epsilon>0$, $\boldsymbol{w}^{0}$ 
		such that $\boldsymbol{w}^{0}$  is feasible,
		\item \textbf{while} $\mbox{obj}\left(\nu\right)-\mbox{obj}\left(\nu-1\right)\geq\epsilon$
		\item $\quad$//\textit{\small{}Solve for  $\boldsymbol{w}^{\nu+1}$ with
			given $\boldsymbol{w}^{\nu}$}{\small \par}
		\item $\quad$\textbf{Step 1}: Compute $\boldsymbol{\widehat{w}}\left(\boldsymbol{w}^{\nu}\right),$
		the solution of  $\boldsymbol{\widehat{w}}\left(\boldsymbol{w}^{\nu}\right)=$$\underset{\boldsymbol{b}}{\text{argmin}}$
		$\boldsymbol{\overline{U}}\left(\boldsymbol{w},\boldsymbol{w}^{\nu}\right)$, s.t. \eqref{formulas}--\eqref{eq:rho_beta_c},\,
		\eqref{eq:t_leq_alpha_nu_e}--\eqref{eq:mgf_beta_e}, using projected gradient descent
		\item $\quad$\textbf{Step 2}: $\ensuremath{\boldsymbol{w}^{\nu+1}=\boldsymbol{w}^{\nu}+\gamma^{\nu}\left(\widehat{\boldsymbol{w}}\left(\boldsymbol{w}^{\nu}\right)-\boldsymbol{w}^{\nu}\right)}$.
		\item $\quad$//\textit{\small{}update index}{\small \par}
		\item \textbf{Set} $\ensuremath{\nu\leftarrow\nu+1}$
		\item \textbf{end while}
		\item \textbf{output: }$\ensuremath{\widehat{\boldsymbol{w}}\left(\boldsymbol{w}^{\nu}\right)}$
	\end{enumerate}
\end{algorithm}

\begin{algorithm}[t]
	\caption{NOVA Algorithm to solve Cache Placement Optimization sub-problem\label{alg:NOVA_Alg5}}
	
	\begin{enumerate}
		\item \textbf{Initialize} $\nu=0$, $\gamma^{\nu}\in\left(0,1\right]$, $\epsilon>0$, $\boldsymbol{L}^{0}$ 
		such that $\boldsymbol{L}^{0}$  is feasible,
		\item \textbf{while} $\mbox{obj}\left(\nu\right)-\mbox{obj}\left(\nu-1\right)\geq\epsilon$
		\item $\quad$//\textit{\small{}Solve for  $\boldsymbol{L}^{\nu+1}$ with
			given $\boldsymbol{L}^{\nu}$}{\small \par}
		\item $\quad$\textbf{Step 1}: Compute $\boldsymbol{\widehat{L}}\left(\boldsymbol{L}^{\nu}\right),$
		the solution of  $\boldsymbol{\widehat{L}}\left(\boldsymbol{L}^{\nu}\right)=$$\underset{\boldsymbol{L}}{\text{argmin}}$
		$\boldsymbol{\overline{U}}\left(\boldsymbol{L},\boldsymbol{L}^{\nu}\right)$, s.t. \eqref{eq:rho_beta_d}--\eqref{eq:rho_nu_e},
\eqref{eq:capConst},
\eqref{eq:mgf_beta_d}--\eqref{eq:mgf_beta_e}, using projected gradient descent
		\item $\quad$\textbf{Step 2}: $\ensuremath{\boldsymbol{L}^{\nu+1}=\boldsymbol{L}^{\nu}+\gamma^{\nu}\left(\widehat{\boldsymbol{L}}\left(\boldsymbol{L}^{\nu}\right)-\boldsymbol{L}^{\nu}\right)}$.
		\item $\quad$//\textit{\small{}update index}{\small \par}
		\item \textbf{Set} $\ensuremath{\nu\leftarrow\nu+1}$
		\item \textbf{end while}
		\item \textbf{output: }$\ensuremath{\widehat{\boldsymbol{L}}\left(\boldsymbol{L}^{\nu}\right)}$
	\end{enumerate}
\end{algorithm}

\subsubsection{Cache Placement Optimization}
Given the auxiliary variables, the server access and PS  selection probabilities, edge-cache window size, and the bandwidth allocation weights, this subproblem can be written as follows.  

\textbf{Input: $\ensuremath{\widetilde{\boldsymbol{\pi}}}$, $\boldsymbol{h}$, $\boldsymbol{\omega}$ and $\boldsymbol{w}$}

\textbf{Objective:} $\qquad\quad\;\;$min $\left(\ref{eq:optfun}\right)$

$\hphantom{\boldsymbol{\text{Objective:}\,}}\qquad\qquad$s.t. 
\eqref{formulas}--
\eqref{eq:rho_nu_e},
\eqref{eq:capConst},
\eqref{eq:mgf_beta_d}-- \eqref{eq:mgf_beta_e}\\
$\hphantom{\boldsymbol{\text{Objective:}\,}}\qquad\qquad$var. $\boldsymbol{L}$

Similar to the aforementioned  Optimization sub-problems, this optimization can be solved using NOVA algorithm. Constraints  \eqref{eq:rho_beta_d}--
\eqref{eq:rho_nu_e}, are linear in $\boldsymbol{L}$, and hence, form a convex domain. Also, Constraint \eqref{eq:capConst} is relaxed to have it convex. Furthermore, the constraints  \eqref{eq:mgf_beta_d}-- \eqref{eq:mgf_beta_e} are convex as shown in the following Lemmas in this subsection.

Algorithm \ref{alg:NOVA_Alg5}  shows the used procedure to solve for $\boldsymbol{L}$. Let ${U_L}\left(\boldsymbol{L};\boldsymbol{L^\nu}\right)$ be the
convex approximation at iterate $\boldsymbol{L^\nu}$ to the original non-convex problem $U\left(\boldsymbol{L}\right)$, where $U\left(\boldsymbol{L}\right)$ is given by (\ref{eq:optfun}), assuming other parameters constant. Then, a valid choice of  ${U_L}\left(\boldsymbol{L};\boldsymbol{L^\nu}\right)$ is the first order approximation of  $U\left(\boldsymbol{L}\right)$, i.e.,  
\begin{equation}
{U_L}\left(\boldsymbol{L},\boldsymbol{L^\nu}\right)=\nabla_{\boldsymbol{L}}U\left(\boldsymbol{L^\nu}\right)^{T}\left(\boldsymbol{L}-\boldsymbol{L^\nu}\right)+\frac{\tau_{L}}{2}\left\Vert \boldsymbol{L}-\boldsymbol{L^\nu}\right\Vert ^{2}.\label{eq:U_L}
\end{equation}
where $\tau_L$ is a regularization parameter. The detailed steps can be seen in Algorithm \ref{alg:NOVA_Alg3}. Since all the constraints  have been shown to be convex in $\boldsymbol{L}$, the optimization problem in Step 1 of  Algorithm \ref{alg:NOVA_Alg3} can be solved by the standard projected gradient descent algorithm. 

\begin{lemma} \label{lem_b}

For fixed $\boldsymbol{h}$, $\ensuremath{\widetilde{\boldsymbol{\pi}}}$, $\boldsymbol{\omega}$ and  $\boldsymbol{w}$, the optimization of our problem over 
$\boldsymbol{L}$ generates a sequence of monotonically decreasing
objective values and therefore is guaranteed to converge to a stationary point.
\end{lemma}

\subsubsection{Edge-cache Window size Optimization}
Given the server access and PS selection probabilities, the bandwidth allocation weights, the cache placement, and the auxiliary variables, this sub-problem can be written as follows. 

\textbf{Input: $\boldsymbol{h}$, $\boldsymbol{w}$, $\boldsymbol{\widetilde{\pi}}$,  and  $\boldsymbol{L}$}

\textbf{Objective:} $\qquad\quad\;\;$min $\left(\ref{eq:optfun}\right)$

$\hphantom{\boldsymbol{\text{Objective:}\,}}\qquad\qquad$s.t. \eqref{formulas}--\eqref{eq:rho_nu_e}, \eqref{eq:mgf_beta_d}--\eqref{eq:mgf_beta_e}\\
$\hphantom{\boldsymbol{\text{Objective:}\,}}\qquad\qquad$var. $\boldsymbol{\omega}$

similarly, this optimization can be solved using NOVA algorithm. It is easy to show that Constraints  \eqref{eq:rho_beta_d}--\eqref{eq:rho_nu_e}, are convex in $\boldsymbol{\omega}$, and hence, form a convex domain. Further, the constraints  \eqref{eq:mgf_beta_d}--\eqref{eq:mgf_beta_e} are convex as shown in Lemma \ref{lem_omega}.

Algorithm \ref{alg:NOVA_Alg5}  shows the used procedure to solve for $\boldsymbol{\omega}$. Let ${U_{\omega}}\left(\boldsymbol{\omega};\boldsymbol{\omega^\nu}\right)$ be the
convex approximation at iterate $\boldsymbol{\omega^\nu}$ to the original non-convex problem $U\left(\boldsymbol{\omega}\right)$, where $U\left(\boldsymbol{\omega}\right)$ is given by (\ref{eq:optfun}), assuming other parameters constant. Then, a valid choice of  ${U_{\omega}}\left(\boldsymbol{\omega};\boldsymbol{\omega^\nu}\right)$ is the first order approximation of  $U\left(\boldsymbol{\omega}\right)$, i.e.,  
\begin{equation}
{U_\omega}\left(\boldsymbol{\omega},\boldsymbol{\omega^\nu}\right)=\nabla_{\boldsymbol{\omega}}U\left(\boldsymbol{\omega^\nu}\right)^{T}\left(\boldsymbol{\omega}-\boldsymbol{\omega^\nu}\right)+\frac{\tau_{\omega}}{2}\left\Vert \boldsymbol{\omega}-\boldsymbol{\omega^\nu}\right\Vert ^{2}.\label{eq:U_L}
\end{equation}
where $\tau_\omega$ is a regularization parameter. The detailed steps can be seen in Algorithm \ref{alg:NOVA_Alg3}. Since all the constraints  have been shown to be convex in $\boldsymbol{\omega}$, the optimization problem in Step 1 of  Algorithm \ref{alg:NOVA_Alg3} can be solved by the standard projected gradient descent algorithm. 
\begin{lemma} \label{lem_omega}
	
	For fixed $\boldsymbol{h}$, $\ensuremath{\widetilde{\boldsymbol{\pi}}}$, $\boldsymbol{\omega}$ and  $\boldsymbol{w}$, the optimization of our problem over 
	$\boldsymbol{\omega}$ generates a sequence of monotonically decreasing
	objective values and therefore is guaranteed to converge to a stationary point.
\end{lemma}
\begin{proof}
The proof is provided in Appendix \ref{apd_ww}. 
\end{proof}

\begin{algorithm}[t]
	\caption{NOVA Algorithm to solve Edge-cache window size optimization sub-problem\label{alg:NOVA_Alg6}}
	
	\begin{enumerate}
		\item \textbf{Initialize} $\nu=0$, $\gamma^{\nu}\in\left(0,1\right]$, $\epsilon>0$, $\boldsymbol{\omega}^{0}$ 
		such that $\boldsymbol{\omega}^{0}$  is feasible,
		\item \textbf{while} $\mbox{obj}\left(\nu\right)-\mbox{obj}\left(\nu-1\right)\geq\epsilon$
		\item $\quad$//\textit{\small{}Solve for  $\boldsymbol{\omega}^{\nu+1}$ with
			given $\boldsymbol{\omega}^{\nu}$}{\small \par}
		\item $\quad$\textbf{Step 1}: Compute $\boldsymbol{\widehat{\omega}}\left(\boldsymbol{\omega}^{\nu}\right),$
		the solution of  $\boldsymbol{\widehat{\omega}}\left(\boldsymbol{\omega}^{\nu}\right)=$$\underset{\boldsymbol{\omega}}{\text{argmin}}$
		$\boldsymbol{\overline{U}}\left(\boldsymbol{\omega},\boldsymbol{\omega}^{\nu}\right)$, s.t. \eqref{eq:rho_beta_d}--\eqref{eq:rho_nu_e},
		\eqref{eq:capConst},
		\eqref{eq:mgf_beta_d}--\eqref{eq:mgf_beta_e}\\
		using projected gradient descent
		\item $\quad$\textbf{Step 2}: $\ensuremath{\boldsymbol{\omega}^{\nu+1}=\boldsymbol{\omega}^{\nu}+\gamma^{\nu}\left(\widehat{\boldsymbol{\omega}}\left(\boldsymbol{\omega}^{\nu}\right)-\boldsymbol{\omega}^{\nu}\right)}$.
		\item $\quad$//\textit{\small{}update index}{\small \par}
		\item \textbf{Set} $\ensuremath{\nu\leftarrow\nu+1}$
		\item \textbf{end while}
		\item \textbf{output: }$\ensuremath{\widehat{\boldsymbol{\omega}}\left(\boldsymbol{\omega}^{\nu}\right)}$
	\end{enumerate}
\end{algorithm}
\section{Proof of Results in Appendix  \ref{opt_mention}} \label{apd_ww}

\subsection{Proof of Lemma \ref{cnv_t}}
The constraints \eqref{eq:mgf_beta_d}--\eqref{eq:mgf_beta_e} are separable for each $h_i$ and due to symmetry of the three constraints it is enough to prove convexity of $E(h)=\sum_{f=1}^{r}\pi_{f,j}q_{f,j,\beta_{j}}\lambda_{f}e^{-\lambda_i \omega_i}\left(\frac{\alpha}{\alpha-h_{i}}\right)^{L_{f}-L_{j,f}}-\left(\Lambda_{j,\beta_{j}}+h_{i}\right)$, assuming that the edge router $\ell$ is unfold, without loss of generality. Thus, it is enough to prove that $E''(h)\ge 0$. 
We further note that it is enough to prove that $D''(h)\ge 0$, where $D(h) = \left(\frac{\alpha}{\alpha-t_{i}}\right)^{L_{f}-L_{j,f}
}$. 
This follows since 
\begin{align}
D^{'}(h) & =(L_{f}-L_{j,f})(1-\frac{h}{\alpha})^{L_{j,f}-L_{f}-1}\times(1/\alpha)\ge0 \\
D^{''}(h) & =(L_{f}^{2}-L_{j,f}^{2}+L_{f}-L_{j,f})(1-\frac{h}{\alpha})^{L_{j,f}-L_{f}-2} \nonumber \\
& \,\,\,\,\,\,\,\,\,\,\,\,\,\,\,\,\,\,\,\,\, \times(1/\alpha^{2})\ge0
\end{align}

%The first derivative of $D(t)$ is given as

%\begin{equation}
%	D'(t) =\frac{\pi_{ij}\lambda_{i}L_{i}\alpha_{j,\nu_j}^{L_i}e^{\beta_{j,\nu_j}tL_{i}}}{\left(\alpha_{j,\nu_j}-t\right)^{L_i+1}}\left[\beta_{j,\nu_j}\left(\alpha_{j,\nu_j}-t\right)+1\right]
%\end{equation}

%Note that $E'(t)\ge0$ since $\alpha_{j,\nu_j}\get$.  Differentiating it again to get the second derivative, we get the second derivative as follows. 
%\begin{equation}
%	E''(t)=\left(1+\frac{1}{\alpha_{j,\nu_j}-t}\right)\left((\alpha_{j,\nu_j}-t)\,L_{i}+L_{i}+1\right)-1 
%	\label{eq:E_22}
%\end{equation}

%Since $\alpha_{j,\nu_j}\get$, $E''(t)$ given in  \eqref{eq:E_22} is non-negative, which proves the Lemma. 
\subsection{Proof of Lemma \ref{cnv_ww}} 
The constraint \eqref{eq:mgf_beta_d}--\eqref{eq:mgf_beta_e}  are separable for each  $\alpha_{j,\beta_j}^{(d)}$, $\alpha_{j,\beta_j}^{(c)}$ and $\alpha_{j,\nu_j}^{(e)}$, respectively. Note that we omit the subscript $\ell$ for simplicity, w.l.o.g. Thus, it is enough to prove convexity of the following three equations
%$E_{1}(\alpha_{j,\beta_{j}}^{(d)})=\sum_{f=1}^{r}\pi_{f,j}\lambda_{f}q_{f,j,\beta_{j}}\left(\frac{\alpha_{j,\beta_{j}}^{(d)}}{\alpha_{j,\beta_{j}}^{(d)}-t}\right)^{L_{f}-L_{j,f}}-\left(\Lambda_{j,\beta_{j}}^{(d)}+t\right)$,
\begin{align*}
\ensuremath{} & E_{1}(\alpha_{j,\beta_{j}}^{(d)})=\\
& \ensuremath{\sum_{f=1}^{r}\pi_{f,j}\lambda_{f}q_{f,j,\beta_{j}}e^{-\lambda_i \omega_i}\left(\frac{\alpha_{j,\beta_{j}}^{(d)}}{\alpha_{j,\beta_{j}}^{(d)}-h}\right)^{L_{f}-L_{j,f}}-\left(\Lambda_{j,\beta_{j}}^{(d)}+h\right)}
\end{align*}

%$E_{2}(\alpha_{j,\beta_{j}}^{(c)})=\sum_{f=1}^{r}\pi_{f,j}\lambda_{f}q_{f,j,\beta_{j}}\left(\frac{\alpha_{j,\beta_{j}}^{(c)}}{\alpha_{j,\beta_{j}}^{(c)}-t}\right)^{L_{f}-L_{j,f}}-\left(\Lambda_{j,\beta_{j}}^{(c)}+t\right)$
\begin{align*}
\ensuremath{} & E_{2}(\alpha_{j,\beta_{j}}^{(\overline{d})})=\\
& \sum_{f=1}^{r}\pi_{f,j}\lambda_{f}q_{f,j,\beta_{j}}e^{-\lambda_i \omega_i}\left(\frac{\alpha_{j,\beta_{j}}^{(\overline{d})}}{\alpha_{j,\beta_{j}}^{(\overline{d})}-h}\right)^{L_{f}-L_{j,f}}-\left(\Lambda_{j,\beta_{j}}^{(\overline{d})}+h\right)\\
& \ensuremath{E_{3}(\alpha_{j,\nu_{j}}^{(e)})=}\\
& \sum_{f=1}^{r}\pi_{f,j}\lambda_{f}p_{f,j,\nu_{j}}e^{-\lambda_i \omega_i}\left(\frac{\alpha_{j,\nu_{j}}^{(e)}}{\alpha_{j,\nu_{j}}^{(e)}-h}\right)^{L_{f}-L_{j,f}}-\left(\Lambda_{j,\nu_{j}}^{(e)}+h\right)
\end{align*}
%and $E_{3}(\alpha_{j,\nu_{j}}^{(e)})=\sum_{f=1}^{r}\pi_{f,j}\lambda_{f}p_{f,j,\nu_{j}}\left(\frac{\alpha_{j,\nu_{j}}^{(e)}}{\alpha_{j,\nu_{j}}^{(e)}-t}\right)^{L_{f}-L_{j,f}}-\left(\Lambda_{j,\nu_{j}}^{(e)}+t\right)$ 
for $h<\alpha_{j,\beta_j}^{(d)}$, $h<\alpha_{j,\beta_j}^{(c)}$, and $h<\alpha_{j,\nu_j}^{(e)}$, respectively. Since there is only a single index $j$, $\beta_j$, and $\nu_j$, here, we ignore the subscripts and superscripts for the
rest of this proof and prove for only one case due to the symmetry. Thus, it is enough to prove that $E_1''(\alpha)\ge 0$  for $h<\alpha$. We further note that it is enough to prove that $D_1''(\alpha)\ge 0$, where $\ensuremath{D_{1}(\alpha)=\left(1-\frac{h}{\alpha}\right)^{L_{j,i}-L_{i}}}$. This holds since, 
\begin{align}
D_{1}^{'}(\alpha) & =(L_{j,i}-L_{i})(1-\frac{h}{\alpha})^{L_{j,i}-L_{i}-1}\times(t/\alpha^{2})\\
D_{1}^{''}(\alpha) & =((L_{j,i}-L_{i})^2-L_{j,i}+L_{i})(1-\frac{h}{\alpha})^{L_{j,i}-L_{i}-2} \times \nonumber \\
& \,\,\,\,\,\,\,\,\,\,\,\,\,\,\,(h/\alpha^{3})\ge0
\end{align}

%\textcolor{blue}{
\section{Key notations used in this paper}\label{notation}
The key used notation in this paper are shown in Table \ref{tab:Key-Notations-Used}. 
\begin{table}[htbp]
	\small{
	\caption{{Key Notations Used in This Paper}\label{tab:Key-Notations-Used}} \resizebox{.48\textwidth}{!}{	
	\begin{tabular}{|l|>{\raggedright}p{6cm}|}
		\hline 
		\textbf{Symbol} & \textbf{Meaning}\tabularnewline
		\hline 
		\hline 
		$r$ & Number of video files in system\tabularnewline
		\hline 
		$m$ & Number of storage nodes \tabularnewline
		\hline 
%		$L_{i}$ & Number of segments for video file $i$\tabularnewline
%		\hline 
		$L_{i}, L_{j,i}$ & Number of video segments and cached chunks for video file $i$, respectively, at server $j$\tabularnewline
		\hline 
		$\lambda_{i}$ & Possion arrival rate of video file $i$\tabularnewline
		\hline 
		$\pi_{i,j,\ell}$ & Probability of retrieving chunk of file $i$ from node $j$ from edge-router $\ell$ using
		probabilistic scheduling algorithm\tabularnewline
		\hline
		$p_{i,j,\nu_j\ell}$ & Probability of retrieving chunk of file $i$ from server $j$ and PS $\nu_j$, if requested through edge-router $\ell$.\tabularnewline
		\hline
		$q_{i,j,\beta_j\ell}$ & Probability of retrieving chunk of file $i$ from server $j$ and PS $\beta_j$, if requested through edge-router $\ell$.\tabularnewline
		\hline 
		$\text{PS}_{\beta_j,\ell}^{(d,j)},\forall \beta_j$ & Set of parallel streams between data center and cache server $j$ which serves edge router $\ell$ \tabularnewline
		\hline 
		$\text{PS}_{\beta_j,\ell}^{(\overline{d},j)},\forall \beta_j$ & Set of parallel streams between cache server $j$ and edge-router $\ell$
		\tabularnewline
		\hline 
		$\text{PS}_{\nu_j,\ell}^{(e,j)},\forall \nu_j$ & Set of parallel streams between cache server $j$ and edge-router $\ell$ assigned to serve cached segments \tabularnewline
		\hline 
		$\left(\alpha_{j}^{(d)},\,\eta_{j}^{(d)}\right)$ & Parameters of Shifted Exponential distribution of service time from data center to cache server $j$\tabularnewline
		\hline 
		$\left(\alpha_{j,\ell}^{(f_j)},\,\eta_{j,\ell}^{(f_j)}\right)$ & Parameters of Shifted Exponential distribution of service time from cache server $j$ to edge-router $\ell$\tabularnewline
		\hline 
		$M_{j,\beta_j,\ell}^{(d)}$ & Moment generating function for the service time of  the parallel stream $\text{PS}_{\beta_j,\ell}^{(d,j)}$ \tabularnewline  
		\hline 
		$M_{j,\beta_j,\ell}^{(\overline{d})}$ & Moment generating function for the service time of  the parallel stream $\text{PS}_{\beta_j,\ell}^{(\overline{d},j)}$ \tabularnewline  
		\hline 
		$M_{j,\beta_j,\ell}^{(e)}$ & Moment generating function for the service time of  the parallel stream $\text{PS}_{\nu_j,\ell}^{(e,j)}$ \tabularnewline  
		\hline
		$\Lambda_{j,\beta_j,\ell}^{(d)}$ & aggregate arrival rate at parallel stream $\text{PS}_{\beta_j,\ell}^{(d,j)}$ \tabularnewline
		\hline
		$\Lambda_{j,\beta_j,\ell}^{(\overline{d})}$ & aggregate arrival rate at parallel stream $\text{PS}_{\beta_j,\ell}^{(\overline{d},j)}$ \tabularnewline
		\hline
		$\Lambda_{j,\beta_j,\ell}^{(e,j)}$ & aggregate arrival rate at parallel stream $\text{PS}_{\beta_j,\ell}^{(e,j)}$ \tabularnewline
		\hline 
		$\sigma$ & Parameter indexing stall duration tail probability \tabularnewline
		\hline 
		$X_{i,\ell}$ &  Random variable corresponding to amount of space in the edge-cache $\ell$ for video file $i$\tabularnewline
		\hline
		$T_{i,j,\beta_j,\nu_j}^{(u)}$ &   Time that chunk $u$ begins to play at client $i$ given that it is downloaded from $\beta_j$ and $\nu_j$.
		\tabularnewline
		\hline
		$\Gamma^{i,j,\beta_j,\nu_j}_{U}$ &  Stall duration for the request of file $i$ from $\beta_j$ and $\nu_j$ queues (not cached at the edge-router).\tabularnewline
		\hline
		$\Gamma^{i,j,\beta_j,\nu_j}_{tot}$ &  Total stall duration for the request of file $i$ from either $\beta_j$ and $\nu_j$ queues or from the edge-cache.\tabularnewline
		\hline
		$D_{i,j,\beta_j,\nu_j}^{(u)}$ & Download time fo chunk $u\in\left\{ 1,\ldots,L_{i}\right\} $
		of file $i$ from node $j$, from $\beta_j$ and $\nu_j$ queues.\tabularnewline
		\hline 
		$R_{j,\nu_j}^{(e)}$ & Service time of the video files at the parallel streams $\text{PS}_{(\nu_j)}^{(e,j)}$ \tabularnewline
		\hline 
		$B_{j,\beta_j}^{(d)}$ & MGF of the Service time of all video files from the parallel stream $\text{PS}_{(\beta_j)}^{(d,j)}$ \tabularnewline
		\hline   
		$\rho_{j,\nu_j}^{(e)}$ & load intensity at the parallel stream $\text{PS}_{(\nu_j)}^{(e,j)}$.\tabularnewline
		\hline   
		$\rho_{j,\beta_j}^{(d)}$ & load intensity at the parallel stream $\text{PS}_{(\beta_j)}^{(d,j)}$.\tabularnewline
		\hline   
		$\rho_{j,\beta_j}^{(\overline{d})}$ & load intensity at the parallel stream $\text{PS}_{(\beta_j)}^{(\overline{d},j)}$.\tabularnewline
		\hline 
		$d_{s}$ & Start-up delay \tabularnewline
		\hline 
		$\tau$ & Chunk size in seconds\tabularnewline
		\hline 
		$\theta$ & Trade off factor between mean stall duration and stall duration tail
		probability\tabularnewline
		\hline 
	\end{tabular}
}
}
\end{table}
\section{Mean Stall Duration}
\label{sec:mean}
In this section, a bound for the mean stall duration, for any video file $i$, is provided. Since probabilistic scheduling is one feasible  strategy, the obtained bound is an upper bound to the optimal strategy. 

Using equation \eqref{stallTime}, the stall duration for the request of file $i$ from $\beta_{j}$ queue,  $\nu_{j}$ queue and server $j$, $\Gamma^{(i,j,\beta_{j},\nu_{j})}_{U}$ is given as
\begin{eqnarray}
\Gamma^{(i,j,\beta_{j},\nu_{j})}_{U}=T_{i,j,\beta_{j},\nu_{j}}^{(L_{i})}-d_{s}-\left(L_{i}-1\right)\tau 
\label{stallTime2}\label{eq:E_T_s_2}
\end{eqnarray}

 An exact evaluation for the play time of segment $L_{i}$ is hard due to the dependencies between  $\mathcal{F}_{i,j,\nu_{j},\beta_{j},z}$ (i.e., equation \eqref{F_1}) random variables for different values of $j$, $\nu_j$, $\beta_j$ and $z$, where $z\in{(1,2,...,L_{i}+1)}$. Hence, we derive an upper-bound on the playtime of the segment $L_{i}$ as follows. Using Jensen's inequality \cite{kuczma2009introduction}, we have for $g_i>0$, 
 
 \begin{equation}
e^{g_{i}\mathbb{E}\left[T_{i}^{\left(L_{i}\right)}\right]} \leq \mathbb{E}\left[e^{g_{i}T_{i}^{\left(L_{i}\right)}}\right]. \label{eq:jensen}
 \end{equation}
 
 Thus, finding an upper bound on the moment generating function for $T_i^{(L_i)}$ will lead to an upper bound on the mean stall duration. Hence, we will now bound the moment generating function for $T_i^{(L_i)}$. Using equation \eqref{eqnSDTP}, we can show that
\begin{align}
\mathbb{E}\left[e^{g_{i}T_{i}^{\left(L_{i}\right)}}\right] & \leq \sum_{j=1}^{m}\pi_{i,j}\times \Biggl[\widetilde{c}+\widetilde{a}\,e^{g_{i}(d_{s}+(L_{i}-1)\tau)}+\overline{a}\,\times\nonumber\nonumber \\
& \sum_{\nu_{j}=1}^{e_{j}}p_{i,j,\nu_{j}}\sum_{\beta_{j}=1}^{d_{j}}q_{i,j,\beta_{j}}e^{g_{i}L_{i}\tau}\times\nonumber \\
& \Biggl(\frac{\widetilde{M}_{j,\nu_{j}}^{(e)}(g_{i})(1-\rho_{j,\beta_{j}}^{(e)})g_{i}((\widetilde{M}_{j,\nu_{j}}^{(e)}(g_{i}))^{L_{j,i}}-1)}{(g_{i}-\Lambda_{j,\beta_{j}}^{(e)}(B_{j,\beta_{j}}^{(e)}(g_{i})-1))(\widetilde{M}_{j,\nu_{j}}^{(e)}(g_{i}))-1)}\nonumber \\
& +\frac{(1-\rho_{j,\beta_{j}}^{(\overline{d})})t_{i}(\widetilde{M}_{j,\nu_{j}}^{(\overline{d})}(g_{i}))^{L_{j,i}-L_{i}}}{g_{i}-\Lambda_{j,\beta_{j}}^{(\overline{d})}(B_{j,\beta_{j}}^{(\overline{d})}(g_{i})-1)}+\nonumber \\
& \frac{(1-\rho_{j,\beta_{j}}^{(d)})t_{i}(\widetilde{M}_{j,\beta_{j}}^{(\overline{d})}(g_{i}))^{L_{i}+1}}{\left[g_{i}-\Lambda_{j,\beta_{j}}^{(d)}(B_{j,\beta_{j}}^{(d)}(g_{i})-1)\right](\widetilde{M}_{j,\beta_{j}}^{(d)}(g_{i}))^{L_{j,i}+1}}\times\nonumber \\
& \Biggl(\frac{(\widetilde{M}_{j,\beta_{j}}^{(d,\overline{d})}(g_{i}))^{L_{i}-L_{j,i}}-(L_{i}-L_{j,i})}{(\widetilde{M}_{j,\beta_{j}}^{(d,\overline{d})}(g_{i}))-1}+\nonumber \\
& +\frac{\widetilde{M}_{j,\beta_{j}}^{(d,\overline{d})}(g_{i})\left((\widetilde{M}_{j,\beta_{j}}^{(d,\overline{d})}(g_{i}))^{L_{i}-L_{j,i}-1}-1\right)}{(\widetilde{M}_{j,\beta_{j}}^{(d,\overline{d})}(g_{i}))-1}\Biggr)\Biggl)\Biggl]\nonumber \\
&
=\sum_{j=1}^{m}\pi_{i,j}\times M_{D}^{(i,j)}
\label{eq:mgf_bound}
\end{align}
 
%
%% Preview source code for paragraph 0
%\begin{eqnarray}
%  \mathbb{E}\left[e^{t_{i}T_{i}^{\left(L_{i}\right)}}\right] & \overset{(a)}{=} & \mathbb{E}\left[\underset{z}{\mbox{max}}  \,\underset{j\in\mathcal{A}_{i}}{\mbox{max}}\,e^{t_{i}p_{ijz}}\right]\nonumber\\
% & = & \mathbb{E}_{\mathcal{A}_{i}}\left[\mathbb{E}\left[\underset{z}{\mbox{max}}\,\underset{j\in{\mathcal{A}_{i}}}{\mbox{max}}\,e^{t_{i}p_{ijz}}|\,\mathcal{A}_{i}\right]\right]\nonumber
%\\
% &\overset{(b)}{\leq} & \mathbb{E}_{\mathcal{A}_{i}}\left[\sum_{j\in \mathcal{A}_{i}}\mathbb{E}\left[\underset{z}{\mbox{max}}\,e^{t_{i}p_{ijz}}\right]\right]\nonumber
%\\
% & = & \mathbb{E}_{\mathcal{A}_{i}}\left[\sum_{j}F_{ij}\mathbf{1}_{\left\{ j\in\mathcal{A}_{i}\right\} }\right]\nonumber
%\\
% & = & \sum_{j}F_{ij}\,\mathbb{E}_{\mathcal{A}_{i}}\left[\mathbf{1}_{\left\{ j\in\mathcal{A}_{i}\right\} }\right]\nonumber
%\\
% & = & \sum_{j}F_{ij}\,\mathbb{P}\left(j\in\mathcal{A}_{i}\right)\nonumber
%\\
% & \overset{(c)}{=} & \sum_{j}F_{ij}\pi_{ij}
%\label{eq:mgf_bound}
%\end{eqnarray}
%where (a) follows from \eqref{eq:pjstart}, (b) follows by upper bounding $\max_{j\in \mathcal{A}_{i}} $ by $\sum_{j\in \mathcal{A}_{i}} $, (c) follows by probabilistic scheduling where $\mathbb{P}\left(j\in\mathcal{A}_{i}\right) = \pi_{ij}$,   and  $F_{ij}=\mathbb{E}\left[\underset{z}{\mbox{max}}  \, e^{t_{i}p_{ijz}}\right]$. We note that the only inequality here is for replacing the maximum by the sum. Since this term will be inside the logarithm for the mean stall latency, the gap between the term and its bound becomes additive rather than multiplicative. 

Substituting \eqref{eq:mgf_bound} in \eqref{eq:jensen},  the  mean stall duration is bounded as follows.

%\begin{equation}
%\mathbb{E}\left[T_{i}^{(L_{i})}\right]\leq\frac{1}{t_{i}}\text{log}\left(\sum_{j=1}^{m}\pi_{ij}M_{D}^{(i,j,\nu_j,\beta_j)}\right).\label{eq:ET_i}
%\end{equation}

% Let $H_{ij}=\sum_{\ell=1}^{L_{i}}e^{-t_{i}\left(d_{s}+\left(\ell-1\right)\tau\right)}Z_{{i,j}}^{(\ell)}(t_{i})$, where $Z_{{i,j}}^{(\ell)}(t)$ is defined in equation \eqref{eq:M_D_ij}. We  note that $H_{ij}$ can be simplified using the geometric series formula as follows.
% 
% \begin{lemma}\label{hijlem}
% 	\begin{equation}
% 	H_{ij} =  \frac{e^{-t_{i}\left(d_{s}-\tau\right)}\left(1-\rho_{j}\right)t_{i}B_{j}(t_{i})\widetilde{M}_{j}(t_{i})}{t_{i}-\Lambda_{j}\left(B_{j}(t_{i})-1\right)}\frac{1-\left(\widetilde{M}_{j}(t_{i})\right)^{L_{i}}}{\left(1-\widetilde{M}_{j}(t_{i})\right)},
% 	\label{eq:H}
% 	\end{equation}
% 	where $\widetilde{M}_{j}(t_{i})=M_{j}(t_{\text{i}})e^{-t_{i}\tau}$, $M_j(t_i)$ is given in (\ref{M_j_t_1}), and $B_j(t_i)$ is given in (\ref{eq:servTimeofFileB_j_i}). 
% \end{lemma}
% \begin{proof}
% 	The proof is provided in Appendix \ref{apdx:hjlem}.
% \end{proof}

% Substituting \eqref{eq:ET_i} in \eqref{eq:E_T_s_2} and some manipulations, the  mean stall duration is bounded as follows.

\begin{theorem}
	The mean stall duration time for file $i$ is bounded by 
\begin{equation}
\mathbb{E}\left[\Gamma^{(i)}\right]\leq\frac{1}{t_{i}}\text{log}\left(\sum_{j=1}^{m}\pi_{ij}\left(1+M_{D}^{(i,j)}\right)\right) \label{eq:T_s_main_stall}
\end{equation}
	for any $t_{i}>0$, $\rho_{j,\nu_{j},\ell}^{(e)}<1$, 
	$\rho_{j,\beta_{j},\ell}^{(\overline{d})}<1$
	, $\text{and}$ $\rho_{j,\nu_{j},\ell}^{(e)}<1$
	\,$\text{and}$ \\
	the involved MGFs exist,\,$\forall j,\nu_j, \beta_j$.\label{meanthm}
\end{theorem}
%\begin{proof}
%	The proof is provided in Appendix \ref{apdx:boundmean2}.
%\end{proof}

%To use  the bound \eqref{eq:ET_i}, $F_{ij}$ needs to be bounded too.

%Further, substituting the bounds \eqref{eq:F_ij} and \eqref{eq:ET_i} in \eqref{eq:E_T_s_2}, the  mean stall duration is bounded as follows.  

%, which is the inner summation in  (\ref{eq:ET_s_i}). 

%Thus, we obtain the following result. 

%\begin{equation}
%\mathbb{E}\left[T_{S}^{(i)}\right]\leq%\frac{1}{t_{i}}\text{log}\left(\sum_{j=1}^{m}\pi_{ij}\left(1+H_{ij}\right)\right)\label{eq:T_s_main_stall}
%\end{equation}

%
%Note that Theorem \ref{meanthm} above holds only in the range of $t_i$ when  $t_i-\Lambda_{j}\left(B_{j}(t_i)-1\right)>0$ which reduces to 
%$\,\sum_{f=1}^r\pi_{fj}\lambda_{f}\left(\frac{\alpha_{j}e^{-\beta_{j}t_{i}}}{\alpha_{j}-t_{i}}\right)^{L_{f}}-\left(\Lambda_{j}+t_{i}\right)<0,\,\forall i, j$,
%and $\alpha_{j}-t_i>0$. Further, the server utilization $\rho_j $ must be less than $1$ for stability of the system.

We note that for the scenario, where the files are downloaded rather than streamed, a metric of interest is the mean download time. This is a special case of our approach when the number of segments of each video is one, or $L_i=1$. Thus, the mean download time of the file follows as a special case of Theorem \ref{meanthm}. 

%We note that the authors of \cite{Xiang:2014:Sigmetrics:2014,Yu_TON} gave an upper bound for mean file download time using probabilistic scheduling. However, the bound in this paper is different since we use moment generating function based bound. The two bounds are compared in Section \ref{sec:num}, and the bounds in this paper are shown to outperform those in \cite{Xiang:2014:Sigmetrics:2014,Yu_TON}.

\section{Online Algorithm for Edge-cache Placement}\label{sec:adapt}
We note that for the setup of the edge cache, we assumed
that the edge-cache has a capacity of $C_{e,\ell}$ seconds (ignoring the
index of the edge cache). However, in the caching policy, we
assumed that a file $f$ is removed from the edge cache $\ell$ if it has not
been requested in the last $\omega_{f,\ell}$ seconds. In the optimization,
we found the parameters $\omega_{f,\ell}$, such that the cache capacity
is exceeded with probability less than $\epsilon_{\ell}$. However, this still
assumes that it is possible to exceed the cache capacity
some times. This is, in practice, not possible. Thus, we will
propose a mechanism to adapt the decision obtained by the
optimization formulation so as to never exceed the edge cache
capacity.

\begin{figure}[t]
	\centering
	\includegraphics[trim=00.0in 0.00in 0.00in 0.00in, clip, width=0.47\textwidth]{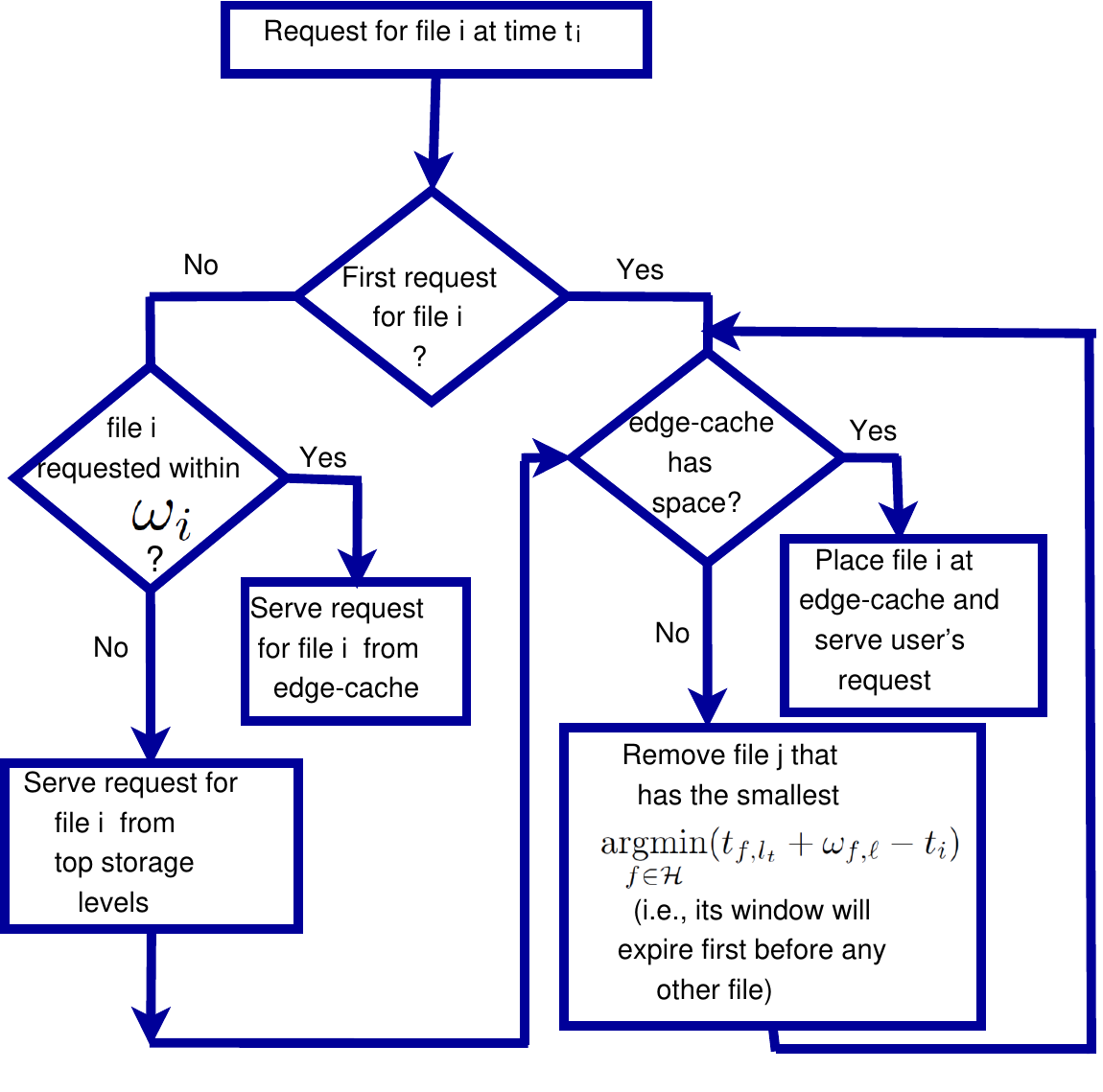}
	%	\vspace{-.45in}
	\caption{\small A flowchart illustrates the online updates for an edge-cache when a file $i$ is requested at time $t_i$. Here, $t_{f,l_t}$ represents the time of the last request of file $i$, and $\mathcal{H}$ is the index set of all video files in the edge-cache. \label{fig:flowchart}}
	%	\vspace{-.25in}
\end{figure}

When a file $i$ is requested, the last request of file $i$ is first
checked. If it has not been requested in the last $\omega_{i,\ell}$ seconds,
it is obtained from the CDN. In order to do that, the space
of the file is reserved in the edge-cache. If this reservation
exceeds the capacity of the edge-cache, certain files have to
be removed. Any file $f$ that has not been requested in the last $\omega_{f,\ell}$ seconds is removed from the cache. If, even after removing
these files, the space in the edge-cache is not enough for
placing file $i$ in the edge-cache, more files must be removed. Assume that $\mathcal{H}$ is the set contains all files in the edge-cache,
and $t_{f,l_t}$ is the last time file $f$ has been requested. Then, if another file needs to be removed to make space for the newly requested file, the file $\text{argmin}_{f\in\mathcal{H}}(t_{f,l_t}+\omega_{f,\ell}-t_i)$ is removed. This
continues till there is enough space for the new incoming file.
Note that multiple files may be removed to make space
for the incoming file, depending on the length of the new file.
This is similar concept to LRU where a complete new file
is added in the cache, and multiple small files may have to be removed to make space. The key part of the online adaptation
so as not to violate the edge-cache capacity constraint is
illustrated in Figure \ref{fig:flowchart}. This flowchart illustrates the online updates for an edge-cache when a file $i$ is requested at time $t_i$.

\section{Edge-cache Performance and  further Evaluation} \label{edgeCacheEval}

\begin{figure}[t]
	\centering
	\includegraphics[trim=0in 0in 4in 0.0in, clip, width=0.45\textwidth]{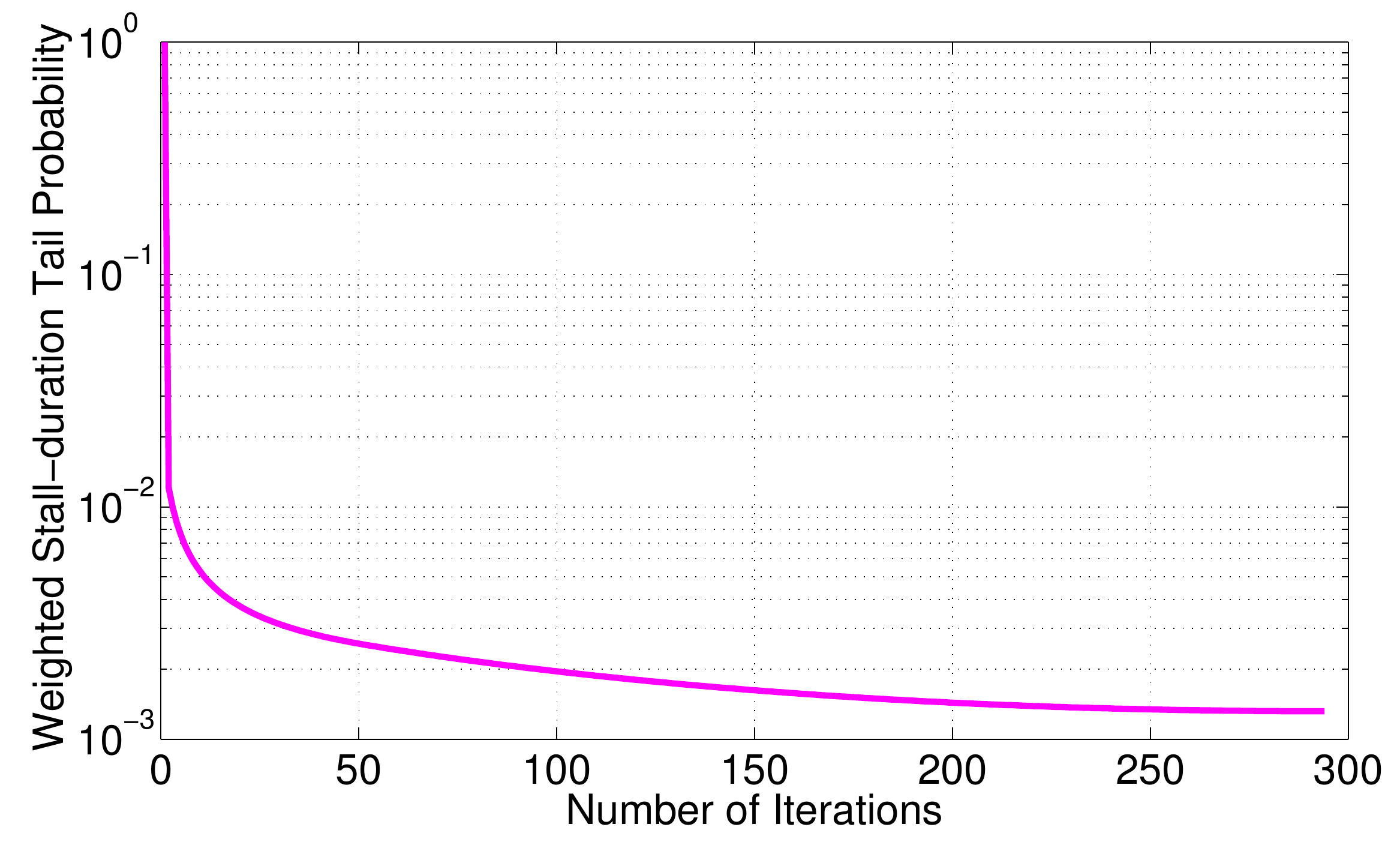}
	\vspace{-.05in}
	\captionof{figure}{ \small Convergence of weighted stall-duration tail probability.}
	\label{fig:SDTP}
\end{figure}%

\textit{Convergence of the proposed algorithm:} 
Figure \ref{fig:SDTP} shows the convergence of our proposed SDTP algorithm, which alternatively optimizes the weighted SDTP of all files over scheduling
probabilities $\widetilde{\boldsymbol{\pi}}$, auxiliary variables $\boldsymbol{t}$, bandwidth allocation weights $\boldsymbol{w}$, cache server placement $\boldsymbol{L}$, and window-size $\omega_i$. We
see that for $r = 500$ video files of size $600$s with
$m = 5$ cache storage nodes, the weighted stall duration tail probability converges  within a few iterations.

\begin{figure}[t]
	\centering\includegraphics[trim=0.01in 0.05in 4.50in 0.0in, clip, width=0.45\textwidth]{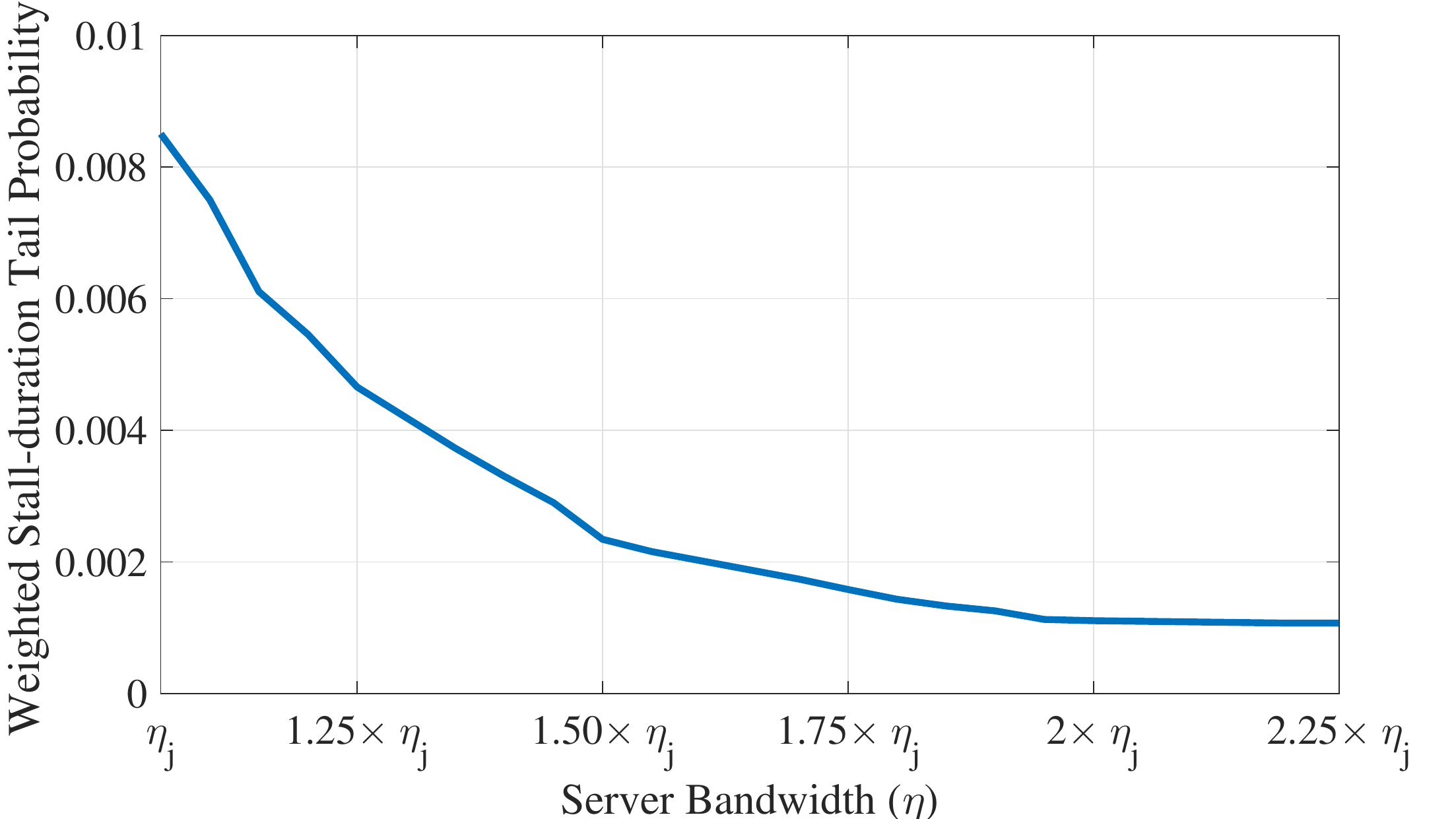}
	%	\vspace{-.45in}
	\caption{\small 
		Weighted SDTP versus the server bandwidth. We vary the server bandwidth from $\eta_j$ to $2.25\eta_j$ with an increment step of $0.25$, where $\eta_j=20$MBps.
		\label{fig:scalingBW}}
	%	\vspace{-.25in}
\end{figure}

\textit{Effect of scaling up the bandwidth of the cache servers and
	datacenter:} The effect of increasing the server bandwidth on the weighted SDTP is plotted in
Figure \ref{fig:scalingBW}. Intuitively, increasing the storage node bandwidth
will increase the service rate of the storage nodes by assigning higher bandwidth to the users,  thus,  reducing
the weighted SDTP.

\begin{figure}[t]
	\centering\includegraphics[trim=0.01in 0.05in 0.150in 0.0in, clip, width=0.45\textwidth]{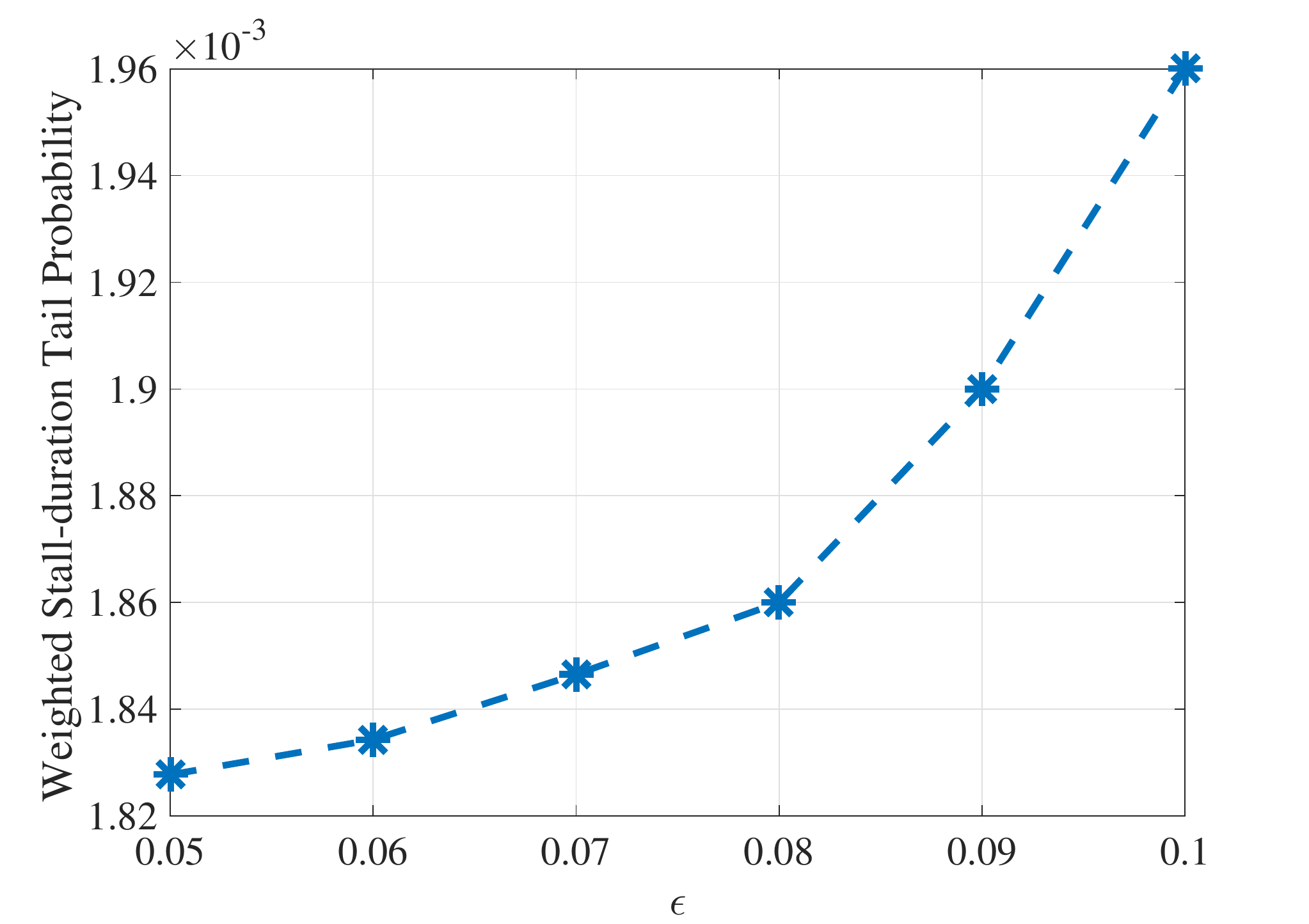}
	%	\vspace{-.45in}
	\caption{\small Weighted SDTP versus 
		the percentage bound on the number of video files (i.e., maximum capacity) in the edge cache $\epsilon$. The percentage of the capacity bound is changed from $0.05$ to $0.1$ for a cache capacity of $0.20\times C_{tot}$. \label{fig:missRateVsepsilon}}
	%	\vspace{-.25in}
\end{figure}

 \textit{Effect of the bound percentage $\epsilon$ in the SDTP:} Figure \ref{fig:missRateVsepsilon} plots the weighted SDTP versus $\epsilon$, i.e., probability that the cache size is exceeded. We see that the SDTP increases significantly with an increase in $\epsilon$. This is because as $\epsilon$ increases, there are more edge capacity constraint violations and the the online adaptations may not remain the optimal choice. 
 
 In the following figures, a trace-based implementation is performed, where the video ID, time requests, video lengths, etc. are obtained from one-week traces of a production system from the major service provider in the US. We note that the arrival process is not Poisson in this case, while the proposed approach still outperform the considered baseline approaches. 
%  We also note that the arrivals of video files in our analysis are estimated from one-week traces obtained from our production system.

\begin{figure}[t]
	%\centering\includegraphics[scale=0.50]{figs_up/cdf_actual_impl_v3}
	\centering\includegraphics[trim=0.01in 0.05in 4.250in 0.0in, clip, width=0.5\textwidth]{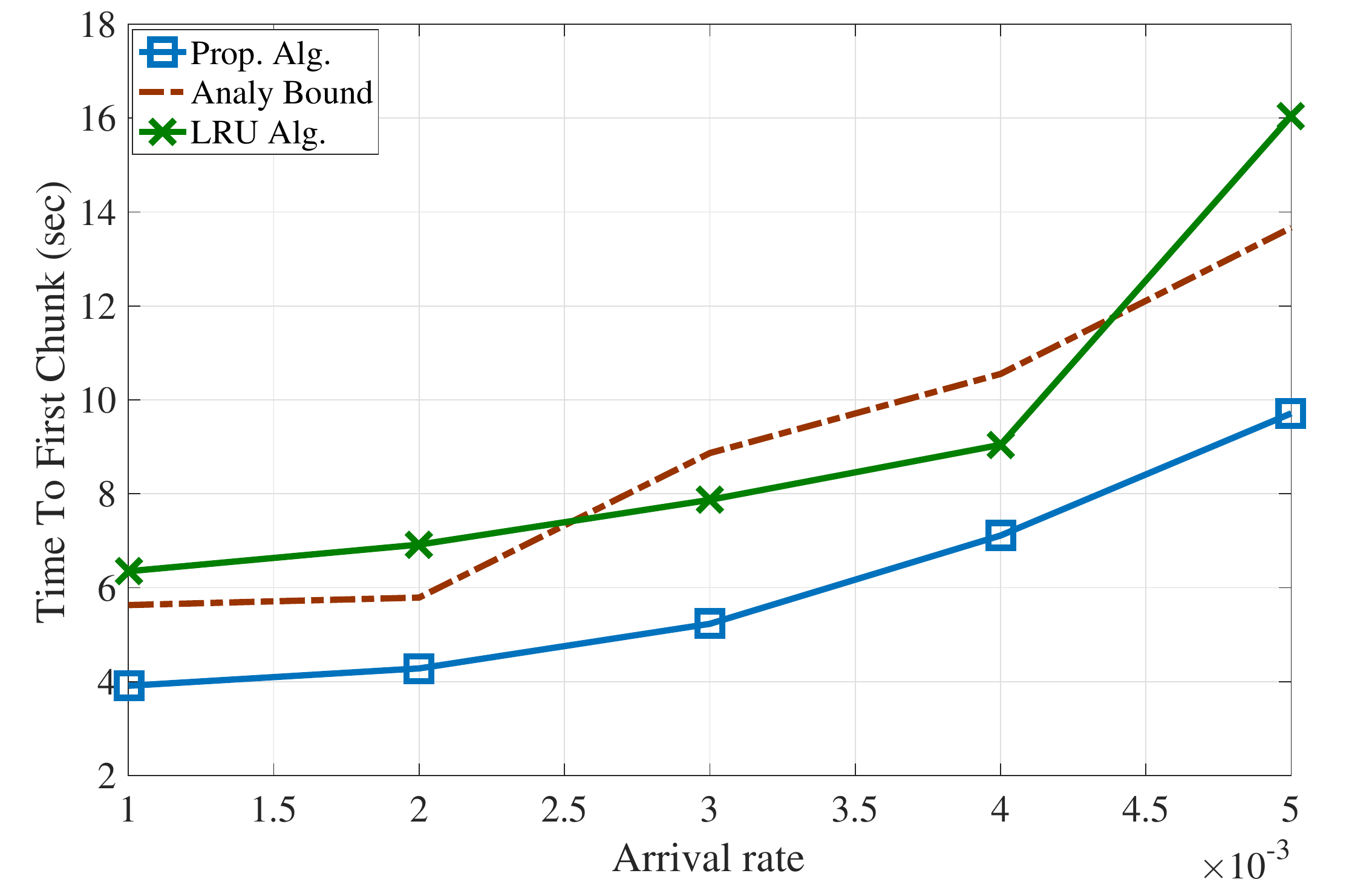}
	%\vspace{-.45in}
	\caption{\small Time to the first chunk for different arrival rates for $1000$ video files.
		\label{fig:tffc}}
	%\vspace{-.25in}
\end{figure}

{\it Effect of the arrival rates on the TTFC:}
Figure \ref{fig:tffc} shows the effect of different video arrival rates on the TTFC for different-size video lengths. The different sizes for video files are obtained from real traces of a major video service provider. We compared our proposed online algorithm with the analytical offline bound and LRU-based (explained in Section IV, B) policies. We see that the TTFC increases with arrival rates, as expected, however, since the TTFC is more significant at high arrival rates, we notice a significant improvement in the download time of the first chunk by about 60\% at the highest arrival rate in Figure \ref{fig:tffc} as compared to the LRU policy.

\begin{figure}[t]
	%\centering\includegraphics[scale=0.50]{figs_up/cdf_actual_impl_v3}
	\centering\includegraphics[trim=0.01in 0.05in 4.250in 0.0in, clip, width=0.5\textwidth]{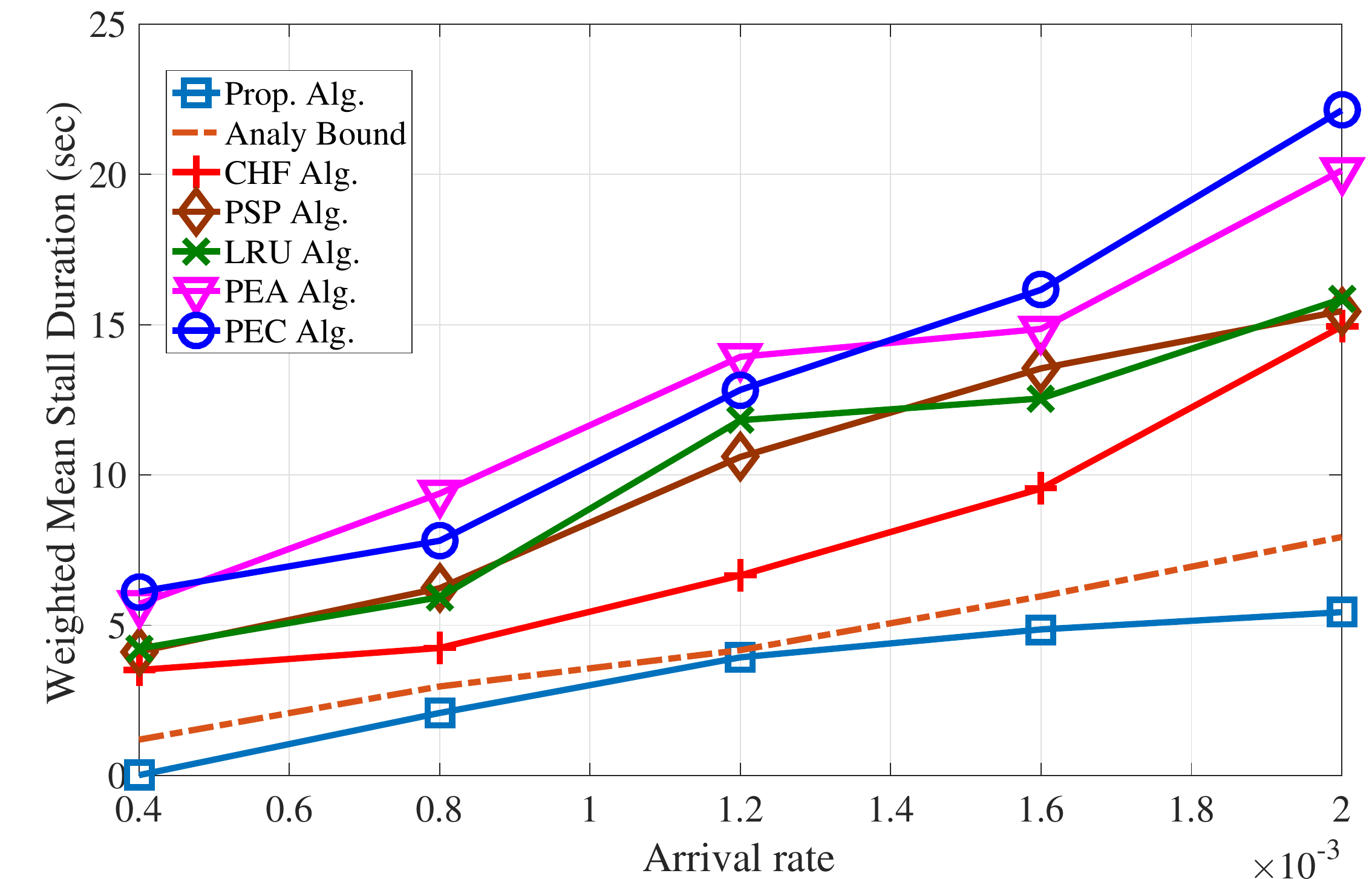}
	%\vspace{-.45in}
	\caption{\small Mean stall duration versus arrival rates.
		\label{fig:wMSDvsRate}}
	%\vspace{-.25in}
\end{figure}

{\it Effect of arrival Rates on the MSD:} The effect of different video arrival rates on the mean stall duration for different-size video length is captured in Figure \ref{fig:wMSDvsRate}. We compared our proposed online algorithm with five baseline policies and we see that the proposed algorithm outperforms all baseline strategies for the QoE metric of mean stall duration. Thus, bandwidth, size of the time-window, access and placement of files in the storage caches are important for the reduction of mean stall duration. Further, obviously, the mean stall duration increases with arrival rates, as expected. Since the mean stall duration is more significant at high arrival rates, we notice a significant improvement in mean stall duration (approximately 15s to about 5s) at the highest arrival rate in Figure \ref{fig:wMSDvsRate} as compared to the LRU policy

\begin{figure}[t]
	%\centering\includegraphics[scale=0.50]{figs_up/cdf_actual_impl_v3}
	\centering\includegraphics[trim=0.01in 0.01in 4.250in 0.0in, clip, width=0.5\textwidth]{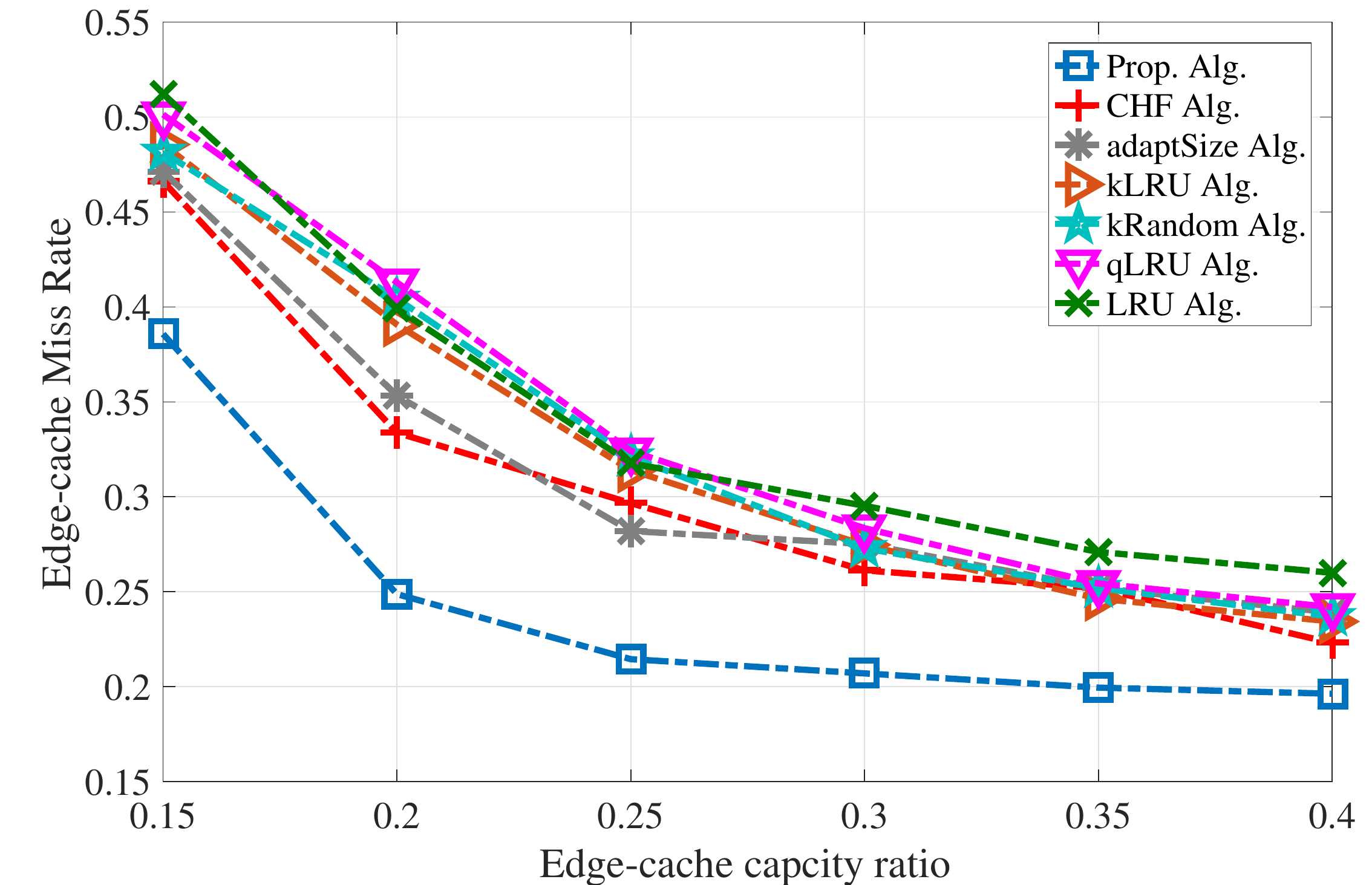}
	%\vspace{-.45in}
	\caption{\small Edge-cache miss rate versus edge-cache capacity ratio.
		\label{fig:edgeCacheMR}}
	%\vspace{-.25in}
\end{figure}

{\it Effect of edge-cache capacity:} We study the miss-rate (percentage of how many video file requests are not served from the edge-cache) performance of the edge-cache. Clearly, the miss-rate decreases with the increasing size of the capacity of the edge-cache. However, when the edge-cache capacity is approximately 35\% of the entire video sizes, the miss-rate is around 20\%.  Further, adaptSize policy does not neither optimize the time to live window of files $\omega_i$'s nor intelligently incorporate the arrival rates in adding/evicting the video files, and thus its performance becomes less sensitive to varying the cache size. The variant versions of LRU (qLRU with $q=0.67$, kLRU and kRandom with $k=6$) obtain closer performance compared to that of the basic LRU where kLRU performs that best among them as it somehow maintains a window (k-requests) for admitting a file into the cache and adapts LRU policy in the eviction process.

\section{Joint Mean-Tail Optimization} \label{tradeoff_sec}

We wish to jointly minimize the two QoE metrics (MSD and SDTP) over the choice of server-PSs scheduling, bandwidth allocation, edge-cache window-size and auxiliary variables. Since this is a multi-objective optimization, the objective can be modeled as a convex combination of the two QoE metrics.

The first objective is the minimization of the mean stall duration, averaged over all the file requests, and is given as $\sum_{i,\ell}\frac{\lambda_{i,\ell}}{\overline{\lambda}}\,\mathbb{E}\left[\Gamma^{\left(i,\ell\right)}\right]$. The second objective is the minimization of stall duration tail probability, averaged over all the video file requests, and is given as $\sum_{i,\ell}\frac{\lambda_{i,\ell}}{\overline{\lambda}}\,{\text{Pr}}\left(\Gamma^{(i,\ell)}\geq x\right)$. Using the expressions for the mean stall duration and the stall duration tail probability, respectively, optimization of a convex combination of the two QoE metrics can be formulated as follows. 
\begin{align}
\sum_{\ell=1}^{R}\sum_{i=1}^{r} & \frac{\lambda_{i,\ell}}{\lambda}\left[\theta\times\text{Pr}(\Gamma^{(i,\ell)}\geq\sigma)+(1-\theta)\times\mathbb{E}\left[\Gamma^{(i,\ell)}\right]\right] \label{jointOptProb}\\
\text{\textbf{s.t.}}\\
& (\ref{formulas})-(\ref{eq:Lij})\label{constQ_1}\\
& g_{i}<\alpha_{j,\beta_{j},\ell}^{(\overline{d})},%%\,
%\,
g_{i}<\alpha_{j,\nu_{j},\ell}^{(e)},\,\,\,\forall i,j,\nu_{j},\ell\\
& 0<g_{i}-\Lambda_{j,\beta_{j}}^{(d)}(B_{j,\beta_{j},\ell}^{(d)}(g_{i})-1),\,\,\,\forall i,j,\beta_{j},\ell\\
& 0<g_{i}-\Lambda_{j,\beta_{j},\ell}^{(\overline{d})}(B_{j,\beta_{j},\ell}^{(\overline{d})}(g_{i})-1),\,\,\,\forall i,j,\beta_{j}\\
& 0<g_{i}-\Lambda_{j,\nu_{j},\ell}^{(e)}(B_{j,\nu_{j},\ell}^{(e)}(g_{i})-1),\,\,\,\forall i,j,\nu_{j},\ell \label{constQ_last}\\
\boldsymbol{\text{var}} & \text{\,\,\,\,\,}\boldsymbol{\pi,}\,\boldsymbol{q},\,\boldsymbol{p},\,\boldsymbol{h},\boldsymbol{g},\,\boldsymbol{w}^{(c)},\,\boldsymbol{w}^{(d)},\,\boldsymbol{w}^{(e)},\,\boldsymbol{L},\boldsymbol{\omega}.\nonumber
\end{align}

Clearly, the above optimization problem is non-convex in all the parameters jointly. This is can be easily seen in the terms which are product of the different variables. Since the problem is non-convex, we propose an iterative algorithm to solve the problem.
This algorithm performs an alternating optimization over the different aforementioned dimensions, such that each sub-problem is shown to have convex constraints and thus can be efficiently solved using NOVA algorithm \cite{scutNOVA}. The subproblems are explained in detailed in Appendix \ref{opt_mention}.

\begin{figure}[t]
	%\centering\includegraphics[scale=0.50]{figs_up/cdf_actual_impl_v3}
	\centering\includegraphics[trim=0.01in 0.05in 4.250in 0.0in, clip, width=0.45\textwidth]{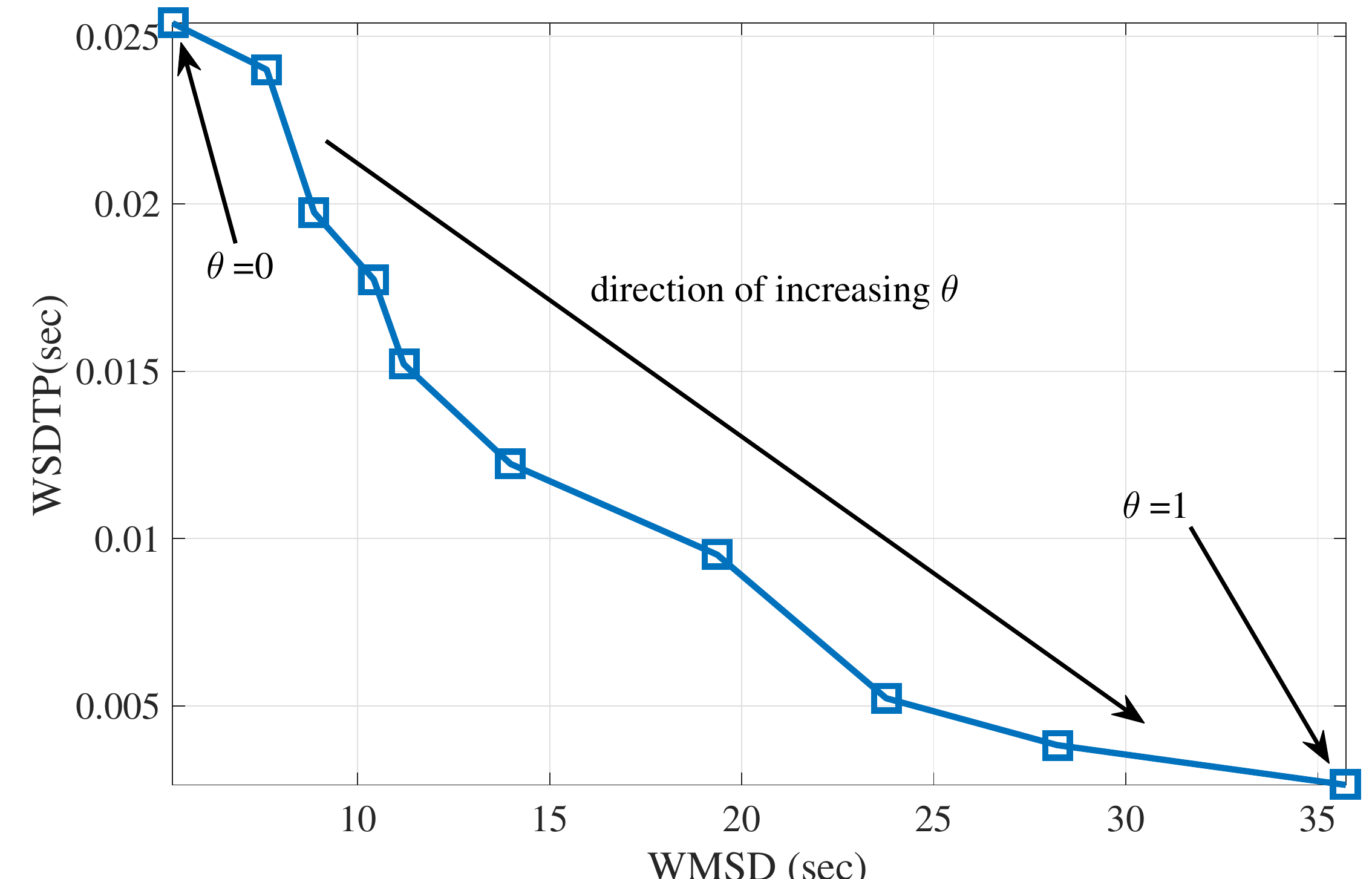}
	%\vspace{-.45in}
	\caption{\small Tradeoff between weighted mean stall duration and weighted stall-duration tail probability.
		\label{fig:tradeoff_SD_TP}}
	%\vspace{-.25in}
\end{figure}
{\it Mean-Tail tradeoff:}
There is a tradeoff between the MSD and SDTP.  Hence, we now investigate this tradeoff in order to get a better understanding of how this traddeoff can be compromised. To do so, we vary $\theta$ in the above optimization problem to get a tradeoff between MSD and SDTP.
Intuitively, if the mean stall duration decreases, the stall duration tail probability also reduces, as depicted in Figure \ref{fig:tradeoff_SD_TP}. Therefore, a question arises whether the optimal point for decreasing the mean stall duration and the stall duration tail probability is the same? Based on our real video traces, we answer this question in negative since we find that at the desgin values that optimize the mean stall duration, the stall duration tail probability is 10+ times higher as compared to the optimal stall duration tail probability. Similarly, the optimal mean stall duration is 7 times lower as compared to the mean stall duration at the design values that optimizes the stall duration tail probability. As a result, an efficient tradeoff point between the two QoE metrics can be chosen based on the point on the curve that is appropriate for the clients.
\section{Extension to Different Quality levels} \label{qltyext}

In this section, we show how our analysis can be extended to cover the scenario when the video can be streamed at different quality levels. We assume that each video file is encoded to different qualities, {\em i.e.}, $Q\in \{1,2, \cdots,V\}$, where $V$ is the number of possible choices for the quality level. The $L_{i}$  chunks of video file $i$ at quality $Q$ are denoted as $G_{i,Q,1}, \cdots, G_{i,Q,L_i}$. We will use a probabilistic quality assignment strategy, where  a chunk of quality $Q$  of size $a_{Q}$  is requested with probability $b_{i,Q}$ for all $Q\in \{1,2, \cdots,V\}$. We further  assume  all the chunks of the video are fetched at the same quality level. From Section \ref{quModel_PS}, we can show that the for a file of quality $Q$ requested from edge-router $\ell$, we choose server $j$ with probability $\pi_{i,j,\ell}^{(Q)}$. Further, we can show that the aggregate arrival rate at $PS^{(d,j)}_{\beta_j}$, $PS^{(\overline{d},j)}_{\beta_j,\ell}$, and $PS^{(e,j)}_{\nu_j,\ell}$, denoted as $\Lambda^{(d)}_{j,\beta_{j}}$,  $\Lambda^{(\overline{d})}_{j,\beta_{j},\ell}$,and  $\Lambda^{(e)}_{j,\nu_{j},\ell}$, respectively are given as follows. 
\begin{eqnarray}
\vspace{-.12in}
\Lambda^{(d)}_{j,\beta_{j}} &=& \sum_{i=1}^r \sum_{Q=1}^V \lambda_{i} \pi_{i,j}^{(Q)}q_{i,j,\beta_{j}}^{(Q)}b_{i,Q}\label{lambdad_betajQ}\\
\Lambda^{(c)}_{j,\beta_{j}} &=& \Lambda^{d}_{j,\beta_{j}} \label{lambdac_betajQ}\\
\Lambda^{(e)}_{j,\nu_{j}} &=& \sum_{i=1}^r \sum_{Q=1}^V \lambda_{i} \pi_{i,j}^{(Q)}p_{i,j,\nu_{j}}^{(Q)}b_{i,Q}\label{lambdae_nujQ}
\vspace{-.12in}
\end{eqnarray}

Similarly, we can define

\begin{eqnarray}
\alpha_{j,\beta_{j}}^{(d,Q)}= w_{j,\beta_{j}}^{(d)} \alpha_j^{(d,Q)}\label{alphadQ},\,\,\,\, \\
\alpha_{j,\beta_{j},\ell}^{(\overline{d},Q)}= w_{j,\beta_{j},\ell}^{(\overline{d})} \alpha_{j,\ell}^{(f_j,Q)}\label{alphacQ},\,\,\,\ 
\alpha_{j,\nu_{j},\ell}^{(e,Q)} = w_{j,\nu_{j},\ell}^{(e)} \alpha_{j}^{(f_j,Q)}\label{alphaeQ},
\end{eqnarray}
for all $\beta_j$, $\nu_j$, $Q$, and  $\ell$.
Note that $\alpha_j^{(.,Q)}=\alpha_j^{(.)}/a_{\ell}$ where $alpha_j^{(.)}$ is a constant service time parameter when $a_{\ell}=1$. We further define the moment generating functions of the service times of $PS^{(d,j,Q)}_{\beta_j}$, $PS^{(\overline{d},j,Q)}_{\beta_j,\ell}$, and $PS^{(e,j,Q)}_{\nu_j,\ell}$ as $M^{(d,Q)}_{j,\beta_j}$, $M^{(\overline{d},Q)}_{j,\beta_j,\ell}$, and $M^{(\overline{d},Q)}_{j,\nu_j,\ell}$, which are defined as follows. 
\begin{align}
&
M_{j,\beta_{j},\ell}^{(d,Q)}  =\frac{\alpha_{j,\beta_{j}}^{(d,Q)}e^{\eta_{j,\beta_j}^{(d)}t}}{\alpha_{j,\beta_{j}}^{(d,Q)}-t},\,\,\,\, \label{MGF_Q_1}\\
&
M_{j,\beta_{j},\ell}^{(\overline{d},Q)}  =\frac{\alpha_{j,\beta_{j},\ell}^{(\overline{d},Q)}e^{\eta_{j,\beta_j,\ell}^{(\overline{d},Q)}t}}{\alpha_{j,\beta_{j},\ell}^{(\overline{d},Q)}-t},\,\,\,\,\\
&
M_{j,\nu_{j},\ell}^{(e,Q)}  =\frac{\alpha_{j,\nu_{j},\ell}^{(e,Q)}e^{\eta_{j,\nu_j,\ell}^{(e,Q)}t}}{\alpha_{j,\nu_{j},\ell}^{(e,Q)}-t}\label{Me_jnujQ}
\end{align}
where $\eta_{j,.}^{(.,Q)}=\eta_{j,.}^{(.)}\times a_{\ell}$ where $eta_{j,.}^{(.)}$ is a constant time shift parameter when $a_{\ell}=1$. Following the same analysis as in Section \ref{sec:StallTailProb}, it is easy to show that the stall duration for the request of file $i$ at quality $Q$ from $\beta_{j}$ queue,  $\nu_{j}$ queue and server $j$, if not in the edge-cache, i.e., $\Gamma^{(i,j,\beta_{j},\nu_{j},Q)}_{U}$ is given as
\begin{equation}
\Gamma^{(i,j,\beta_{j},\nu_{j},Q)}_{U}=T_{i,j,\beta_{j},\nu_{j}}^{(L_{i},Q)}-d_{s}-\left(L_{i}-1\right)\tau.
\label{stallTime_Q}
\end{equation}
This expression is used to derive tight bounds on the QoE metrics. By simplifications and some algebraic manipulation,the following theorems can be derived.
%
%Using these expressions, the following theorem summarizes the stall duration tail probability for file $i$. We include the edge router index $\ell$ in all the expressions in the result for the ease of using it in the following section.
\begin{theorem}
	\label{MSD_qlty}
	The mean stall duration for video
	file $i$ streamed with quality $Q$ requested through edge router $\ell$ is bounded by
\begin{equation}
	\mathbb{E}\left[\Gamma^{(i,\ell,Q)}\right]\leq\frac{1}{g_{i}}\text{log}\left(\sum_{j=1}^{m}\pi_{i,j}^{(\ell)}\left(1+M_{D}^{(i,j,\ell)}\right)\right) \label{eq:T_s_main_stall_qlty}
\end{equation}
where:	\begin{align}
M_{D}^{(i,j,\ell,Q)} & =  \widetilde{c}_{\ell}+\widetilde{a}_{\ell}\,e^{g_{i}(d_{s}+(L_{i}-1)\tau)}+\overline{a}_{\ell}\,\times\nonumber\nonumber \\
& \sum_{\nu_{j}=1}^{e_{j}}p_{i,j,\nu_{j},{\ell}}^{(Q)}\sum_{\beta_{j}=1}^{d_{j}}q_{i,j,\beta_{j},{\ell}}^{(Q)}e^{g_{i}L_{i}\tau}\times\nonumber \\
& \Biggl(\frac{\widetilde{M}_{j,\nu_{j},\ell}^{(e,Q)}(g_{i})(1-\rho_{j,\beta_{j}}^{(e)})g_{i}((\widetilde{M}_{j,\nu_{j},\ell}^{(e.,Q)}(g_{i}))^{L_{j,i}}-1)}{(g_{i}-\Lambda_{j,\beta_{j}}^{(e)}(B_{j,\beta_{j},\ell}^{(e)}(g_{i})-1))(\widetilde{M}_{j,\nu_{j},\ell}^{(e,Q)}(g_{i}))-1)}\nonumber \\
& +\frac{(1-\rho_{j,\beta_{j}}^{(\overline{d})})g_{i}(\widetilde{M}_{j,\nu_{j},\ell}^{(\overline{d},Q)}(g_{i}))^{L_{j,i}-L_{i}}}{g_{i}-\Lambda_{j,\beta_{j}}^{(\overline{d})}(B_{j,\beta_{j},\ell}^{(\overline{d})}(g_{i})-1)}+\nonumber \\
& \frac{(1-\rho_{j,\beta_{j}}^{(d)})g_{i}(\widetilde{M}_{j,\beta_{j},\ell}^{(\overline{d},Q)}(g_{i}))^{L_{i}+1}}{\left[g_{i}-\Lambda_{j,\beta_{j}}^{(d)}(B_{j,\beta_{j},\ell}^{(d)}(g_{i})-1)\right](\widetilde{M}_{j,\beta_{j},\ell}^{(d,Q)}(g_{i}))^{L_{j,i}+1}}\times\nonumber \\
& \Biggl(\frac{(\widetilde{M}_{j,\beta_{j},\ell}^{(d,\overline{d},Q)}(g_{i}))^{L_{i}-L_{j,i}}-(L_{i}-L_{j,i})}{(\widetilde{M}_{j,\beta_{j},\ell}^{(d,\overline{d},Q)}(g_{i}))-1}+\nonumber \\
& \frac{\widetilde{M}_{j,\beta_{j},\ell}^{(d,\overline{d},Q)}(g_{i})\left((\widetilde{M}_{j,\beta_{j},\ell}^{(d,\overline{d},Q)}(g_{i}))^{L_{i}-L_{j,i}-1}-1\right)}{(\widetilde{M}_{j,\beta_{j},\ell}^{(d,\overline{d},Q)}(g_{i}))-1}\Biggr)\Biggl)
\label{eq:mgf_bound_qlty}
\end{align}
and $\widetilde{c}$, $\widetilde{a}$, and $\overline{a}$ are defined earlier in Section \ref{sec:StallTailProb}, e.g., equation \eqref{E_stall_tot}.

\end{theorem}

\begin{theorem}
	\label{tailLemmaQlty}
	The stall duration tail probability for video
	file $i$ requested at quality $Q$ and through edge router $\ell$ is bounded by
\begin{align}
	& \text{\text{Pr}}\left(\Gamma_{tot}^{(i,\ell,Q)}\geq\sigma\right)\leq\nonumber\\
	& \sum_{j=1}^{m}\pi_{i,j,\ell}^{(Q)}\times\Biggl[\overline{c}_{\ell}+\widetilde{a}_{\ell}\,e^{-h_{i}\sigma}+\overline{a}_{\ell}\sum_{\nu_{j}=1}^{e_{j}^{(\ell)}}p_{i,j,\nu_{j},\ell}^{(Q)}\times\nonumber\nonumber\\
	& \sum_{\beta_{j}=1}^{d_{j}}q_{i,j,\beta_{j},\ell}^{(Q)}e^{h_{i}L_{i}\tau}\times\left(\delta^{(e,\ell,Q)}+\delta^{(\overline{d},\ell,Q)}+\delta^{(d,\overline{d},\ell,Q)}\right)\label{deltaAllQ}
	\end{align}
	for  $\rho_{j,\beta_{j}}^{(d)}<1$, $\rho_{j,\beta_{j},\ell }^{(\overline{d})}<1$, $\rho_{j,\nu_{j},\ell}^{(e)}<1$, 
	where the auxiliary variables in the statement of the Theorem are similarly defined as those in  equations \eqref{deltae}-\eqref{eq:abar}.
%	Theorem \ref{tailLemmaQlty}.
	\end{theorem}

Having derived the MSD and SDTP, one can formulate a constrained optimization problem to jointly optimize a convex combination of all QoE metrics as follows

\begin{eqnarray}
& \ensuremath{\text{min}\,\,\,\sum_{\ell=1}^{R}\sum_{i=1}^{r}\frac{\lambda_{i,\ell}}{\overline{\lambda}_{i}}\left[\theta_{1}\left(\sum_{Q=1}^{V}-b_{i,\ell}L_{i}a_{\ell}\right)+\right.}\nonumber\nonumber \\
& \ensuremath{\left.\sum_{Q=1}^{V}\frac{b_{i,Q}}{t_{i}}\left(\theta_{2}\times\text{Pr}(\Gamma^{(i,\ell,Q)}\geq\sigma)+\theta_{3}\times\mathbb{E}\left[\Gamma^{(i,\ell,Q)}\right]\right)\right]}\nonumber \\ \label{jointOpt}\\
\text{\textbf{s.t.}}\nonumber \\
& \eqref{constQ_1}-\eqref{constQ_last}\\
& \theta_{1}+\theta_{2}+\theta_{3}=1\\
\nonumber 
\end{eqnarray}

To solve this porblem, we still have to use alternating minization based approaches since the problem is not jointly convex in all the optimized parametes. Thus, we propose an iterative algorithm (similar to that explained in \ref{opt_mention}) to solve the
problem. The proposed algorithm divides the problem into sub-problems that optimizes one variable at a time while fixing the other. We refer the interested reader to \cite{Abubakr_SPCM2}  for detailed treatment of this problem.\\

We also note that similar methodology can be used to handle the scenarios where the chunks have different sizes.  If the video files have different chunk size (in MB), our analysis can be easily extended to handle such cases as follows. Since a video file has different chunk sizes, the service time will be different from one chunk to another. However,  one can still get the MGF of the service time in a similar fashion to those in   \eqref{MGF_Q_1}-\eqref{Me_jnujQ}. Thus, the service time of a chunk will be related to its size. For example, for a chunk indexed by $\kappa$ and requested from the PS $PS^{(d,j,\kappa)}_{\beta_j}$, the MGF of the servce time will be $M_{j,\beta_{j}}^{(d,\kappa)}$. Hence, similar formula to that in \eqref{eq:T_s_main_stall_qlty} for the stall duration under different sizes for the chunks can be obtained.

%}

%\newpage
%\input{downLoadPlayTimeApdx}
%\pagebreak{}
%\newpage
%\clearpage
%\input{apdx2}

\end{document}